\documentclass[11pt,letterpaper]{article}
\pdfoutput=1
\usepackage[top=3cm, left=2.5cm, bottom=3cm, right=2.5cm, includefoot,letterpaper]{geometry}
\usepackage{amsmath,amsfonts,graphicx,epsfig}
\usepackage{multirow}
\usepackage[normalem]{ulem}
\usepackage{pxfonts}

\DeclareMathAlphabet   {\mathsc}{OT1}{cmr}{m}{sc}

\def\[{\left [}
\def\]{\right ]}
\def\({\left (}
\def\){\right )}

\newcommand{\lbr}{\left\{}
\newcommand{\rbr}{\right\}}
\newcommand{\beq}{\begin{equation}}
\newcommand{\eeq}{\end{equation}}
\newcommand{\bea}{\begin{eqnarray}}
\newcommand{\eea}{\end{eqnarray}}

\newcommand{\GeV}      {~\mathrm{GeV}}
\newcommand{\TeV}      {~\mathrm{TeV}}

\newcommand{\EW}       {\mathsc{ew}}

\newcommand{\SUSY}     {\mathsc{susy}}

\newcommand{\gappeq}{\mathrel{\rlap {\raise.5ex\hbox{$>$}}
{\lower.5ex\hbox{$\sim$}}}}
\newcommand{\lappeq}{\mathrel{\rlap{\raise.5ex\hbox{$<$}}
{\lower.5ex\hbox{$\sim$}}}}

\def\igraph#1{\vtop{\null\hbox{\includegraphics[height=46mm]{figures/#1}}}}
\def\igraphnew#1{\vtop{\null\hbox{\includegraphics[height=65mm]{figures/#1}}}}

\begin{document}

\author{\textbf{Baris~Altunkaynak}, \textbf{Brent~D.~Nelson}  \\
{\small Department of Physics, Northeastern University,
Boston, MA 02115}
\and
\textbf{Lisa~L.~Everett}, \textbf{Yongyan~Rao}\\
{\small Department of Physics, University of Wisconsin,
Madison, WI 53706}
\and
\textbf{Ian-Woo~Kim}\\
{\small Department of Physics, University of Michigan,
Ann Arbor, MI 48109}\\
{\small Department of Physics, University of Wisconsin,
Madison, WI 53706}}

\title{\emph{Landscape of Supersymmetric Particle Mass Hierarchies  in Deflected Mirage Mediation}}


\maketitle

\abstract{With the aim of uncovering viable regions of parameter space in deflected mirage mediation (DMM) models of  supersymmetry breaking, we study the landscape of particle mass hierarchies for the lightest four non-Standard Model states for DMM models and compare the results to that of minimal supergravity/constrained MSSM (mSUGRA/CMSSM) models, building on previous studies of Feldman, Liu, and Nath. Deflected mirage mediation is a string-motivated scenario in which the soft terms include comparable contributions from gravity mediation, gauge mediation, and anomaly mediation.  DMM allows a wide variety of phenomenologically preferred models with light charginos and neutralinos, including novel patterns in which the heavy Higgs particles are lighter than the lightest superpartner.  We use this analysis to motivate two DMM benchmark points to be used for more detailed collider studies.    One model point has a higgsino-dominated lightest superpartner and a compressed yet heavy spectrum, while the other has a  stau NLSP and similar features to mSUGRA/CMSSM models, but with a slightly less stretched spectrum.}

\section{Introduction}
\label{sec:intro}

In the next decade, the hypothesis that physics beyond the Standard Model (SM) is given by TeV-scale softly broken supersymmetry (for reviews, see \cite{Martin:1997ns,Chung:2003fi}) will soon face unprecedented experimental tests at the Large Hadron Collider (LHC).  These tests will have important implications for the supersymmetry breaking sector of these theories, which include the minimal supersymmetric standard model (MSSM) and its extensions.
The parameter space associated with supersymmetry breaking is known to be vast; under the general (and phenomenologically preferable) assumptions of minimal flavor violation and CP conservation, there are approximately 30 parameters in the MSSM that are expected to be of relevance for collider physics.  One approach is to undertake a general analysis of this large parameter space, as was done recently by Berger {\it et al.}~\cite{Berger:2008cq}, or to perform a model-independent study of the LHC phenomenology of low energy supersymmetry in which all mass parameters are taken as free inputs at the weak scale \cite{Konar:2010bi}.  Another (standard) approach is to build theoretical models of the supersymmetry breaking terms and select benchmark parameter points for further detailed study, as in e.g.~\cite{Allanach:2002nj}. 

To this end, in previous work a subset of us proposed a class of string-motivated supersymmetry breaking models known as {\it deflected mirage mediation} (DMM) models \cite{Everett:2008qy,Everett:2008ey,Nakamura:2008ey}.  Deflected mirage mediation is a generalization of the well-known mirage mediation framework  \cite{mirage,Choi:2005uz,miragepheno} to include the effects of gauge mediation. Hence, it is a ``mixed" supersymmetry breaking scenario in which the three standard mediation mechanisms  -- gravity mediation, gauge mediation, and anomaly mediation --  are all present and provide comparable contributions to the soft terms.  DMM models in general have a larger parameter space than that of models such as minimal supergravity/constrained MSSM (mSUGRA/CMSSM) models, but with certain model assumptions, DMM can have only a few additional parameters, making it a simple and practical framework for benchmark studies.   In previous studies of DMM phenomenology  \cite{Choi:2009jn,Holmes:2009mx,Altunkaynak:2010xe}), the benchmark points were motivated primarily within a top-down approach, e.g.~by choosing parameter values suggested by what might be expected in an underlying string embedding.  However, to get a more complete picture of the viable regions of DMM parameter space, we also wish to seek a more ``bottom-up" approach for selecting DMM benchmark models.

In this paper, we explore this idea by employing an alternative classification of supersymmetry breaking models in which one studies the landscape of phenomenologically allowed mass patterns of the lightest four new particles (superpartners and/or additional Higgs states) of the theory.  This strategy was used by Feldman, Liu, and Nath (FLN) and applied primarily to mSUGRA/CMSSM models in a series of papers \cite{Feldman:2007zn,Feldman:2007fq,Feldman:2008hs}, and has also been employed in the general MSSM  \cite{Berger:2008cq}.  As with all landscape studies, the results are of course quite sensitive to the (unknown {\it a priori}) measure that is imposed on the parameter space ({\it i.e.}, flat priors, log priors, etc.), and hence the results must be considered with a grain of salt.  Nevertheless, the approach has utility in that it provides an alternative to top-down methods to seeking novel benchmark points for further studies by considering the notion of ``preferred" regions of parameter space based solely on phenomenological constraints.  As an example, the FLN studies showed that even in mSUGRA/CMSSM models,  there is a significant diversity of possible mass hierarchy patterns among the lightest four states, which in turn results in a broader diversity of collider phenomenology than might have been anticipated by considering standard benchmark points.

Hence, we here follow the general procedure of FLN and enumerate the light mass hierarchy patterns for a specific subset of DMM models after imposing a series of phenomenological bounds, and compare the outcome when relevant to mSUGRA/CMSSM models.   As the LHC has made great progress in setting limits on the parameter space of supersymmetric models just in the past year, in this current version of our work we incorporate these exciting developments.  Although our primary focus is on  repeat the analysis for mSUGRA/CMSSM models both to reflect updated bounds and also because we use {\tt SoftSUSY} \cite{softsusy} while the FLN studies employed {\tt SuSpect} \cite{suspect}. These codes differ in their treatment of the radiative corrections of the chargino and neutralino sectors, with {\tt SoftSUSY} including the full set of radiative corrections and {\tt SuSpect} using an approximate treatment.  Although the differences between the codes are negligible in many case, the differences do matter when there is a high degree of degeneracy in these sectors and the precise mass ordering is affected. The cuts we use include current direct collider bounds on the superparticles and Higgs bosons (including recent LHC limits on the masses of colored superpartners), indirect bounds (including the updated bound on $B_s\rightarrow \mu^+\mu^-$), and several options for the dark matter constraints, including the most recent bounds from WMAP7.  

 The results of our DMM analysis show several striking features that were not known from previous phenomenological studies of this framework, which were motivated primarily from the top-down perspective.  A main result is the emergence of a hierarchy pattern most preferred by dark matter-preferred constraints, which we call the ``Higgs LSP" pattern just for notational purposes, though the true lightest superpartner (LSP) is still of course the lightest neutralino.    In this pattern, the lightest superpartner (LSP) is a ``well-tempered"/mixed composition neutralino with a mass of the order of the TeV scale, the heavy Higgs bosons are lighter than the LSP, and the colored superpartners have masses at the 2-3 TeV scale.  The resulting squeezed yet heavy spectrum, which is also found in another prevalent DMM hierarchy pattern in which the next-to-lightest superpartner (NLSP) is a chargino,  is characterized by a relatively low value of the DMM gaugino mirage unification scale.  Another class of prominent DMM hierarchy patterns more closely resemble typical patterns found in mSUGRA/CMSSM models, in which the LSP is bino-like, and the lightest stau is often the NLSP.  Such patterns are characterized by higher values of the mirage unification scale and thus more stretched spectra than the Higgs LSP and chargino NLSP spectra, but they are generally not as stretched as typical mSUGRA/CMSSM spectra (for which the effective mirage unification scale is just the true unification scale of $M_G\sim 10^{16} \GeV$).  Each of these general categories of patterns tends to favor certain ranges of the underlying DMM parameter space, and has shown that some of the specific top-down motivated parameter values are not the preferred values from the perspective of this study.  In particular, the model points with a gaugino mirage unification scale in the TeV range that were a focus of previous top-down motivated studies of DMM  phenomenology,  though a very small subset of all possibilities, are not particularly favored within this landscape approach.

 This paper is structured as follows.  In Section 2, we provide a brief review of deflected mirage mediation.  We discuss the specific cuts and describe our methodology in Section 3, and present our results in Section 4.  We begin with an analysis of the impact of selected cuts and a characterization of the NLSP's in Section 4.1.  In Section 4.2, we present the detailed classification of the mass hierarchy patterns for the lightest four new states for a subset of DMM models, and compare the results with mSUGRA/CMSSM models.  In Section 4.3, we discuss the physics of several of the most prevalent hierarchy patterns within DMM models, focusing in particular on two cases: (1) the Higgs LSP and other DMM hierarchy patterns with higgsino-dominated LSP's and compressed but heavy spectra, and (2) stau NLSP patterns that resemble mSUGRA/CMSSM hierarchy patterns.  We present benchmark points for the two  general cases.  In Section 5, we turn to our summary and conclusions.

\section{Theoretical Background}
\label{sec:hier}
We consider two classes of models, minimal supergravity and deflected mirage mediation.  The two scenarios have the common feature that with certain assumptions, they satisfy the general criteria of minimal flavor violation (MFV) and CP conservation, such that the relevant soft breaking terms have the following general forms at the electroweak scale:
\begin{itemize}
\item Gaugino masses:  $M_a$, $a=1,2,3$.

\item Third generation trilinear couplings: $A_t$, $A_b$, and $A_\tau$.

\item Third generation scalar masses:  $m^2_{Q_3}$,  $m^2_{u_3}$,  $m^2_{d_3}$,  $m^2_{L_3}$,  and $m^2_{e_3}$.

\item Light generation scalar masses: $m^2_{Q}$,  $m^2_{u}$,  $m^2_{d}$,  $m^2_{L}$,  and $m^2_{e}$.
\item Higgs scalar masses $m^2_{H_u}$, $m^2_{H_d}$.
\end{itemize}
Here the Yukawa couplings of the light generations and the neutrino sector are neglected for simplicity and the magnitude of the supersymmetric Higgs mass parameter $\mu$ and the associated soft supersymmetry breaking term $b=B\mu$ are fixed in terms of $m_Z$ and $\tan\beta$.  Throughout our analysis, we will fix the sign of $\mu$ to be positive ($\mu>0$).  

MSUGRA/CMSSM models are particularly straightforward models of this type.  Supersymmetry breaking is assumed to arise in the observable sector at the traditional grand unification (GUT) scale of $M_{G} = 2 \times 10^{16} \GeV$, and the soft breaking terms are fixed at $M_G$  as follows:
a unified gaugino mass $m_{1/2}$, a unified trilinear scalar coupling $A_0$, and a unified scalar mass $m_0$,
The mass parameters $m_0$ and $m_{1/2}$ range from the electroweak to TeV scale, while $A_0$ can range from zero to the TeV scale.  The soft breaking terms take on the general forms described above upon renormalization group evolution to the electroweak scale.  The resulting low energy phenomenology has been studied extensively in the literature (see e.g. the reviews \cite{Martin:1997ns,Chung:2003fi}).  

In contrast, in deflected mirage mediation, the parameters are governed by two mass scales: an overall scale for the soft terms (in contrast to the two separate scales governing the gaugino and sfermion masses in mSUGRA) and the messenger scale associated with gauge mediation. Under certain assumptions,  deflected mirage mediation models can be in accordance with minimal flavor violation and the soft terms can be CP conserving.  To see this more clearly, we will now provide a brief review of DMM models (we refer the reader to \cite{Everett:2008qy,Everett:2008ey} for more details).

The deflected mirage mediation framework is a string-motivated supergravity framework in which gravity mediation, anomaly mediation, and gauge mediation all can contribute sizably to the soft terms.  The scenario is motivated by the Kachru-Kallosh-Linde-Trivedi (KKLT) \cite{kklt} approach to moduli stabilization in Type IIB string theory and the resulting phenomenological scenario of mirage mediation \cite{mirage}, in which it was demonstrated that the anomaly-mediated terms are comparable to the gravity-mediated terms resulting from the K\"{a}hler modulus associated with the compactification volume.  DMM generalizes mirage mediation to include effects of gauge mediation, which occur due to the generic presence of an additional SM singlet that couples to vectorlike messenger pairs.  As shown in \cite{Everett:2008qy,Everett:2008ey}, depending on the details of the stabilization of this singlet field, the resulting gauge-mediated terms can be comparable to the anomaly-mediated and modulus/gravity-mediated terms.

Deflected mirage mediation derives its name, as does mirage mediation, from the phenomenon that the gauginos appear to unify not at the GUT scale, but at a lower scale called the mirage unification scale (in mirage mediation, this also occurs for the scalar masses).  The ``deflection"  occurs because  the mirage unification scale can slide up or down for fixed $\alpha_m$.  As we will see, the mirage unification scale plays an important role in governing the mass spectrum of DMM models, since it fixes the low energy values of the gaugino mass ratios.  The general expressions for the soft supersymmetry breaking terms in deflected mirage mediation have been presented in \cite{Everett:2008qy,Everett:2008ey,Altunkaynak:2010xe}.   However, since we will consider only a particular subset of the possible parameter space, here we will rewrite them in a slightly more transparent form.   

We first will comment about the model parameters in DMM.   The dimensionful parameters are
a common mass scale $M_0$, which sets the overall scale of the soft terms and 
the messenger scale $M_{\rm mess}$, which can range from the $10-100$ TeV scale to the GUT scale.
The dimensionless DMM parameters are
the ratio of anomaly mediation to gravity mediation $\alpha_m$, the ratio of gauge mediation to anomaly mediation $\alpha_g$, the number of messenger pairs $N$, and the modular weights $\{n_i\}$ of the MSSM multiplets (as well as $\tan\beta$ and the sign of $\mu$).  The theoretically motivated ranges for $\alpha_m$ and $\alpha_g$ are $0\leq \alpha_m \leq 2$ and  $-1\leq \alpha_g \leq 2$. The messengers are assumed to be five-plets under a unified $SU(5)$ gauge symmetry that includes the SM gauge group.
The modular weights, which describe the matter field couplings to the K\"{a}hler modulus, can in principle be flavor and generation-dependent.  To avoid this issue and to reduce the number of independent parameters, we will fix the modular weights to the following standard values: $n_i=1/2$ for the matter fields and $n_i=1$ for the Higgs fields.\footnote{This was also done in \cite{Everett:2008qy,Everett:2008ey,Altunkaynak:2010xe}.  See  \cite{Choi:2005uz} for examples in which the modular weights are varied in mirage mediation.}   It is this subset of DMM models that we will focus on in this work.  Note, however, that it is not necessary to fix the modular weights of the MSSM matter fields to universal values to avoid flavor-changing effects.  Since the scalar masses depend in detail on the modular weights, whether one makes our specific choice or allows for more general possibilities will have an important impact on the outcome for the preferred hierarchy patterns.  We will defer the study of alternate choices of modular weights and their resulting effect on the DMM landscape for future work.

Under these assumptions, and defining the quantity
\begin{equation}
\widetilde{\alpha}_m=\alpha_m \left (\frac{1}{16\pi^2} \ln\frac{M_P}{m_{3/2}} \right),
\end{equation}
where $M_P$ is the Planck mass and $m_{3/2}$ is the gravitino mass, the gaugino masses are 
\begin{eqnarray}
M_a(M_G)=M_0[1+g_0^2(b_a+N)\widetilde{\alpha}_m],
\label{gauginomasses}
\end{eqnarray}
and the threshold corrections are given by  $M_a (M_{\rm mess}^-) = M_a(M_{\rm mess}^+)
+ \Delta M_a$, in which
\begin{eqnarray}
 \Delta M_a=  - N M_0 g_a^2(M_{\rm mess}) \widetilde{\alpha}_m(1+\alpha_g). \label{gauginothresh}
\end{eqnarray}
In the above,  $g_0=g_a(M_G)$ is the unified gauge coupling at $M_G$, and $b_a$ are the MSSM beta functions ($b_3=-3$, $b_2=1$, $b_1=33/5$).   The trilinear couplings for the third generation, which receive no threshold corrections, are given by
\begin{eqnarray}
A_{t,b,\tau}(M_G)=M_0 [1+b_{t,b,\tau} \widetilde{\alpha}_m],
\label{aterms}
\end{eqnarray}
in which $b_{t,b,\tau}$ are given by $b_t=6y_t^2+y_b^2-\frac{16}{3} g_3^2-3 g_2^2-\frac{13}{15}g_1^2$, $b_b=6y_b^2+y_t^2+y_{\tau}^2- \frac{16}{3} g_3^2-3 g_2^2-\frac{7}{15}g_1^2$, and $b_{\tau}=3y_b^2+4y_{\tau}^2-3 g_2^2-\frac{9}{5}g_1^2$,
where the third generation Yukawas $y_{t,b,\tau}$ and the  gauge couplings are all evaluated at the GUT scale.  The soft scalar mass-squares of the matter ($m^2_{\widetilde{f}_i}$) and Higgs fields at the GUT scale are given by
\begin{eqnarray}
m^2_{\widetilde{f}_i}(M_G)&=& M_0^2\left [\frac{1}{2}-\theta_{\widetilde{f}_i}\widetilde{\alpha}_m -\dot{\gamma}_{\widetilde{f}_i}^\prime\widetilde{\alpha}_m^2  \right ]\nonumber \\
m^2_{H_{u,d}}(M_G)&=& -M_0^2\left [\theta_{H_{u,d}}\widetilde{\alpha}_m +\dot{\gamma}_{H_{u,d}}^\prime \widetilde{\alpha}_m^2 \right ],
\label{scalarmasses}
\end{eqnarray}
in which
$\theta_i = 4 \sum_a g_a^2 c_a^{(i)} - \sum_{i,j,k} |y_{ijk}|^2$
and 
$\dot{\gamma}_i'=2\sum_a g_a^4(b_a+N) c_a^{(i)} - \sum_{lm} |y_{ilm}|^2b_{y_{ilm}}$,
where $c_a^{(i)}$ are the quadratic Casimirs and $y_{ijk}$ denote the relevant Yukawa couplings, which are also evaluated at $M_G$.  The soft scalar mass-squared parameters have threshold corrections, $m_i^2 ( M_{\rm mess}^- ) = m_i^2 (M_{\rm mess}^+) + \Delta m_i^2$, which are
\begin{eqnarray}
\Delta m^2_{i}=M_0^2\sum_a 2 c_a^{(i)}N g_a^4(M_{\rm mess})\widetilde{\alpha}_m(1+\alpha_g).
\label{msqrthresh}
\end{eqnarray}
We list  these corrections, as well as the $b_{t,b,\tau}$, $\theta_i$, and $\dot{\gamma}'_i$ parameters, in the Appendix.

Eqs. (\ref{gauginomasses}), (\ref{aterms}), and (\ref{scalarmasses}) show that when $\alpha_m\rightarrow 0$, the anomaly mediation contributions vanish.  The gauge mediation contributions, which enter the threshold corrections of Eqs.~(\ref{gauginothresh}) and (\ref{msqrthresh}), clearly vanish when $N\rightarrow 0$.  If both  $\alpha_m$ and $N$ are zero, the theory is thus a modulus/supergravity theory  (akin to dilaton domination \cite{Brignole:1993dj}) that is more restrictive than mSUGRA/CMSSM.  In this limit the gaugino masses and $A$ terms are unified, the MSSM matter fields have a common scalar mass that is related to the gaugino masses, and the Higgs soft mass-squares vanish ({\it i.e.}, it is a subset of non-universal Higgs (NUH) extensions of mSUGRA/CMSSM models).  Hence, although the subset of DMM models considered here have a larger parameter space,  they do not contain mSUGRA/CMSSM as a specific limit.

The threshold corrections that result from integrating out the messengers have significant effects on the low energy spectrum.  For nonzero $N$, these corrections are governed by  $\alpha_g$.  The threshold corrections from gauge mediation and anomaly mediation interfere destructively when $\alpha_g<0$, with perfect cancellation when $\alpha_g=-1$. Even if $\alpha_g=0$, the effects of the messengers still are present if $N\neq 0$.

We close this section with a discussion of mirage unification in DMM models, which plays a primary role in governing the superpartner masses.  The DMM mirage unification scale, which is the scale at which the gauginos appear to unify, is
\begin{equation}
\label{miraged}
M_{\rm mir}=M_G \left (\frac{m_{3/2}}{M_{Pl}}\right )^{\alpha_m \rho/2},
\end{equation}
in which the parameter $\rho$ is
\begin{equation}
\label{rhodef}
\rho= \left (1+ \frac{2Ng_0^2}{16\pi^2}  \ln \frac{M_{\rm GUT} }{M_{\rm mess}}\right )
\left (1- \frac{\alpha_{\rm m} \alpha_{\rm g} Ng_0^2}{16\pi^2}
 \ln \frac{M_P}{m_{3/2}}\right )^{-1}.
\end{equation}
In the mirage mediation limit of $N=0$, the mirage unification scale is at an intermediate scale for the standard KKLT value of $\alpha_m=1$, and is at the TeV scale for $\alpha_m=2$.  For nonzero $N$, this scale can be pushed up or down depending on the magnitude and sign of $\alpha_g$.   In general, the lowering of the mirage unification scale from that of the pure supergravity limit of $M_G$ results in a more compressed spectrum in the gaugino sector, such that one expects DMM models to be characterized by distinctive gaugino mass ratios and a greater possibility of mixed-composition lightest neutralinos.  As we will see, it will also be useful to characterize the hierarchy patterns in this subset of DMM models according to the mirage unification scale.

\section{Methodology and Bounds}
\label{sec:bounds}
To get an adequate survey of the hierarchy patterns, we have generated large sets of model parameters and corresponding low energy spectra, then study how the patterns of the lightest four non-SM particles (superpartners and non-SM Higgs fields) vary as we impose a series of phenomenological bounds.   In this section, we will provide a detailed overview of our general procedure and methodology for this analysis.

The first step is the generation of models.   For the subset of DMM models under consideration, we generated a very large model set consisting of 24.75 million models.  The models were obtained by randomly scanning over the parameter ranges:
\begin{equation} \lbr \begin{array}{c} 1 \leq N \leq 5 \\ 4
\leq \log(M_{\rm mess}/{\rm GeV}) \leq 16 \\ 50 \GeV \leq M_0 \leq
2000 \GeV \\ 0 \leq \alpha_m \leq 2 \\ -1 \leq \alpha_g \leq 2
\\ 1 \leq \tan\beta \leq 60 \end{array} \rbr\, . \label{ranges}
\end{equation}
As described in the previous section, the ranges chosen for the dimensionless ratio parameters $\alpha_m$ and $\alpha_g$ are the theoretically-motivated ranges studied in previous literature \cite{Everett:2008ey,mirage}. For simplicity, flat priors were used for all parameters except the messenger scale, where we scanned over its logarithm with flat priors.  We also made the specific choice always to include effects of gauge mediation messenger pairs by restricting their number $N$ to take nonzero values.

For purposes of comparison,  we generated 2~million mSUGRA/CMSSM models by scanning input parameters from the ranges used by  FLN  \cite{Feldman:2007zn}) with flat distributions in the parameters:
\begin{equation} \lbr \begin{array}{c} 0 \leq m_0 \leq 4\TeV \\ 0
\leq m_{1/2} \leq 2 \TeV \\ |\frac{A_0}{m_0}| \leq 10 \\
1 \leq \tan\beta \leq 60 \end{array} \rbr \, . \label{mSUGRA}
\end{equation}
After generating our model sets, the parameters are then evolved to the electroweak symmetry-breaking scale using a modified\footnote{(The modifications are needed for DMM models due to the presence of the messenger scale; see \cite{Altunkaynak:2010xe} for further details.  They do not affect the analysis of the mSUGRA/CMSSM models.} version of {\tt SoftSUSY 3.0.7} \cite{softsusy} with a default scale of $m_{\EW} = \sqrt{m_{\tilde{t}_1} m_{\tilde{t}_2}}$.    Having obtained the parameters at the electroweak scale, we use {\tt SoftSUSY} and {\tt MicrOmegas} to determine the mass spectrum as well as collider and cosmological observables.

Before proceeding, we first note a well-known important difference between {\tt SoftSUSY}, which we use in this work, and {\tt SuSpect} \cite{suspect}, which was used by FLN and Berger et al., regarding the ways in which the chargino and neutralino sectors are incorporated.  {\tt SoftSUSY} includes the full radiative corrections to the charginos and neutralinos as given in the seminal paper of Pierce, Bagger, Matchev, and Zhang (PBMZ) \cite{pbmz}, while {\tt SuSpect} simply includes the radiative corrections in the gaugino and higgsino limits, as discussed in PBMZ.\footnote{In this approximation  \cite{pbmz}, the corrections are only made to the diagonal entries in the undiagonalized tree level chargino and neutralino mass matrices.  In addition, all loop masses are set to their diagonal values, the quark masses are set to zero, the squarks and sleptons are taken to be degenerate, and $m_h=m_Z$, $m_H=m_H^\pm=m_A$, as summarized in Figures 11-12 of PBMZ \cite{pbmz}.} 
  Hence, we expect that in situations in which the charginos and neutralinos are highly degenerate, the differences in the codes can result in different outcomes for the precise mass ordering of the lightest four non-SM states (and where there are differences, {\tt SoftSUSY} should be used since it incorporates the full set of corrections).  This is an important fact to keep in mind when comparing our results to previous landscape studies of this kind.

At this point we impose a sequence of phenomenological requirements on the candidate models. The first requirement is the presence of radiative electroweak symmetry-breaking, which in both model sets reduces the sample size by a large fraction.  The second requirement is that the lightest R-parity odd state (the LSP) is a neutralino.   As we will see, in this class of DMM models quite often the lightest non-SM particle is one of the heavier Higgs states.  As stated, we call such patterns ``Higgs LSP" patterns, but in such cases it is important to note that the true LSP is still the lightest R-parity odd particle.  

Our third requirement is a series of bounds that we will call the ``direct" bounds, in that they reflect conservative direct search limits for new states beyond the Standard Model. These constraints are summarized in Table~\ref{tbl:direct}.  Some are updated limits based on Djouadi {\it et al}.~\cite{Djouadi:2006be} and the FLN studies.  However, we include direct limits on the pseudoscalar and charged Higgs masses, and employ the direct limit on the SM Higgs boson mass of $m_h>114.4 \GeV$, though we also investigate how the results change with the tightening of the lightest Higgs mass by occasionally considering the weaker limit of $m_h>100 \GeV$ employed by FLN.  For the charginos, it is necessary to modify the standard LEP bound of $m_{\tilde{\chi}^{\pm}_1}>104.5 \GeV$ in the case of a squeezed spectrum, which often occurs in DMM models.  As discussed in \cite{Berger:2008cq}, the chargino bound degrades to 95 GeV in the case in which the lightest chargino and the LSP have a high degree of degeneracy ($m_{\tilde{\chi}^{\pm}_1}-m_{\tilde{\chi}^{0}_1} < 2 \GeV$).   

We have also included bounds on the gluino and squark masses from very recent LHC results.   Null results for superpartner searches  by ATLAS \cite{atlas1} and CMS \cite{cms} at $\sqrt{s}=7$ TeV with 35 pb$^{-1}$, as well as new ATLAS results with 165 pb$^{-1}$ \cite{atlas2} place strong constraints on SUSY models.  To obtain limits on SUSY parameters, they use the mSUGRA/CMSSM framework and study the $m_0-m_{1/2}$ plane for fixed $\tan\beta$, $A_0$, and the sign of $\mu$, or alternatively, consider simplified models with only first and second generation squarks, a gluino octet, and the lightest neutralino.  For example, the ATLAS data with 35 pb$^{-1}$ (165 pb$^{-1}$) result in $m_{\tilde{g}}>500 \GeV$ ($m_{\tilde{g}}>725 \GeV$)  at 95 \% confidence level for the simplified models, with the bound increasing to 870 GeV (1025 GeV) for degerate squarks and gluinos.  Within mSUGRA/CMSSM models, the corresponding ATLAS limits for degenerate squarks and gluinos are  775 GeV (950 GeV) for $\tan\beta =3$ ($\tan\beta=10$), $A_0=0$, and $\mu >0$.  

To interpret these bounds, it is important to keep in mind that previous studies based on the Tevatron gluino mass bound of $m_{\tilde{g}} \geq 309\GeV$ (obtained for mSUGRA/CMSSM models)~\cite{:2007ww} have shown the gluino bound is degraded when the gauginos are squeezed, as shown in \cite{Alwall:2008ve,Alves:2010za}. To account at least partially for this effect, we take a relatively conservative approach and restrict $m_{\tilde{g}}>800 \GeV$ and $m_{\tilde{q}}>800 \GeV$ (for the first and second generation squarks) in all cases.  To understand the impact of these new LHC bounds in our analysis, we will occasionally compare these results with those obtained 
with the prior Tevatron limits, for which the degradation of the bounds for a squeezed gaugino spectrum is taken into account by considering a weaker bound of $m_{\tilde{g}}>125 \GeV$ for cases with $m_{\tilde{g}}/m_{\tilde{\chi}_1^0}<5$.

\begin{table}[t]
{\begin{center}
\begin{tabular}{|l|c|}\cline{1-2}
Condition & Bound \\
\cline{1-2}
Higgs & $m_h>114.4 \GeV$ 
\\
Chargino* & $m_{\chi_1^{\pm}} > 104.5 \GeV$\\
Stop & $m_{\tilde{t}_1} > 101.5 \GeV$ \\
Stau & $m_{\tilde{\tau}_1} > 98.8 \GeV$ \\
%
%
Gluino & $m_{\tilde{g}} > 800 \GeV$ \\ 
1st+2nd Family Squarks & $m_{\tilde{q}}>800 \GeV$ \\
Pseudoscalar Higgs & $m_A > 85\GeV$ \\
Charged Higgs & $m_{H^{\pm}} > 79.3\GeV$ \\
\cline{1-2}
\end{tabular}
\end{center}}
{\caption{\label{tbl:direct}\footnotesize {\bf Direct (collider) mass bounds}.  In addition to the FLN limits, we take the SM Higgs mass limit of 114.4 GeV and include the pseudoscalar and charged Higgs mass limits.   The new LHC bounds on the gluino and 1st/2nd family squarks are also included; see the discussion in the text.  The asterisk indicates that the chargino limit is degraded to 95 GeV when the mass difference between the lightest chargino and the LSP is less than 2 GeV.}}
\end{table}
\begin{table}[t]
{\begin{center}
\begin{tabular}{|c|c|}\cline{1-2}
Condition & Bound \\
\cline{1-2}
$b \to s \gamma$ & $229 \times 10^{-6} \leq {\rm Br}(b\to s \gamma) \leq 481 \times 10^{-6}$ \\
$B_s \to \mu^+ \mu^-$ & ${\rm Br}(B_s \to \mu^+ \mu^-) < 4.0 \times 10^{-8}$ \\
$(g_{\mu} -2)_{\SUSY}$ & $-11.4\times10^{-10} \leq (g_{\mu}-2)_{\SUSY} \leq 9.4 \times 10^{-9}$ \\
\cline{1-2}
\end{tabular}
\end{center}}
{\caption{\label{tbl:indirect}\footnotesize{\bf Indirect bounds}. Our indirect limits include the recent update to ${\rm Br}(B_s \to
\mu^+ \mu^-)$ rate  \cite{Collaboration:2011fi}. 
We use the HFAG range for $b\rightarrow s \gamma$ \cite{Barberio:2007cr}, and the FLN and Djouadi range for $(g_{\mu}-2)_{\rm SUSY}$.  }}
\end{table}

Our fourth requirement involves indirect constraints arising from the rare decays $b\rightarrow s\gamma$ and $B_s\rightarrow \mu^+\mu^-$, as well as the new physics contributions to the
anomalous magnetic moment of the muon. For the inclusive $b\to s
\gamma$ rate, we consider the average value as
derived by the Heavy Flavor Averaging Group~\cite{Barberio:2007cr}
\begin{equation} {\rm Br}(b\to s \gamma) = (355\pm 24^{+9}_{-10} \pm 3) \times
10^{-6} \, .\label{HFAG} \end{equation}
As shown in Table~\ref{tbl:indirect}, the range we use is a $3.5\sigma$ range about the best fit value:
\begin{equation}
229 \times 10^{-6} \leq {\rm Br}(b\to s \gamma) \leq 481 \times 10^{-6},
\end{equation}
which was also used in \cite{Feldman:2008hs}.  
For $B_s \to \mu^+ \mu^-$, we use the very recent CDF bound~\cite{Collaboration:2011fi}:
%
\begin{equation} {\rm Br}(B_s \to \mu^+ \mu^-) < 4.0 \times 10^{-8}
 \, ,\label{Bmumu} \end{equation}
which is a 95\% confidence level upper bound. 
Finally, for the anomalous magnetic moment of the
muon, we constrain the new physics contribution to $g_\mu -2$
to lie within the conservative range of FLN and Djouadi {\it et al.}:
\begin{equation}
-11.4\times10^{-10} \leq (g_{\mu}-2)_{\SUSY} \leq 9.4 \times 10^{-9}
\label{amubound}. \end{equation}
%
\begin{table}[t]
{\begin{center}
\begin{tabular}{|c|c|}\cline{1-2}
Condition & Bound \\
\cline{1-2}
WMAP7 & $0.0997<\Omega_{\tilde{\chi}} h^2<0.1221$ \\
WMAP Preferred &$0.07<\Omega_{\tilde{\chi}} h^2<0.14$ \\
WMAP Upper & $\Omega_{\tilde{\chi}} h^2<0.121$ \\
\cline{1-2}
\end{tabular}
\end{center}}
{\caption{\label{tbl:darkmatter}\footnotesize{\bf Dark matter bounds}. We consider three options for the dark matter constraints: the WMAP7 $2\sigma$ range \cite{Larson:2010gs},  a broader ``WMAP Preferred" range, and the WMAP5 upper bound, as used in \cite{Berger:2008cq}. }}
\end{table}

Our final requirement is to impose that the
thermal relic density of neutralinos is consistent with the measured
cosmological density as inferred from the cosmic microwave
background.  It is standard to assume that the LSP neutralinos are the entirety of the dark matter.  The resulting relic abundance, which is determined by standard cosmology, is then compared to the current constraints from WMAP. 
WMAP has now reported  $1\sigma$ results from five~\cite{Dunkley:2008ie} and seven~\cite{Larson:2010gs} years of
observation:
\begin{eqnarray} \Omega_{\tilde{\chi}} h^2|_{\rm 5yr} = 0.1099 \pm 0.0062
\label{Ochi5yr}\, ,\\  \Omega_{\tilde{\chi}} h^2|_{\rm 7yr} = 0.1109 \pm
0.0056 \label{Ochi7yr}\, .
\end{eqnarray}
In imposing these bounds, it is important to note that the cosmological assumptions can be questioned: for example, the LSP may not be the only dark candidate, or there may be effects from non-standard cosmology that alter the relic abundance (such as in kination-dominated quintessence theories \cite{kination}).  Furthermore, the precision quoted in the WMAP results far out-paces that of the
theoretical tools available to calculate the value of $\Omega_{\tilde{\chi}}
h^2$ from first principles.  For our computation we use the package {\tt MicrOmegas 2.2 CPC}, 
but results from {\tt DarkSUSY} often give values that differ by more than the quoted errors in the WMAP
measurements (for example, due to uncertainties in the halo models).  

For these reasons, we will consider three options for the dark matter constraints, which are summarized in Table~\ref{tbl:darkmatter}.  The first is the $2\sigma$ range from WMAP7 \cite{Larson:2010gs}:
\begin{equation}
0.0997<\Omega_{\tilde{\chi}} h^2<0.1221.
\label{DM7}
\end{equation}
The second option is to consider a broader range that we call ``WMAP Preferred'':
\begin{equation} 0.07 \leq \Omega_{\tilde{\chi}} h^2 \leq 0.14\, ,
\label{DMwide} \end{equation}
which is slightly more expansive than the WMAP3 $2\sigma$ range utilized by FLN of
\begin{equation} 0.0855 \leq \Omega_{\tilde{\chi}} h^2 \leq 0.121.
\label{DMBA} \end{equation}
The third option, which is even less restrictive, is to follow the procedure of \cite{Berger:2008cq}, and only impose an upper limit on the relic abundance obtained from WMAP5:
\begin{equation}
\Omega_{\tilde{\chi}} h^2 \leq 0.121.
\label{DMupper}
\end{equation}
In our analysis, we will primarily focus on the ``WMAP Preferred'' range, both for statistics reasons and also to get a sense on the resulting constraints on the model sets due to both upper and lower bounds on the dark matter relic abundance.  However, we will also investigate the WMAP5 upper bound and the WMAP7 limits for purposes of comparison.

\section{Hierarchy Patterns}
\subsection{Hierarchy Patterns in DMM and mSUGRA/CMSSM Models}
\label{sec:hierarchy}

We now discuss the allowed hierarchy patterns of the lightest four non-SM states within the subset of DMM models under consideration, which are specified entirely by the six parameters $\{N,\, M_{\rm mess},\, M_0,\, \alpha_m,\, \alpha_g,\, \tan\beta\}$ with the ranges specified in Eq.~(\ref{ranges}).\footnote{Recall that we set the modular weights to fixed values, as discussed in Section~\ref{sec:hier}, and we take $\mu >0$.}  These results will be compared to the subset of mSUGRA/CMSSM models as specified by $\{m_0,\,m_{1/2},\, A_0,\, \tan\beta\}$  with the ranges of Eq.~(\ref{mSUGRA}), though we must once again emphasize that as with any landscape study, interpreting such comparisons should be done with great care, since the impact of the sample sizes as well as the choice of priors and parameter ranges can be very different in the two cases.   First we will investigate the impact of different phenomenological bounds and the distribution of the NLSP's.  Second, we will present the enumeration of the most prevalent hierarchy patterns in each case, focusing on the impact of the different dark matter bounds.  

The first object of our consideration is the impact of a progressive application of the phenomenological bounds on the number of surviving models and the distribution of the NLSP's.  We begin with the simple requirement of  proper electroweak symmetry breaking, and then add the requirement of a neutralino LSP.  Next, the direct and indirect bounds as given in Tables 1 and 2 are included, and finally one of the dark matter limits of Table 3. In Table 4, we present the number of cases for a particular NLSP (expressed in terms of percentage of the total number of surviving models) for our DMM and mSUGRA/CMSSM model sets that result from this progressive application of cuts with the WMAP Preferred limit of Eq.~(\ref{DMwide}).  Table 5 reproduces this analysis not only for the WMAP Preferred limit, but also the WMAP7 limit of Eq.~(\ref{DM7}) and the WMAP upper limit of Eq.~(\ref{DMupper}).

From these results, we see that for our set of DMM models, the electroweak symmetry breaking and neutralino LSP requirements result in a relatively equal distribution of neutralino, chargino, and stau NLSP's, as well as a smaller subset of models in which the heavy Higgs particles are lighter than the neutralino LSP -- this is what we call the ``Higgs LSP" pattern.  The application of the direct and indirect limits disfavor the chargino NLSP cases and favor the neutralino and stau NLSP cases.  As shown in Table 4, the WMAP preferred dark matter constraint then drastically alters this pattern, leading to the resurgence of chargino NLSP cases as well as the dramatic emergence of the Higgs LSP pattern as the dominant DMM pattern.  For the mSUGRA/CMSSM models,  we see that neutralino NLSP's (and to a lesser extent, stau NLSP's) are preferred by each progressive application of cuts except for the dark matter constraints, which again raise the relative number of chargino NLSP cases and, most dramatically, strongly prefer stau NLSP patterns.  Table 5 demonstrates that the WMAP Preferred and WMAP7 limits result in similar percentages of NLSP's, while the WMAP5 upper limit strongly prefers cases with chargino NLSP's for both the DMM and mSUGRA/CMSSM model sets.  The preference for chargino NLSP patterns that results from imposing only an upper limit on the dark matter relic abundance is as expected since the NLSP is typically a chargino when the LSP has strong wino or higgsino components, and in such cases the relic abundance is typically too low unless the LSP is sufficiently heavy (at or around the TeV scale).

\begin{table}
{\begin{center}\begin{footnotesize}
\begin{tabular}{|l||r|r||r|r||r|r||r|r|}\cline{1-9}
 & \multicolumn{2}{|c||}{Proper EWSB} &
\multicolumn{2}{|c||}{Neutralino LSP} & \multicolumn{2}{|c||}{Exper.
Bounds}
& \multicolumn{2}{|c|}{WMAP Preferred} \\
NLSP & DMM & mSUGRA & DMM & mSUGRA & DMM & mSUGRA & DMM & mSUGRA \\
\cline{1-9}
$\chi^0_2$ &19.9 & 50.8 & 25.6 & 66.1 & 36.7 & 67.8
& 17.2 & 28.7 \\
$\chi^{\pm}_1$ & 26.2
& 3.0 & 33.8 & 3.9 & 14.8 & 1.3
& 26.9 & 20.1 \\
$\tilde{\tau}_1$ & 18.6 & 19.2 & 24.0 & 25.0 & 35.4 & 27.1
& 13.9 & 46.1 \\
$\tilde{g}$ & 0.1 & 0.0 & 0.1 & 0.0 & 0.0 & -- & -- & -- \\
$\tilde{t}_1$ & 0.0 & 2.3 & 0.0 & 3.0 & 0.0 & 1.9 & 0.0 & 2.1 \\
%
%
$\tilde{\ell}_R$ & 0.2 & 0.0 & 0.3 & 0.0 & 0.4 & -- & 0.1 & -- \\
$H$ or $A$ & 1.4 & 1.2 & 1.8 & 1.6 & 1.6 & 1.6 & 2.7 & 2.8 \\
%
%
\cline{1-9}
``Higgs LSP" & 17.7 & 0.4 & 14.4 & 0.4 & 10.9 & 0.2
& 39.2 & 0.2 \\
%
%
Other LSP & 15.9 & 23.0 & -- & -- & -- & -- & -- & -- \\
\cline{1-9}
\end{tabular}
\end{footnotesize}\end{center}} {\caption{\label{tbl:dmmflow1}\footnotesize{\bf  Percentage of NLSP's and impact of progressive cuts.}
The percentage of NLSPs of the surviving DMM and mSUGRA/CMSSM (labeled here simply as ``mSUGRA") models for a progressive application of phenomenological bounds, including the direct and indirect limits as given in Tables 1-2 (grouped under ``Exper. Bounds") and the WMAP Preferred limit of Eq.~(\ref{DMwide}).  Our notation is that ``0.0" indicates a statistically insignificant set of models and ``--"  indicates that there were no surviving models.  The ``Higgs LSP" cases are those in which some or all of the heavy Higgs particles are lighter than the neutralino LSP, while the ``Other LSP" cases are those in which the LSP is not the lightest neutralino but instead another R-parity odd particle.}}
\end{table}
\begin{table}
{\begin{center}\begin{footnotesize}
\begin{tabular}{|l||r|r||r|r||r|r|}\cline{1-7}
 & \multicolumn{2}{|c||}{WMAP Preferred} &
\multicolumn{2}{|c||}{WMAP 7-Year} & \multicolumn{2}{|c|}{Upper
Bound Only} \\
NLSP &DMM &  mSUGRA & DMM & mSUGRA & DMM &mSUGRA \\
\cline{1-7}
$\chi^0_2$ & 17.2 & 28.7 & 20.0 & 28.8 & 8.9 & 24.6 \\
$\chi^{\pm}_1$ & 26.9 & 20.1 & 27.1 & 18.3 & 45.1 & 44.7 \\
$\tilde{\tau}_1$ & 13.9 & 46.1 & 16.7 & 48.7 & 8.2 & 27.1 \\
$\tilde{g}$ & -- & -- & -- & -- &  0.1 &-- \\
$\tilde{t}_1$ & 0.0 & 2.1 & 0.0 & 2.0 & 0.0 & 2.3 \\
%
%
%
$H$ or $A$ & 2.7 & 2.8 & 2.7& 2.2 & 2.2 & 1.4 \\
%
%
\cline{1-7}
Higgs LSP & 39.2 & 0.2 &  33.6 &--& 35.6 & 0.0 \\
\cline{1-7}
%
%
%
%
%
%
\end{tabular}
\end{footnotesize}\end{center}} {\caption{\label{tbl:dm114.4}\footnotesize{\bf
NLSP's for Different Dark Matter
Assumptions}.  The percentage of NLSP's for different dark matter limits as discussed in Section~\ref{sec:bounds}. The notation is as in the previous table.}}
\end{table}

Clearly, the imposition of a relic density constraint with both upper and lower bounds -- even a relatively generous one as in Eq.~(\ref{DMwide}) -- has a dramatic impact on the size of the allowed parameter space and on the nature of the hierarchy patterns that survive, indicating the sensitivity of this analysis to the standard cosmological assumptions inherent in these limits.  In the case of mSUGRA/CMSSM models, it is well known that most of the viable parameter space results in a bino-like LSP; co-annihilation channels or resonance effects are then needed to reduce the resulting abundance at freeze-out to acceptable levels, favoring scenarios with relatively light staus (or stops).  The exception is the hyperbolic branch/focus point region, in which the LSP has a nontrivial higgsino component -- as we will see, such cases have a chargino NLSP.  In contrast, in DMM models, we will see that while the LSP can be bino-like and thus require stau co-annihilation or another dilution effect,  it can instead be purely higgsino with a mass in the TeV range (to satisfy the lower limit on the relic abundance for the WMAP Preferred or WMAP7 limits), or it can be a mixture of gauginos and higgsinos, {\it i.e.},  a ``well-tempered" neutralino~\cite{ArkaniHamed:2006mb}.   The WMAP preferred DMM patterns of Table 4, which are the Higgs LSP and chargino NLSP patterns, are indeed scenarios with a higgsino dominated or mixed-composition LSP, as discussed in the next subsection.

As opposed to the mSUGRA/CMSSM case,  stau NLSP patterns are not particularly favored in our subset of DMM models, although a significant number of them remain.  One reason is that there are additional avenues in DMM for obtaining a roughly correct relic density, as mentioned previously.  Another reason is that the overall mass scale for the scalars and gauginos are controlled by the same mass scale $M_0$ in DMM and it is the gauginos that can be more easily deflected by threshold corrections to lower values.   Furthermore, in DMM models the trilinear couplings ($A$ terms) are not separately adjustable, which also affects the low energy spectrum.  In the mSUGRA/CMSSM case, the masses of the scalars and the gauginos are governed by two separately adjustable parameters ($m_0$ and $m_{1/2}$), and the left-right scalar mixing terms also have contributions from the independent parameter $A_0$.   For this reason, patterns in which sfermions are the NLSP are relatively disfavored within the subset of DMM models considered here.  We see this most strikingly in the paucity of stop NLSP patterns.   It is of course important to keep in mind that this does not preclude the possibility that viable DMM models have stop NLSP's,  as this should certainly be possible with alternative sets of modular weights.

Note that gluino NLSP patterns are absent once the dark matter cuts are imposed.
While the absence of gluino NLSPs is expected in mSUGRA/CMSSM, one might think that this would not be the case in DMM, given the generic possibility of relatively light
gluinos \cite{Altunkaynak:2010xe}.   While it is indeed true that
lighter gluinos are more prevalent in DMM, it is difficult to have
the gluino as the NLSP. One reason is of course the new LHC bound on the gluino mass; as this bound increases, clearly this possiblity diminishes.  However, even with the weaker Tevatron limits on the gluino mass, this feature persists.  In this case, the primary reason is that it requires a precise conspiracy of parameters to
make the gluino lighter than the lightest chargino and the second-lightest neutralino, {\it i.e.}, a mirage unification scale
at very particular (very low scale) values.   Gluino NLSP patterns typically
predict too low of a relic abundance because the LSP would then be is generically
a mixed-composition state with a relatively low mass, and hence such
patterns are completely eliminated by the WMAP preferred constraint.\\

Given this intuition about the prevalence of different NLSP's for the different cases, we now present the detailed hierarchy patterns of the lightest four non-Standard Model states obtained in our model sets, using the cuts of Section~\ref{sec:bounds} and considering each of the three dark  matter limits separately.  Here we will use the FLN notation for labeling the patterns (mSP1, mSP2, etc.), and use primes when the ordering of the lightest chargino and second-lightest neutralino is reversed from that of FLN, which occurs because we use  {\tt SoftSUSY} as opposed to {\tt SuSpect}.  For example, in mSUGRA/CMSSM models it is found that in the case of near degeneracy between the  lightest chargino $\chi_1^\pm$  and the second-lightest neutralino $\chi_2^0$, $\chi_1^\pm$ is always lighter than $\chi_2^0$ when {\tt SuSpect} is used while either outcome can occur when {\tt SoftSUSY} is used, depending on whether these two states are wino-dominated or whether they include a nontrivial higgsino component.

\begin{table}[p]
{\begin{center}\begin{footnotesize}
\begin{tabular}{|c|l|l|l|l||c|c||c|c||c|c|}\cline{1-11}
\multicolumn{5}{|c||}{Hierarchy} & \multicolumn{2}{|c||}{ WMAP Preferred} & \multicolumn{2}{|c||}{WMAP 7-year}& \multicolumn{2}{|c|}{ Upper Bound Only}  \\ \cline{1-11}
mSP & 1 & 2 & 3 & 4 &  DMM& mSUGRA & DMM & mSUGRA  & DMM & mSUGRA \\
\cline{1-11}
mSP3$'$ & $\chi_1^0$ & $\chi_2^0$ & $\chi_1^{\pm}$ & $\tilde{\tau}_1$ &
 8.1 & 2.9 & 9.6 & 2.4 & 5.5 & 1.8 \\
mSP2$'$ & $\chi_1^0$ & $\chi_2^0$ & $\chi_1^{\pm}$ & $H$,$A$ &
8.2 & 12.4 & 9.5 & 12.9 & 2.9 & 12.0 \\
%
mSP4$'$  & $\chi_1^0$ & $\chi_2^0$ & $\chi_1^{\pm}$ & $\tilde{g}$ &
 0.5 & -- & 0.5 & -- & 0.3 & -- \\
 & $\chi_1^0$ & $\chi_2^0$ & $\chi_1^{\pm}$ & $\tilde{t}_1$ &
0.1 & -- & 0.1 & --& 0.1 & -- \\
 & $\chi_1^0$ & $\chi_2^0$ & $\chi_1^{\pm}$ & $\tilde{\ell}_R$ &
 0.1 &-- & 0.1 & -- & 0.1 & -- \\
mSP1$'$  & $\chi_1^0$ & $\chi_2^0$ & $\chi_1^{\pm}$ & $\chi_3^0$ &
0.0 & 12.4 & 0.0 & 13.0 & 0.0 & 10.1 \\
 & $\chi_1^0$ & $\chi_2^0$ & $\chi_3^0$ & $\chi_1^{\pm}$ &
--& 0.8 & -- & 0.4 & -- & 0.7  \\
\cline{1-11}
mSP3 & $\chi_1^0$ & $\chi_1^{\pm}$ & $\chi_2^0$ & $\tilde{\tau}_1$ &
5.8& --& 6.8 & -- &17.4 & --  \\
mSP2 & $\chi_1^0$ & $\chi_1^{\pm}$ & $\chi_2^0$ & $H$,$A$ &
12.3& -- & 11.0 & -- & 12.1 & 0.1  \\
mSP4 & $\chi_1^0$ & $\chi_1^{\pm}$ & $\chi_2^0$ & $\tilde{g}$ &
2.5 & --& 2.5 & -- & 4.6 & --  \\
 & $\chi_1^0$ & $\chi_1^{\pm}$ & $\tilde{g}$ & $\chi_2^0$ &
-- & -- & -- &-- & 0.9 & -- \\
 & $\chi_1^0$ & $\chi_1^{\pm}$ &  \multicolumn{2}{|c||}{$H$,$A$}  &
1.0& -- & 0.8 & -- & 3.1& -- \\
 & $\chi_1^0$ & $\chi_1^{\pm}$ & $\chi_2^0$ & $\tilde{\ell}_R$ &
--& -- & -- & -- &  2.5& --\\
mSP1  & $\chi_1^0$ & $\chi_1^{\pm}$ & $\chi_2^0$ & $\chi_3^0$ &
4.1&  20.3 & 4.7 & 18.3 & 2.9 & 44.6  \\
 & $\chi_1^0$ & $\chi_1^{\pm}$ & $\tilde{\ell}_R$ & $\tilde{\tau}_1$ &
-- & -- & -- & -- & 0.6 & --\\
 & $\chi_1^0$ & $\chi_1^{\pm}$ & $\tilde{g}$ & $H$,$A$ &
-- & -- & -- & -- & 0.2 & -- \\
 & $\chi_1^0$ & $\chi_1^{\pm}$ & $\tilde{\tau}_1$ & $\chi_2^0$ &
 0.6 & -- &  0.7 & -- & 0.3& -- \\
 & $\chi_1^0$ & $\chi_1^{\pm}$ & $\tilde{\tau}_1$ & $\tilde{\ell}_R$ &
0.3 & -- & 0.3 & --& 0.2 & -- \\
 & $\chi_1^0$ & $\chi_1^{\pm}$ & $\tilde{\ell}_R$ & $\chi_2^0$ &
 -- & -- & -- & -- & 0.1 & -- \\
 & $\chi_1^0$ & $\chi_1^{\pm}$ & $\chi_2^0$ & $\tilde{t}_1$ &
  0.1 & -- & 0.1 & -- & 0.0& -- \\
 & $\chi_1^0$ & $\chi_1^{\pm}$ & $\tilde{\tau}_1$ & $H$,$A$ &
 0.1 & -- & 0.1 & -- & 0.0 & -- \\
 & $\chi_1^0$ & $\chi_1^{\pm}$ & $H$,$A$ & $\chi_2^0$ &
0.1 & -- & 0.1 & -- & 0.0& -- \\
\cline{1-11}
mSP6$'$ & $\chi_1^0$ & $\tilde{\tau}_1$ & $\chi_2^0$ & $\chi_1^{\pm}$ &
 6.6& 23.5 & 7.6& 24.8 & 4.6 & 16.1  \\
mSP7$'$ & $\chi_1^0$ & $\tilde{\tau}_1$ & $\tilde{\ell}_R$ & $\chi_2^0$ &
3.2&  5.7 & 3.8 & 5.1 & 2.3&  4.1 \\
mSP8 & $\chi_1^0$ & $\tilde{\tau}_1$ & \multicolumn{2}{|c||}{$H$,$A$} &
 1.4 & 1.6 & 1.9 & 2.0 & 0.4 & 0.6\\
mSP7 & $\chi_1^0$ & $\tilde{\tau}_1$ & $\tilde{\ell}_R$ & $\chi_1^{\pm}$ &
 1.0& -- & 1.3 & -- & 0.2 & --  \\
mSP6 & $\chi_1^0$ & $\tilde{\tau}_1$ & $\chi_1^{\pm}$ & $\chi_2^0$ &
 0.5& -- & 0.7 & -- & 0.2 & --  \\
 & $\chi_1^0$ & $\tilde{\tau}_1$ & $\chi_1^{\pm}$ & $\tilde{\ell}_R$ &
 0.6 & -- & 0.7 & -- & 0.2 & -- \\
mSP5 & $\chi_1^0$ & $\tilde{\tau}_1$ & $\tilde{\ell}_R$ & $\tilde{\nu}_3$ &
 0.3& 14.6 & 0.4 &15.7 & 0.1 & 6.0  \\
 & $\chi_1^0$ & $\tilde{\tau}_1$ & $\chi_1^{\pm}$ & $H$,$A$ &
0.2 & -- & 0.2 & -- & 0.1& -- \\
mSP9 & $\chi_1^0$ & $\tilde{\tau}_1$ & $\tilde{\ell}_R$ & $H$,$A$ &
 0.1 & 0.3 & 0.1 & 0.5 & 0.1 & 0.1 \\
%
%
%
\cline{1-11}
%
mSP14 & $\chi_1^0$ & \multicolumn{2}{|c|}{$H$,$A$} & $H^{\pm}$ &
 1.9& 2.1 & 1.9 & 1.7 & 1.2 & 0.8 \\
 & $\chi_1^0$ & \multicolumn{2}{|c|}{$H$,$A$} & $\chi_1^{\pm}$ &
 0.6 & -- & 0.3 & -- & 0.8 & 0.0 \\
mSP15$'$ & $\chi_1^0$ & \multicolumn{2}{|c|}{$H$,$A$} & $\chi_2^0$ &
 0.2 & 0.6 & 0.2 & 0.4& 0.2 & 0.4 \\
mSP16 & $\chi_1^0$ & \multicolumn{2}{|c|}{$H$,$A$} & $\tilde{\tau}_1$ &
 0.1& --&  0.0 &--&  0.1& --   \\
%
\cline{1-11}
 & $\chi_1^0$ & $\tilde{g}$ & $\chi_1^{\pm}$ & $\chi_2^0$ &
 -- & -- & -- & -- & 0.1 & --  \\
mSP11$'$ & $\chi_1^0$ & $\tilde{t}_1$ & $\chi_2^0$ & $\chi_1^{\pm}$ &
-- & 1.7 &  -- & 1.8 & -- & 2.1  \\
mSP12$'$ & $\chi_1^0$ & $\tilde{t}_1$ & $\tilde{\tau}_1$ & $\chi_2^0$ &
 --& 0.3 & -- & 0.1 & -- & 0.2  \\
%
%
\cline{1-11}
%
 & \multicolumn{2}{|c|}{$H$,$A$} & $H^{\pm}$ & $\chi_1^0$ &
 38.4 & 0.2 &  33.4 & 0.0 & 34.6 &0.0 \\
 & \multicolumn{2}{|c|}{$H$,$A$} & $\chi_1^0$ & $H^{\pm}$ &
0.7 &  -- & 0.6 & -- & 1.1 & -- \\
\cline{1-11}
 & \multicolumn{4}{|c||}{TOTAL} & 99.7 & 99.4 & 100.0 & 99.1 & 99.0 & 99.7 \\
\cline{1-11}
\end{tabular}
\end{footnotesize}\end{center}} {\caption{\label{tbl:hier}\footnotesize{\bf
Hierarchy Patterns and Relative Percentages}. The dominant hierarchy patterns of the four lightest non-SM states for our DMM and mSUGRA/CMSSM model sets for the direct and indirect bounds of Tables 1 and 2 and the dark matter limits of Table 3.  The numbers given are the relative percentages of models for the given pattern.}}
\end{table}

Table 6 shows that for the WMAP Preferred and WMAP7 bounds result in similar orderings of the most
common patterns, but the pattern ordering is quite different for the WMAP upper bound.  This cut preferentially selects patterns with light charginos, which include the Higgs LSP pattern and the chargino NLSP patterns.
This result is expected since such patterns typically have LSPs with a
significant fraction of wino and/or higgsino components, and thus a lowered relic abundance as compared to bino-dominated cases. 

As expected from the analysis of the NLSP's,
the favored pattern in our subset of DMM models is the Higgs LSP pattern ($H,A<H^\pm<\chi_1^0$) for all three dark matter limits.
This pattern has a relatively squeezed yet heavy spectrum, an LSP with a mixed composition of bino, wino,
and higgsino states, and  a high degree of degeneracy between the
lightest chargino and the lightest two neutralinos, as we will
discuss in greater detail in the next subsection.  The other favored
patterns in our subset of DMM models typically have light charginos
and neutralinos compared to the sfermion masses, which is
expected since the gaugino mass ratios at low energies are
adjustable depending on the size of the mirage unification scale and
the threshold effects from integrating out the messenger fields.  For the WMAP Preferred bound, the dominant six hierarchy patterns are as follows (in order):
\begin{center}
\textbf{DMM Hierarchies}
\end{center}
 \begin{itemize}
 \item Higgs LSP patterns: $H,A<H^\pm<\chi_1^0$
 \item Chargino NLSP pattern:  $\chi_1^0<\chi_1^\pm<\chi_2^0<H,A$ (mSP2)
 \item Neutralino NLSP patterns:  $\chi_1^0<\chi_2^0<\chi_1^\pm<H,A$ (mSP2$'$) and  $\chi_1^0<\chi_2^0<\chi_1^\pm<\tilde{\tau}_1$ (mSP3$'$)
 \item Stau NLSP pattern $\chi_1^0<\tilde{\tau}_1<\chi_2^0<\chi_1^\pm$ (mSP6$'$) and chargino NLSP pattern $\chi_1^0<\chi_1^\pm<\chi_2^0<\tilde{\tau}_1$ (mSP3).
 \end{itemize}
These are also the top six DMM patterns for the WMAP7 bound, with an interchange of the ordering of the mSP3$'$ and mSP2$'$ patterns.  For the WMAP5 upper limit, the top DMM patterns that emerge from our study are the Higgs LSP, mSP3, mSP2, mSP3$'$, mSP6$'$, and mSP4 ($\chi_1^0<\chi_1^\pm<\chi_2^0<\tilde{g}$), demonstrating the increased prevalence for chargino NSLP patterns.


For the mSUGRA/CMSSM model set, the top five mSUGRA/CMSSM hierarchy patterns for the WMAP Preferred constraint are as follows (in order):
\begin{center}
\textbf{mSUGRA/CMSSM Hierarchies}
\end{center}
\begin{itemize}
\item Stau NLSP pattern: $\chi_1^0<\tilde{\tau}_1<\chi_2^0<\chi_1^\pm$ (mSP6$'$) 
\item Chargino NLSP pattern: $\chi_1^0<\chi_1^\pm<\chi_2^0<\chi_3^0$ (mSP1)
\item Stau NLSP pattern: $\chi_1^0<\tilde{\tau}_1<\tilde{\ell}_R<\tilde{\nu}_3$ (mSP5) 
\item Neutralino NLSP patterns:  $\chi_1^0<\chi_2^0<\chi_1^\pm<\chi_3^0$ (mSP1$'$) and  $\chi_1^0<\chi_2^0<\chi_1^\pm<H,A$ (mSP2$'$). 
\end{itemize}
Taking into account the existence of the primed scenarios here due to the use of {\tt SoftSUSY} as opposed to {\tt SuSpect}, these are the same top five patterns found by FLN \cite{Feldman:2007zn}. The precise ordering of the FLN results differs from what is found here because the FLN bounds on the lightest Higgs mass and the superpartner masses are weaker.  In general, for both mSUGRA/CMSSM and DMM cases, relaxing the bound on the lightest Higgs to 100 GeV as done by FLN will favor scenarios with lighter sfermion masses, since a tightening of the Higgs limit tends to require an increase in the third generation scalar masses.  As for DMM, the WMAP7 constraint results in the same top five hierarchy patterns (again with a slight reordering, here of mSP1$'$ and mSP2$'$).  The WMAP5 upper limit overwhelmingly prefers the mSP1 pattern, with mSP6$'$, mSP2$'$, mSP1$'$, and mSP7$'$ as the next most prevalent patterns. 

The set of the most prevalent hierarchy patterns in both the DMM and mSUGRA/CMSSM cases obtained with this set of direct, indirect, and cosmological bounds, are quite robust with respect to modifications of the direct limits on the Higgs and superpartner masses.  As previously mentioned, a weaker Higgs mass bound tends to allow patterns with lighter scalars, such as mSP5.  Weaker limits on the gluino and squark masses as obtained by the Tevatron also favor models with lighter scalars such as the mSP3 scenario, but for both the DMM and mSUGRA/CMSSM model sets, the same set of prevalent patterns is obtained using these bounds as what was obtained here using stricter limits.  As we will see in the next subsection, a further increase of the LHC limits on the gluino and squark masses (such as the full 1025 GeV bounds for the simplified models as obtained in the most recent ATLAS results) for example in the DMM case will tend to prefer the Higgs LSP and chargino NLSP patterns like the mSP2 pattern, since these are patterns that tend to have squeezed yet heavy spectra. 

\begin{figure}[!b]
\begin{center}
\begin{tabular}{cc}
\igraph{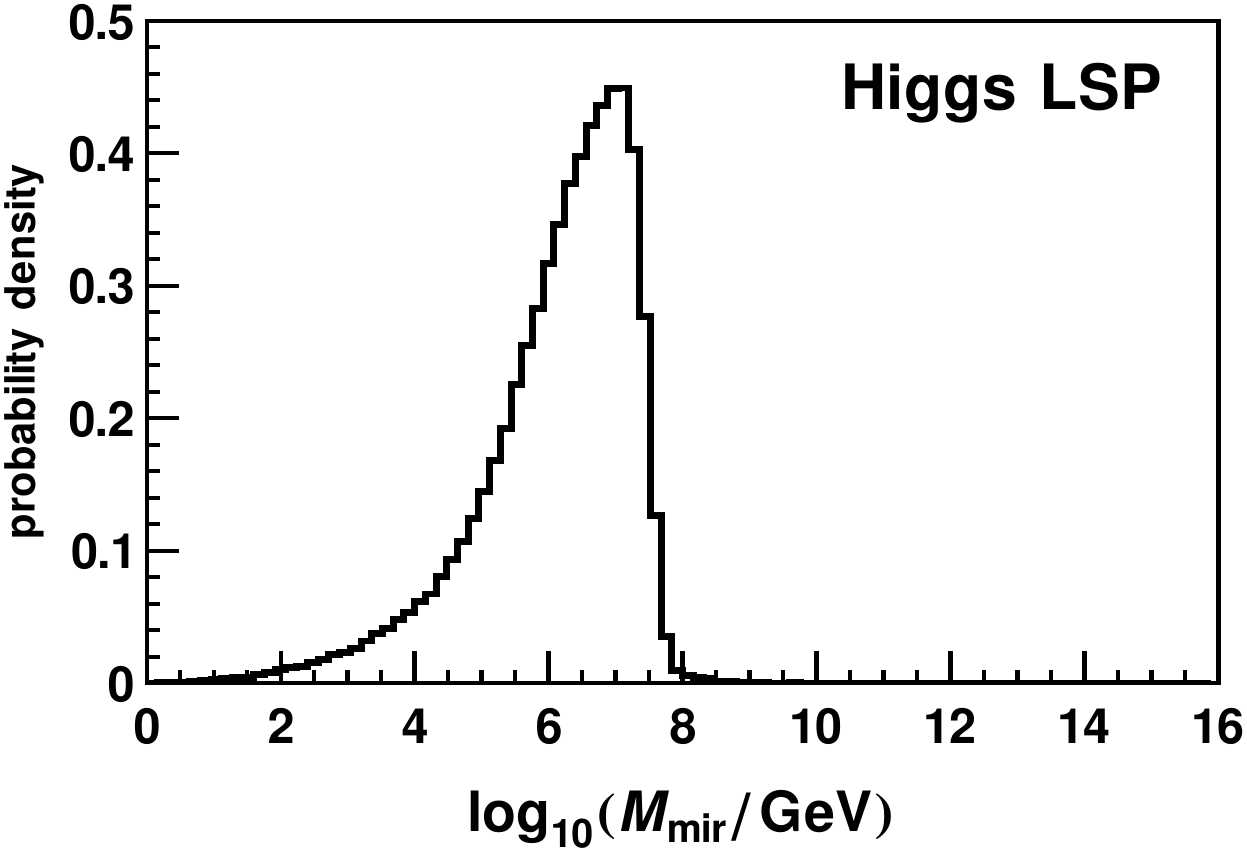} & \igraph{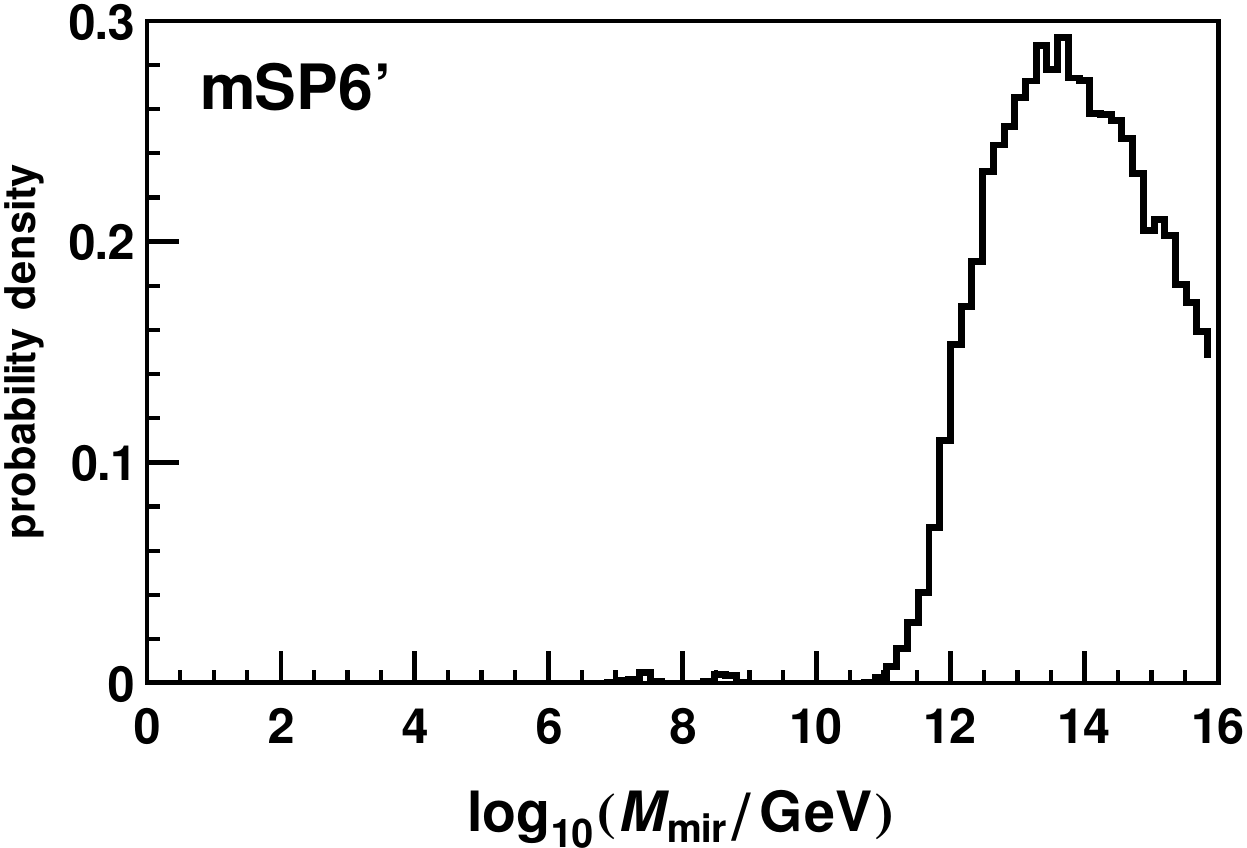} \\
\igraph{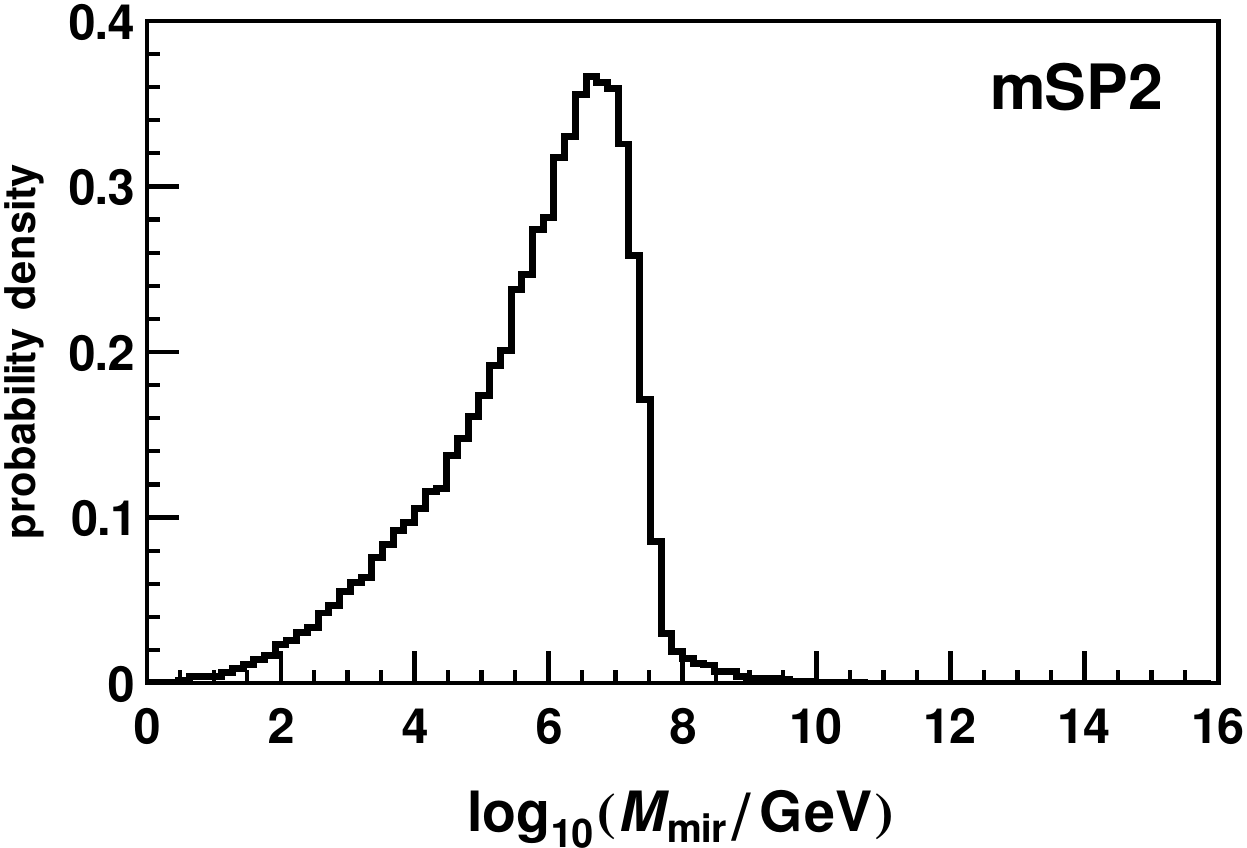} & \igraph{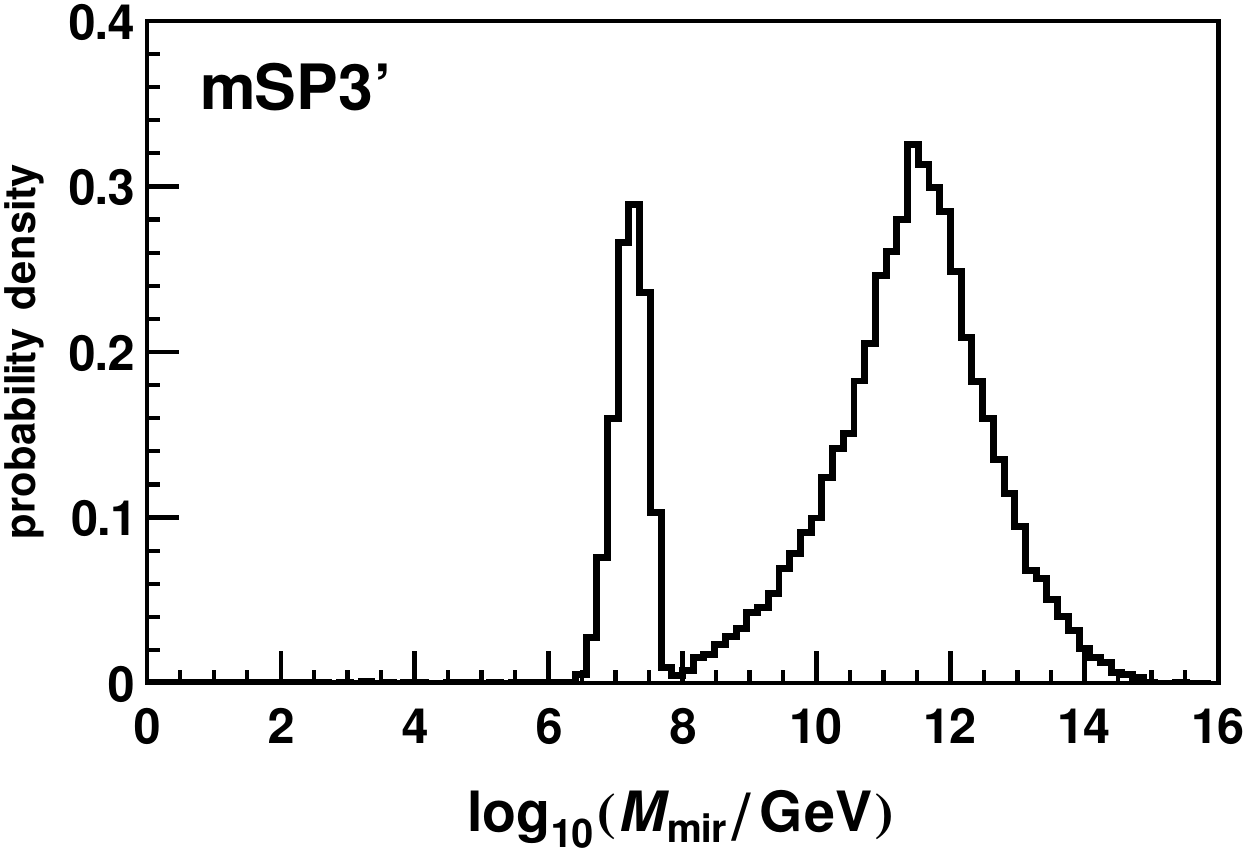} \\
\end{tabular}
\caption{\textbf{Mirage unification scale}.
 Histograms of the gaugino sector mirage unification
scale (in~GeV) for the Higgs LSP (upper left panel), mSP6$'$ (upper right panel), mSP2 (lower left panel) and mSP3$'$ (lower right panel) patterns, with the direct and indirect limits of Tables 1-2 and the WMAP Preferred dark matter constraints.}
\end{center}
\label{fig:mirage3}
\end{figure}

In summary, the dominant DMM pattern in our model set is the Higgs LSP pattern, for which the heavy Higgs particles are lighter than the LSP.  For two other sets of prevalent DMM patterns, the only difference in the lightest four non-SM particles is the ordering of the lightest chargino and the second-lightest neutralino as either the NLSP or the third-lightest particle.  There are also DMM patterns in which the stau is the NLSP, which more closely resemble similar mSUGRA/CMSSM patterns.  The top five patterns of the mSUGRA/CMSSM model set overlap with what was found by FLN, with only minimal reordering that reflects the more stringent collider limits used in this analysis.   
As our primary goal is to use this categorization as a means to get a handle on the DMM parameter space, in what follows we will focus on several of the most prevalent DMM hierarchy patterns and investigate the resulting distributions of the particle masses, comparing  the outcome to analogous mSUGRA/CMSSM patterns when relevant.  We will use this information to determine phenomenologically motivated DMM benchmark model points and comment upon their implications for LHC searches.

\subsection{DMM Hierarchy Patterns: Pattern Groupings and Mass Spectra}
\label{sec:corr}

We now investigate the characteristic mass spectra of several of the prominent DMM hierarchy patterns discussed in Section~\ref{sec:hierarchy}.  To begin, we provide a brief discussion of collective properties of several classes of hierarchy patterns by pointing out similarities in their underlying DMM parameter space.  We focus on the dependence on the gaugino mirage unification scale as given in Eq.~(\ref{miraged}), which is the scale at which the gaugino mass parameters unify at one-loop.  Although the mirage unification scale is not a physical scale in the sense that no new physics enters at that scale, it has a strong impact on the gaugino mass ratios at low energies. In mSUGRA/CMSSM models, the unification of the gaugino masses arises at the GUT scale, leading to a very specific set of splittings between the gauginos at the TeV scale.  However, the ability in DMM models to ``slide" this unification feature of the gaugino masses strongly affects the gaugino mass ratios at low energies, which in turn has a strong impact on both dark matter and collider signatures.

\begin{figure}[!t]
\begin{center}
\begin{tabular}{cc}
\igraph{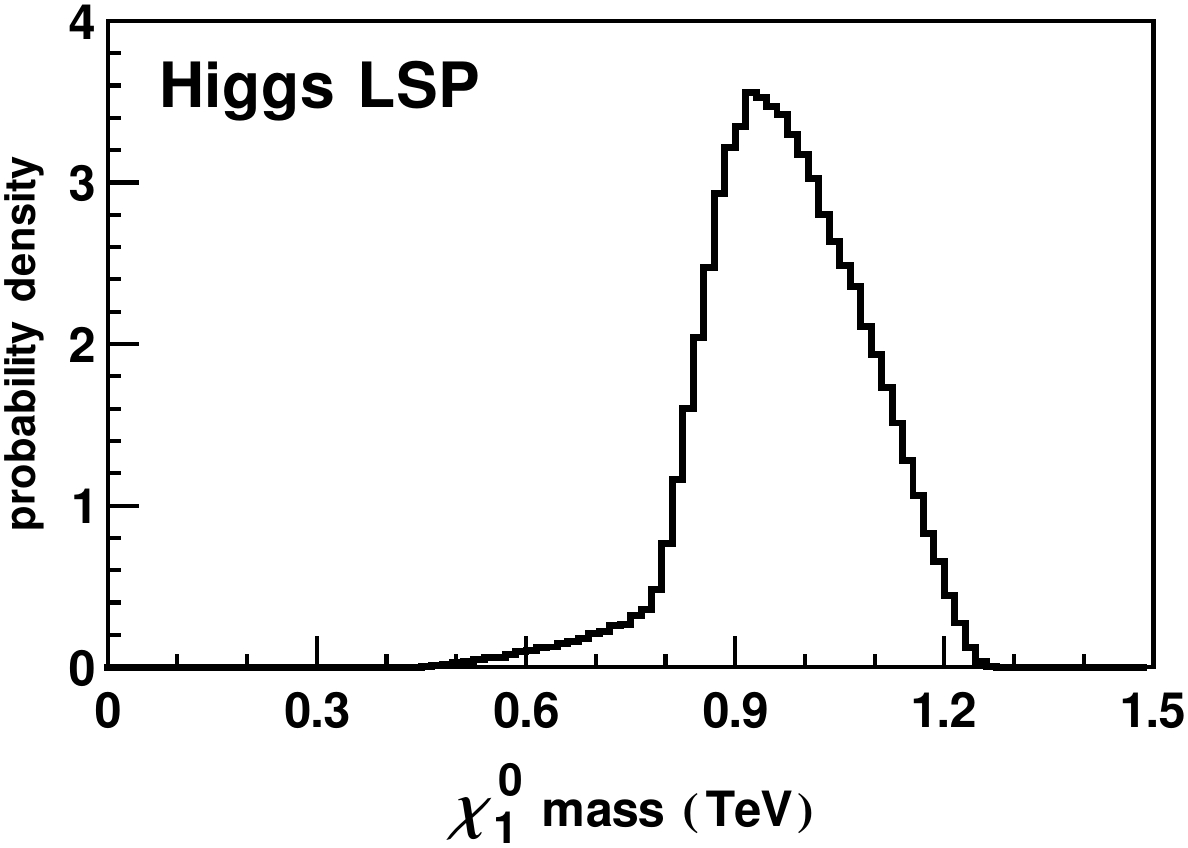} & \igraph{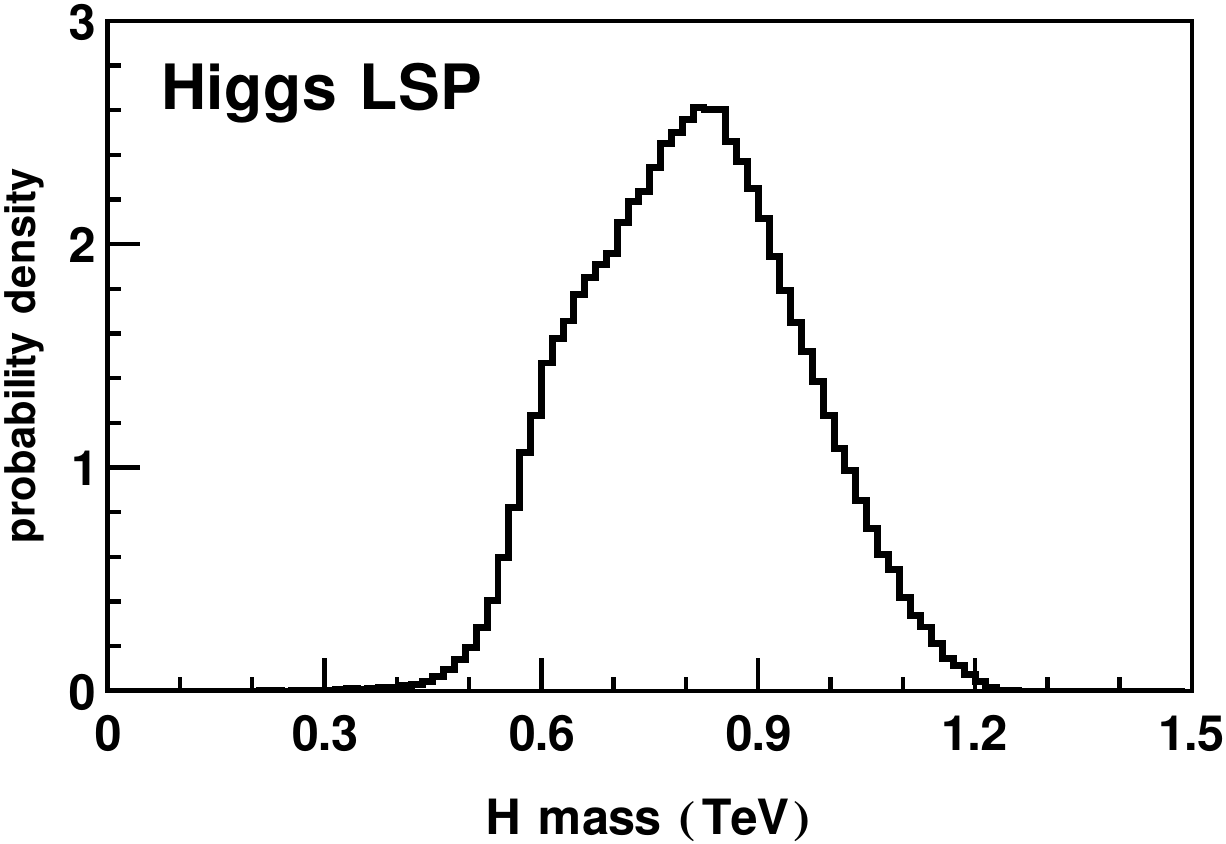}
\end{tabular}
\caption{\textbf{Higgs LSP pattern: $\chi_1^0$ mass and $H$ mass}.
 Histograms of the mass of the neutralino LSP $\chi_1^0$  (left panel) and the mass of the heavy Higgs $H$  (right panel) for the Higgs LSP pattern, with the direct and indirect limits of Tables 1-2 and the WMAP Preferred dark matter constraints.}
\end{center}
\label{fig:higgsLSPneut1mass}
\end{figure}
\begin{figure}[!t]
\begin{center}
\begin{tabular}{cc}
\igraph{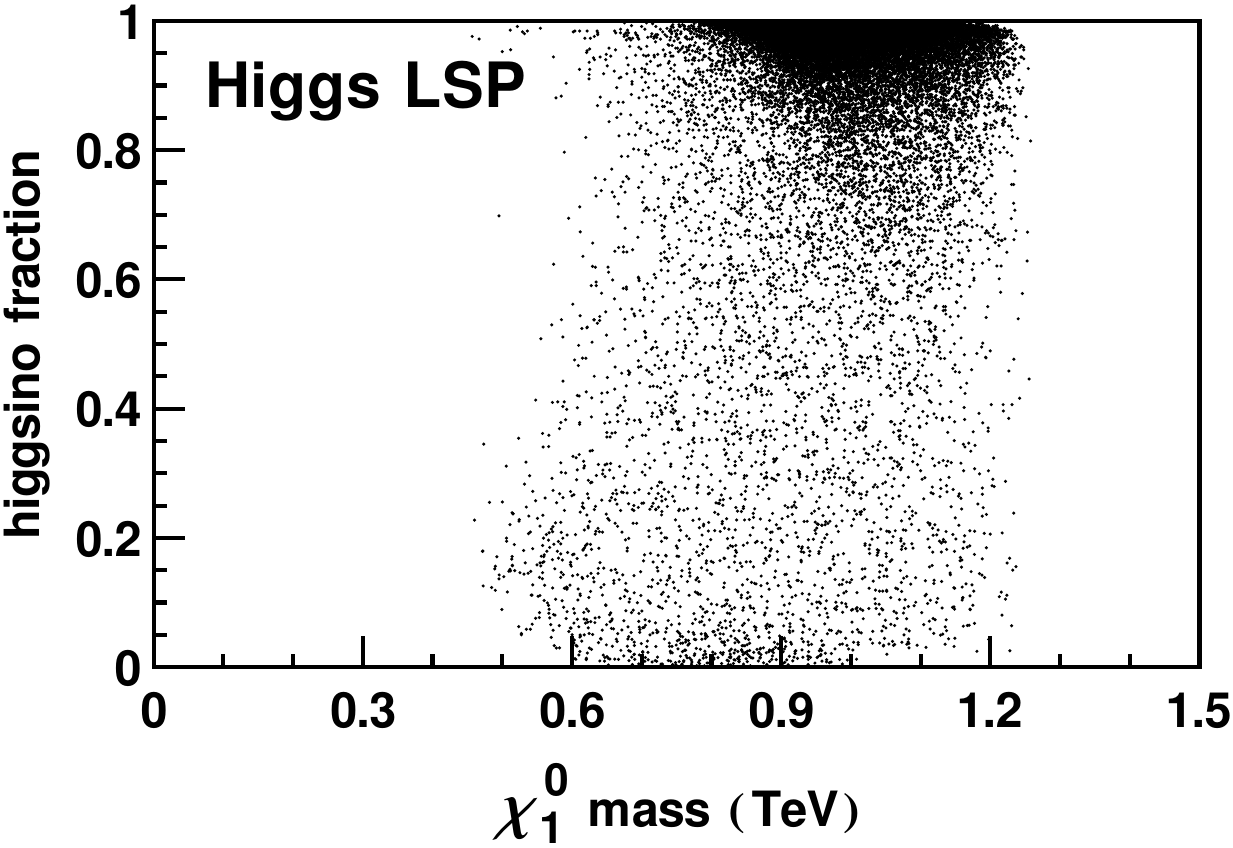} & \igraph{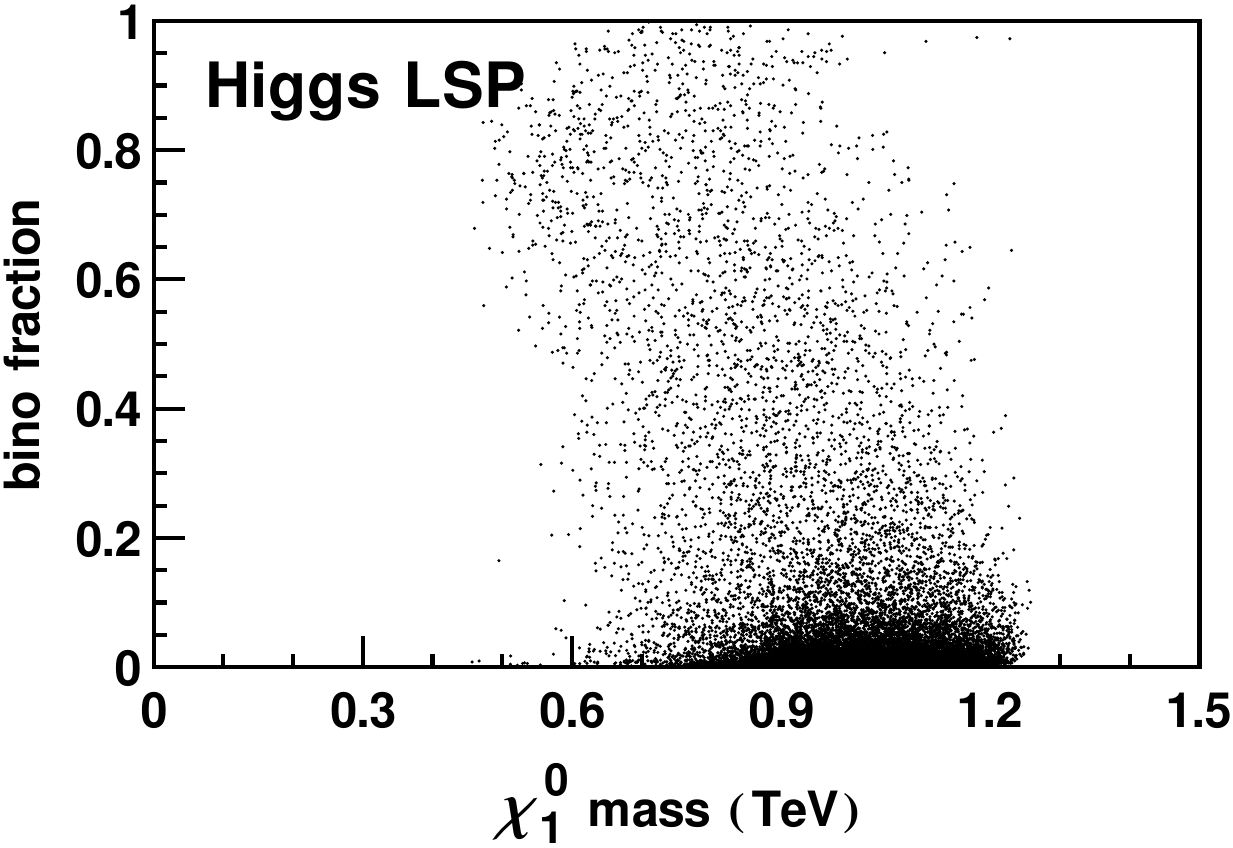}
\end{tabular}
\caption{\textbf{Higgs LSP pattern: higgsino and bino fraction}.
Two-dimensional histograms of the higgsino fraction (left panel) and bino fraction (right panel) $\chi_1^0$ as a function of the $\chi_1^0$ mass for the Higgs LSP pattern, with the direct and indirect limits of Tables 1-2 and the WMAP Preferred dark matter bound.}
\end{center}
\label{fig:higgsLSPhiggsinobino}
\end{figure}

To see this explicitly,  in Fig.~1 we show the distribution of the mirage unification scale for four of the prevalent DMM hierarchy patterns discussed in the previous subsection:  the Higgs LSP ($H,A<H^\pm<\chi_1^0$), mSP6$'$ ($\chi_1^0<\tilde{\tau}_1<\chi_2^0<\chi_1^\pm$), mSP2 ($\chi_1^0<\chi_1^\pm<\chi_2^0<H,A$), and mSP3$'$  ($\chi_1^0<\chi_2^0<\chi_1^\pm<\tilde{\tau}_1$).\footnote{We do not display the distributions for mSP2$'$, which strongly resembles that of mSP2, and mSP3, which resembles mSP3$'$ except that the large peak at higher values of the mirage unification scale is dramatically reduced.} The distribution of the mirage unification scale peaks at lower values and has a sharp cutoff at $10^8$ GeV for the Higgs LSP and mSP2 patterns, as well as for the narrower peak of the mSP3$'$ pattern.  The relatively low typical value of the mirage unification scale for these patterns will indicate a mass spectrum characterized by higgsino-dominated and/or mixed-composition LSP's.  The similar structure of the distributions also suggests that these patterns share similar phenomenological features, as explored below.  In contrast, the peak of the mSP6$'$ distribution and the right peak of the mSP3$'$ distribution have much higher values of the mirage unification scale, with the peak for the mSP6$'$ pattern approaching the mSUGRA/CMSSM limit of $M_{\rm G}\sim 10^{16}$ GeV.  Hence, in these cases the mass spectra are more reminiscent of mSUGRA/CMSSM models, with in bino-dominated neutralino LSP's.

\begin{figure}[!t]
\begin{center}
\begin{tabular}{cc}
\igraph{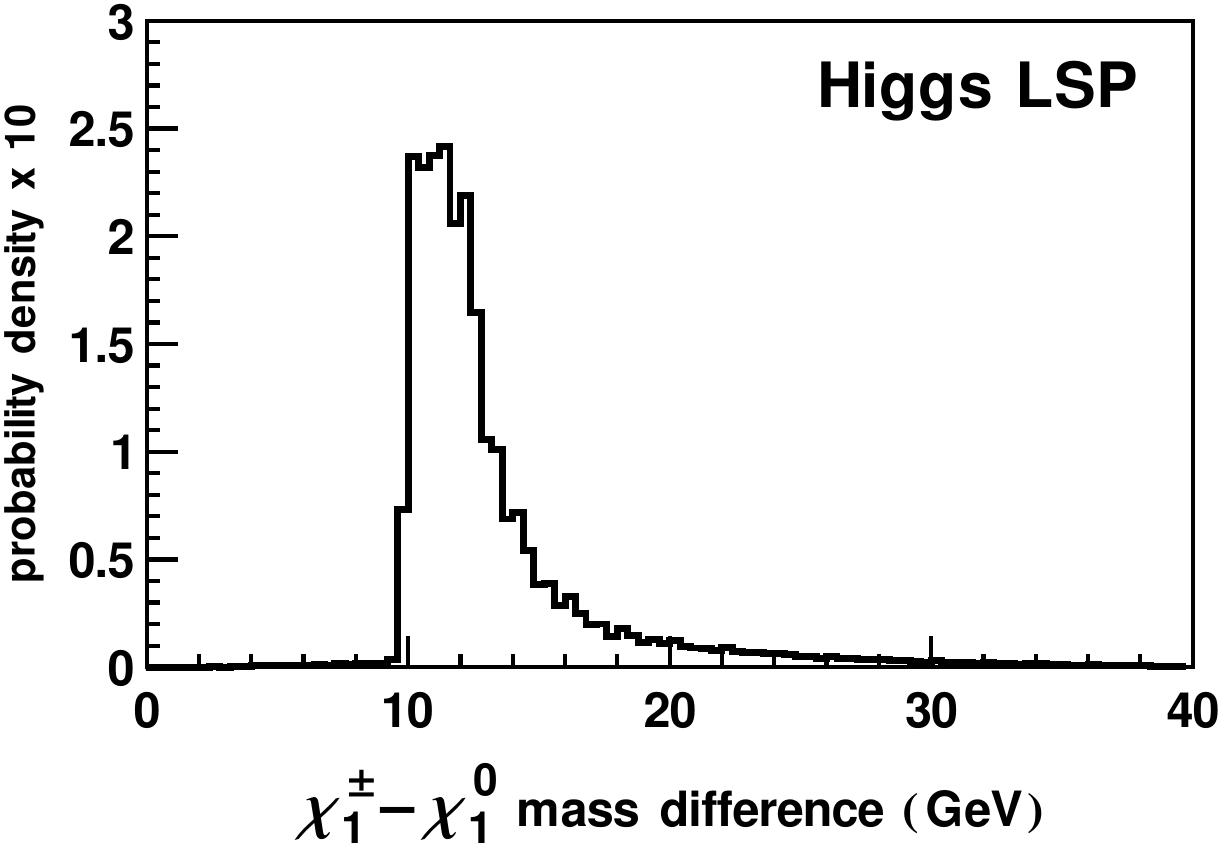} & \igraph{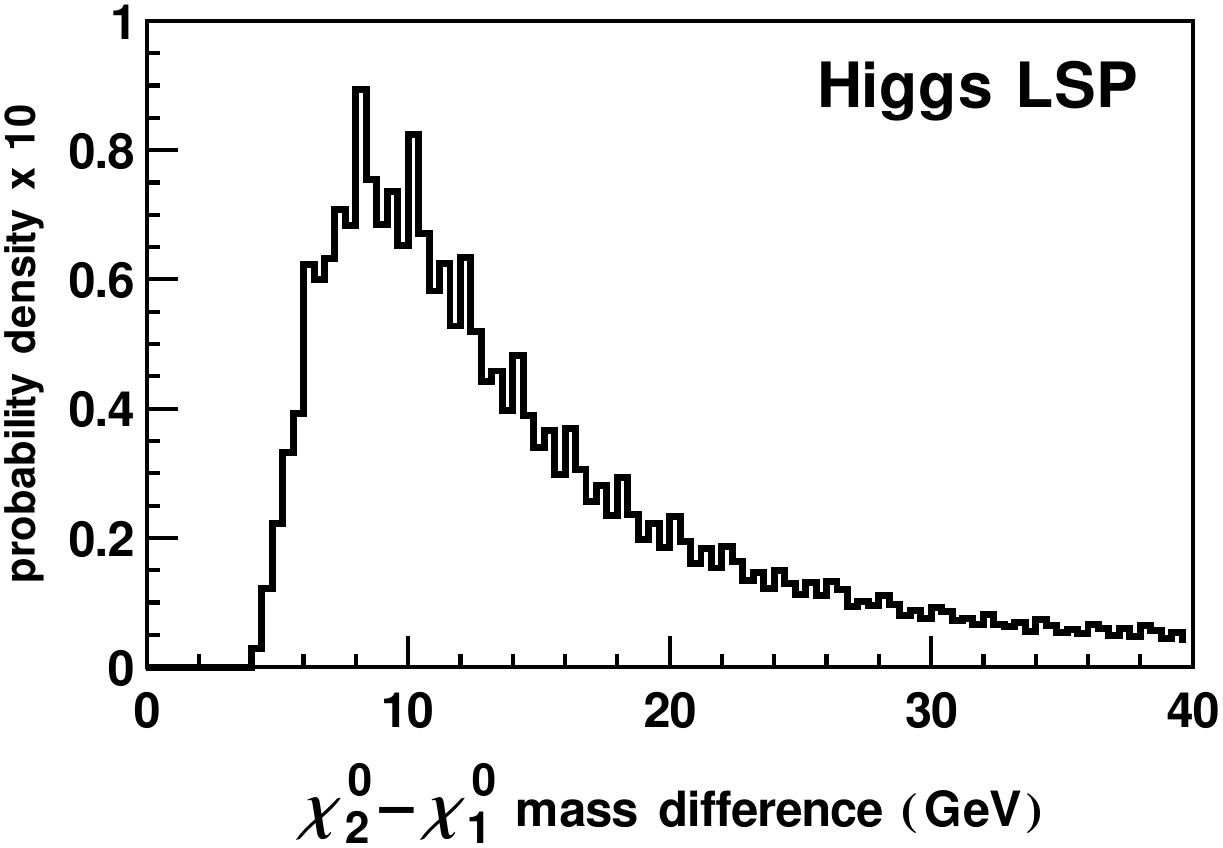}
\end{tabular}
\caption{\textbf{Higgs LSP pattern: $\chi_1^\pm-\chi_1^0$ and $\chi_2^0-\chi_1^0$ mass differences}.
Histograms of the mass difference between $\chi_1^\pm$ and $\chi_1^0$  (left panel) and the mass difference between $\chi_2^0$ and $\chi_1^0$ (right panel) for the Higgs LSP pattern, with the direct and indirect limits of Tables 1-2 and the WMAP Preferred dark matter constraints. }
\end{center}
\label{fig:higgsLSPstuff1}
\end{figure}
\begin{figure}[!t]
\begin{center}
\begin{tabular}{cc}
\igraph{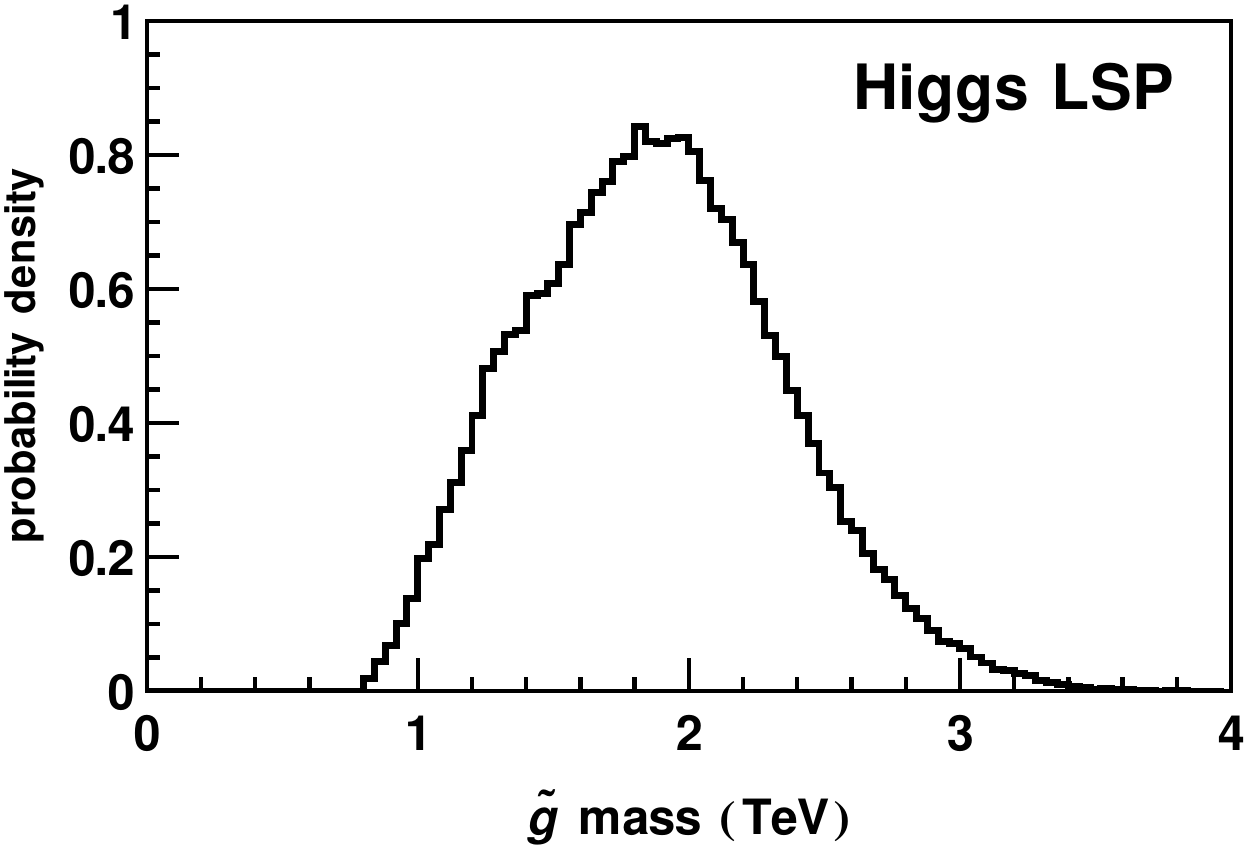} & \igraph{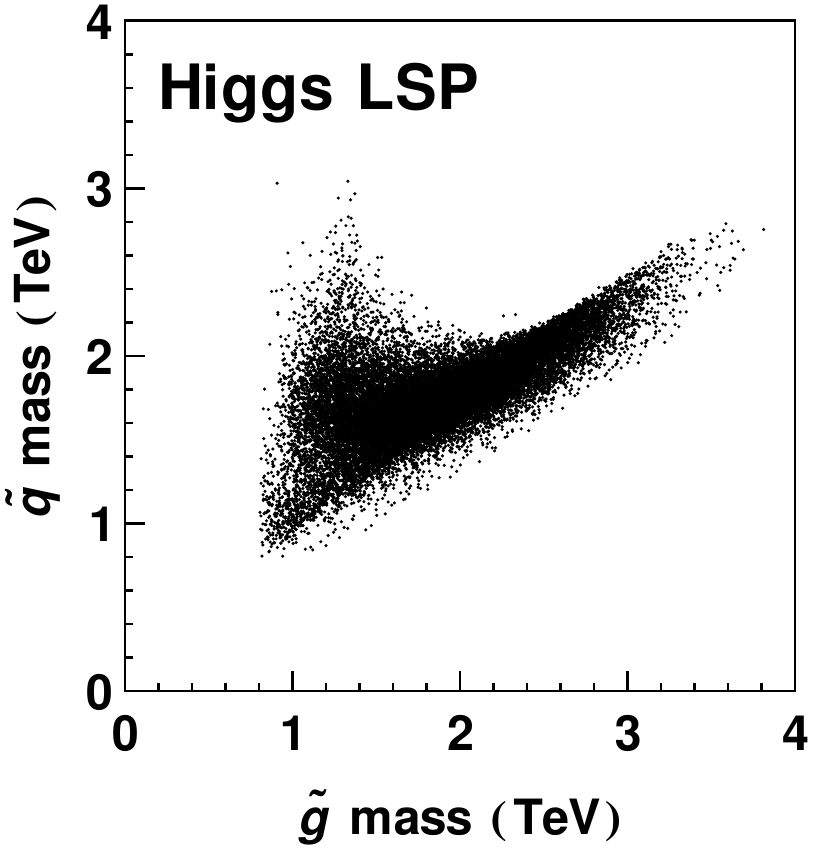}
\end{tabular}
\caption{\textbf{Higgs LSP pattern: gluino and squark masses}.
Histograms of the mass of the gluino (left panel) and the first generation squark versus gluino masses for the Higgs LSP pattern, 
with the direct and indirect limits of Tables 1-2 and the WMAP Preferred dark matter constraints. }
\end{center}
\label{fig:higgsLSPchargino}
\end{figure}

Our procedure is thus to consider several patterns sequentially according to whether they typically have a low or high mirage unification scale.  A rough grouping of patterns is fruitful not only because it gives less weight to the specific rankings of the hierarchy patterns obtained in our landscape study (as it depended on assumptions such as naive flat priors), but also because the DMM mass spectra tend to have significant degeneracies in the lightest four new particle masses, such that the lightest four states are not nearly enough to characterize the LHC physics.   To this end, we will begin with the Higgs LSP pattern and study the typical mass spectra, then compare it the mSP2 pattern, which also has a low mirage unification scale.  We will then study the mSP3$'$ pattern, which is a mixed pattern with both features.  Finally, we will study the mSP6$'$ pattern, and compare it to the mSUGRA/CMSSM expectations.

\subsubsection{DMM Higgsino/mixed LSP patterns: Higgs LSP}

We begin with the Higgs LSP pattern  ($H,A<H^\pm<\chi_1^0$), since this pattern is the most prominent outcome from our study once the dark matter constraints are taken into account, and because of the rarity of this pattern in mSUGRA.  In Fig.~2, we show the distribution of the mass of the lightest neutralino $\chi_1^0$ (the true LSP) and the heavy Higgs $H$.  The heavy Higgses $H$, $A$, and $H^\pm$ are strongly degenerate, with typical masses of order 800 GeV.  We see that the heavy Higgs particles are not particularly light in the Higgs LSP pattern; these particles are lighter than the lightest neutralino  simply because $\chi_1^0$ must be of the order of the TeV scale to satisfy the dark matter constraints. 

The higgsino and bino fraction of $\chi_1^0$ as a function of the mass of $\chi_1^0$ is shown in Fig.~3.
Clearly, this pattern results predominantly in higgsino-dominated LSP's with masses of order the TeV scale.  As we will see, this type of neutralino LSP will be characteristic of all DMM patterns with mirage unification scales less than $10^8$ GeV.  In addition, the lightest chargino $\chi_1^\pm$  and second-lightest neutralino $\chi_2^0$ are very close in mass to $\chi_1^0$, with $(m_{\chi_1^\pm}-m_{\chi_1^0})/m_{\chi_1^0}\sim 0.01$ and
$(m_{\chi_2^0}-m_{\chi_1^0})/m_{\chi_1^0}\sim 0.02$ on the average, as shown in Fig.~4.

\begin{figure}[!t]
\begin{center}
\begin{tabular}{cc}
\igraph{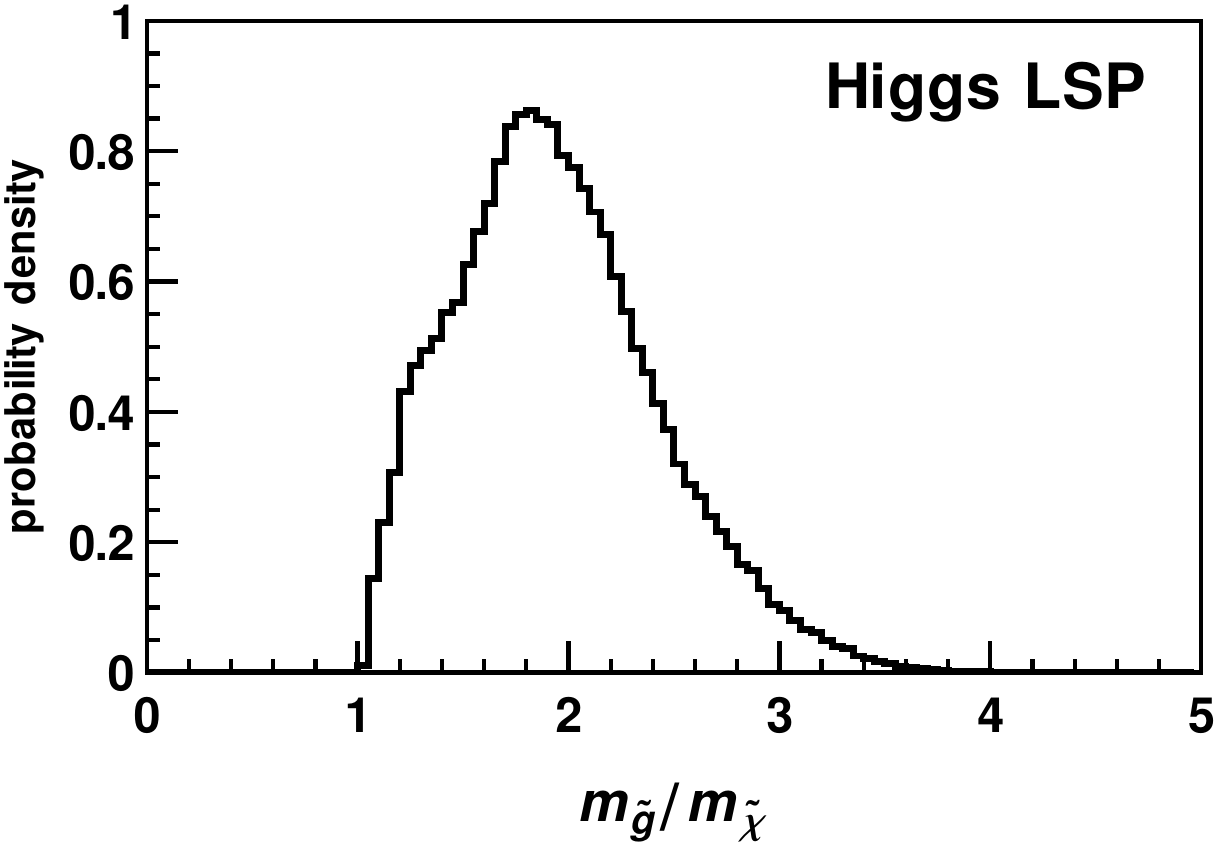} & \igraph{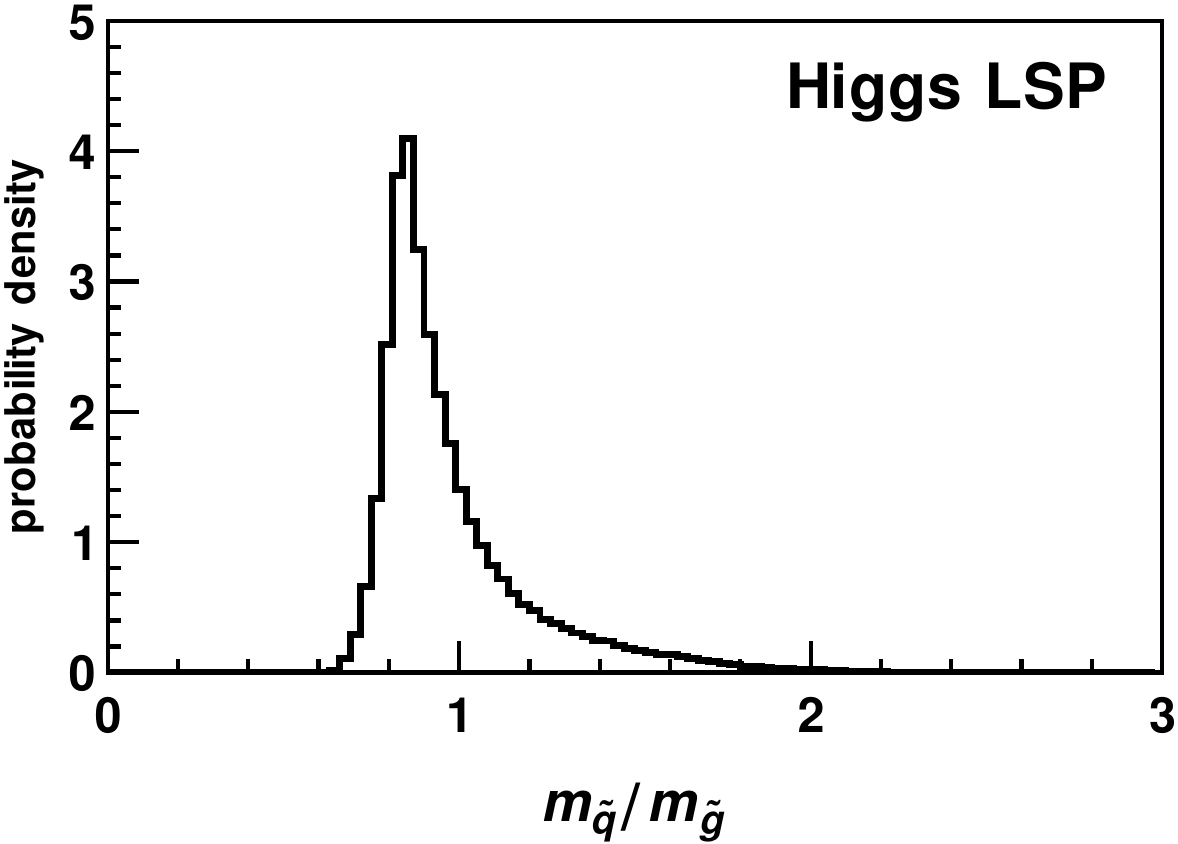}
\end{tabular}
\caption{\textbf{Higgs LSP pattern: gluino/neutralino mass ratio and squark/gluino mass ratio}.
One-dimensional histogram of the ratio of the gluino mass to the mass of the lightest neutralino  (left panel) and the ratio of the first generation up-type squark mass to the gluino mass (right panel) for the Higgs LSP pattern, 
with the direct and indirect limits of Tables 1-2 and the WMAP Preferred dark matter constraints. }
\end{center}
\label{fig:higgsLSPgluino}
\end{figure}

The distribution of the gluino mass and the gluino versus lightest up-type squark masses are shown in Fig.~5.  We see that the gluino is typically of order ~1-3 TeV or so, as are the lighter generation squarks, so they easily escape the LHC bounds as used in this analysis. 
In Fig.~6, we show the ratio of the gluino
mass to the mass of the lightest superpartner in the left panel, and
the ratio of the first generation up-type squark mass to the
gluino mass in the right panel.  The peak value of the
gluino to neutralino mass ratio is about 2, as opposed to the much
higher values found in mSUGRA/CMSSM models because of the characteristic
splitting of the low energy values of the three gaugino masses of $M_1:M_2:M_3\sim 1:2:6$.  The first
generation squark to gluino mass ratio distribution shows that the
gluino and squarks are typically comparable in size.  
Overall,
the spectrum is relatively heavy due to the need for a
TeV-scale $\chi_1^0$, but quite compressed, which helps
to ensure that the light Higgs is heavy enough to evade the
114.4 GeV bound. We note that relaxing the Higgs bound to 100 GeV
does not change the number of viable models with this pattern, so it
is quite robust with respect to this constraint.

\begin{figure}[!t]
\begin{center}
\begin{tabular}{c}
\igraphnew{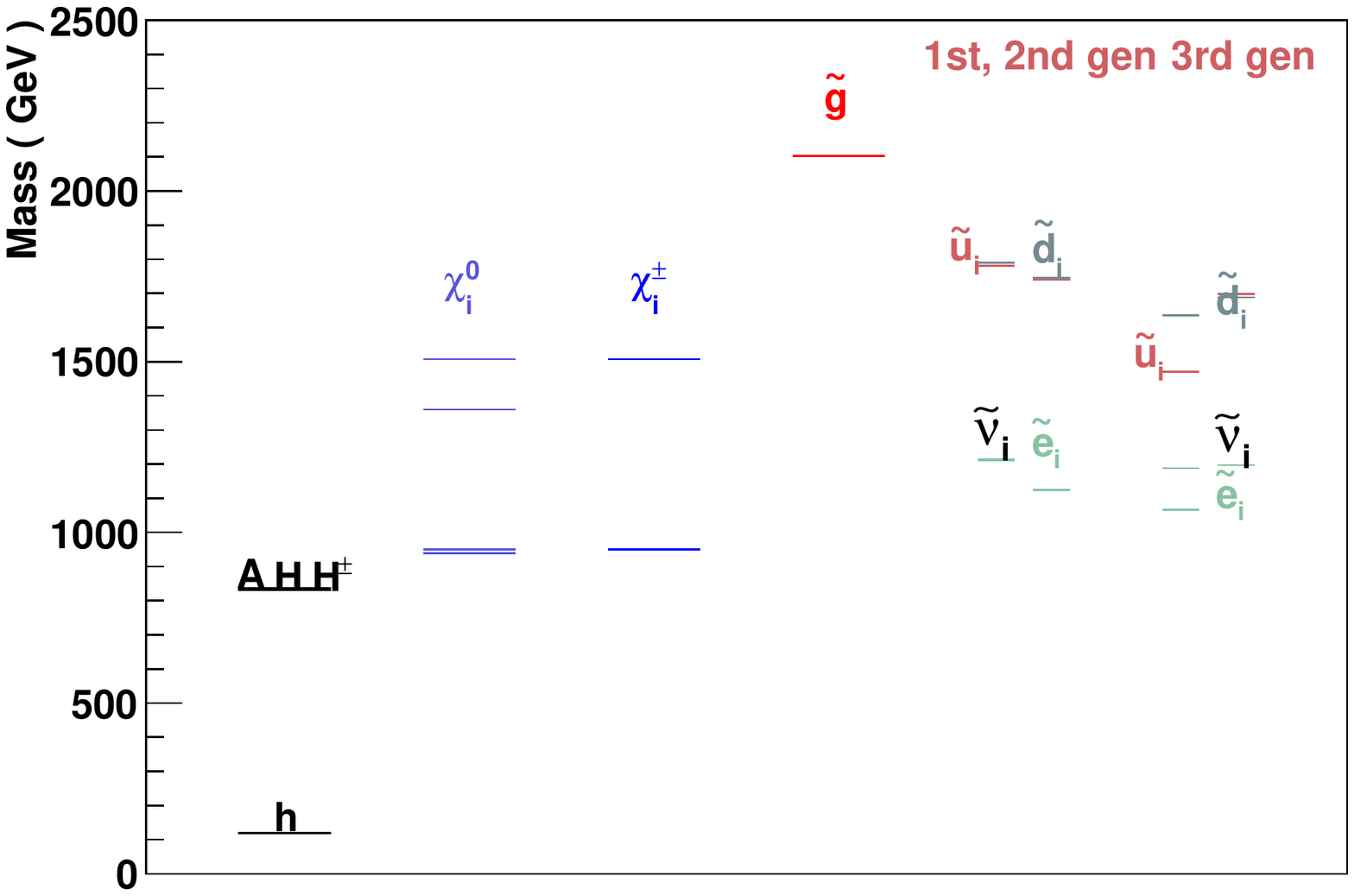}
\end{tabular}
\caption{\textbf{DMM benchmark point \#1 (Higgs LSP): mass spectrum.}
The mass spectrum for DMM benchmark point \#1. The explicit masses are also given in Table 7.}
\end{center}
\label{fig:higgsLSPspectrum}
\end{figure}
\begin{table}
\begin{center}
\begin{tabular}{|cc|cc|cc|cc|}
\hline
\hline
Particle & Mass & Particle & Mass & Particle & Mass & Particle & Mass \\
\hline
$h$ & 120 & $H$ &831 &$\tilde{u}_L$ & 1781 & $\tilde{u}_R$ & 1740 \\
$A$ & 832 & $H^\pm$ & 836& $\tilde{d}_L$ & 1790 & $\tilde{d}_R$ & 1746 \\
$\chi^0_1$ & 939 & $\chi^0_2$ & 950 & $\tilde{\ell}_L$ & 1215 & $\tilde{\ell}_R$ & 1125 \\
$\chi^0_3$ & 1360 & $\chi^0_4$ & 1507 &$\tilde{t}_1$ & 1471 & $\tilde{t}_2$ & 1699 \\
$\chi^\pm_1$ & 951 & $\chi^\pm_2$ & 1507 & $\tilde{b}_1$ & 1636 & $\tilde{b}_2$ & 1689 \\
$\tilde{g}$ & 2103 & $\tilde{\tau}_1$ & 1067 & $\tilde{\tau}_2$ & 1197 & &  \\
 \hline 
 \hline 
\end{tabular}
\end{center}
\caption{The particle masses (in GeV) for the DMM benchmark point \#1, as shown in Fig.~7.}
\end{table}


As a concrete example that shows these features, we have selected a DMM benchmark point (Point \#1), for which the mass spectrum is shown in Fig.~7 and the particle masses are given in Table 7. This point corresponds to the following input parameters: $N = 1$, $M_{\rm mess} = 5.0\times 10^{12}$, $M_0 = 1570$ GeV, $\alpha_m = 1.46$, $\alpha_g = -0.70$, and $\tan\beta = 34.5$.  The gluino is the heaviest superpartner, with a mass upwards of 2 TeV, while the other squarks have masses of about 1.7 TeV.  The mass of the lightest Higgs boson is $119.6$ GeV, and the heavy Higgs bosons have masses of about $830$ GeV.  Furthermore, the LSP mass is around 940 GeV and the mass of the lightest chargino is about 950 GeV, reflecting the strong higgsino component of the lighest neutralino.

\begin{figure}[!t]
\begin{center}
\begin{tabular}{cc}
\igraph{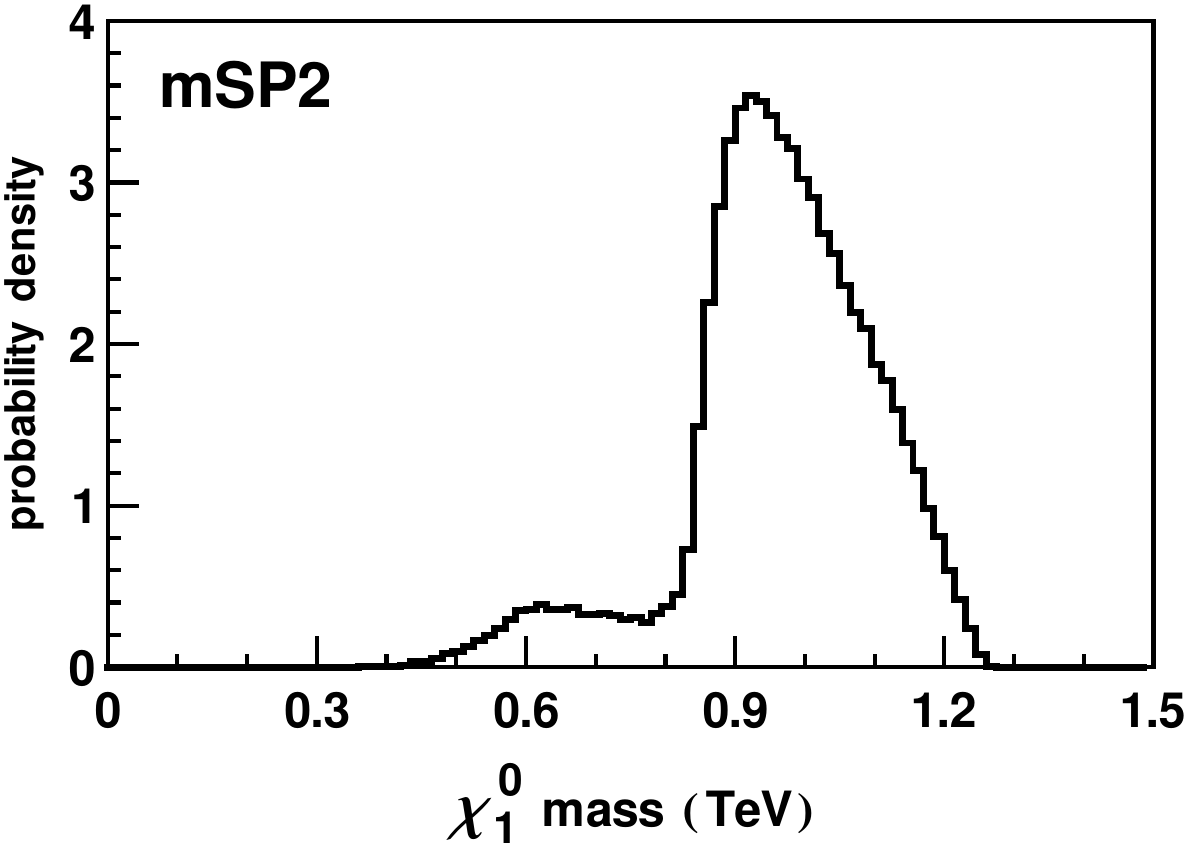} & \igraph{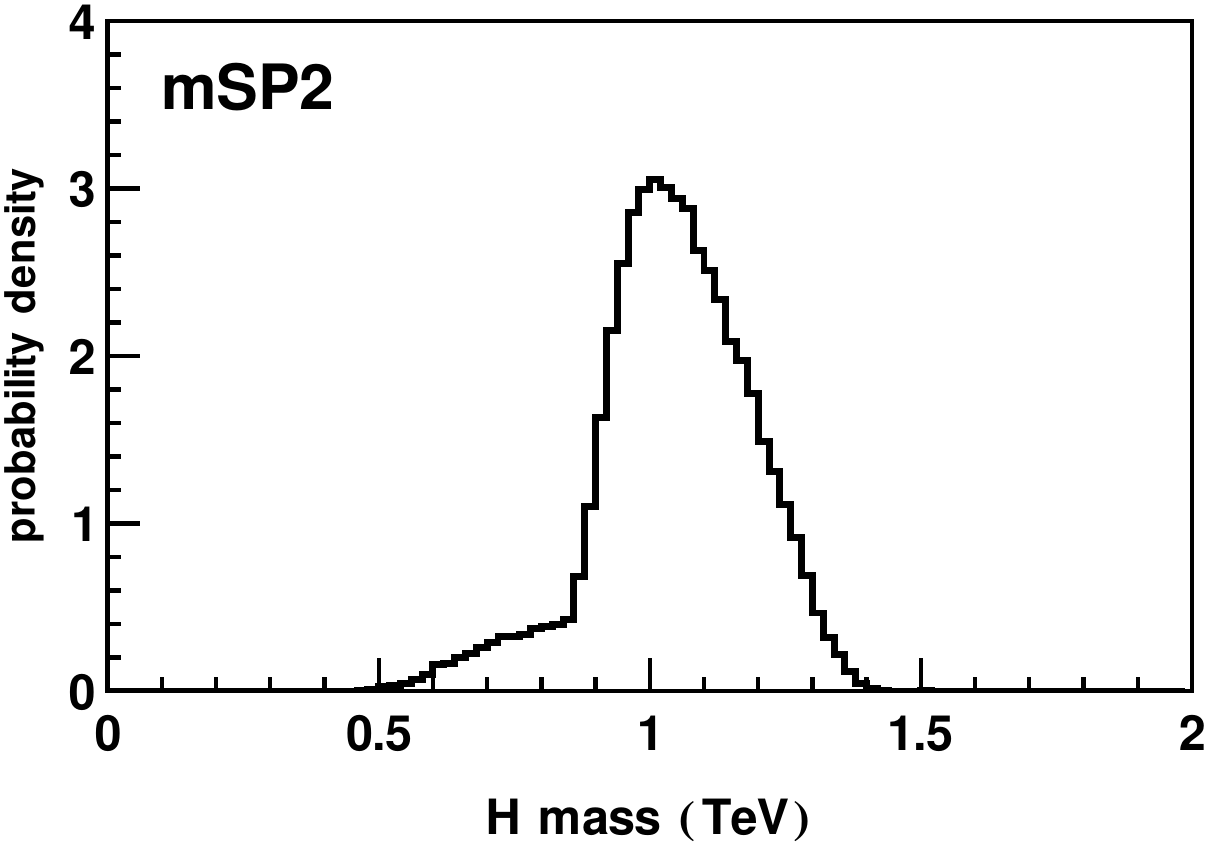}
\end{tabular}
\caption{\textbf{mSP2 pattern: $\chi_1^0$ mass and $H$ mass}.
Histograms of the mass of the neutralino LSP $\chi_1^0$  (left panel) and the mass of the $H$ boson (right panel) for the mSP2 pattern, with the direct and indirect limits of Tables 1-2 and the WMAP Preferred dark matter constraints.}
\end{center}
\label{fig:msp2stuff1}
\end{figure}
\begin{figure}[!t]
\begin{center}
\begin{tabular}{cc}
\igraph{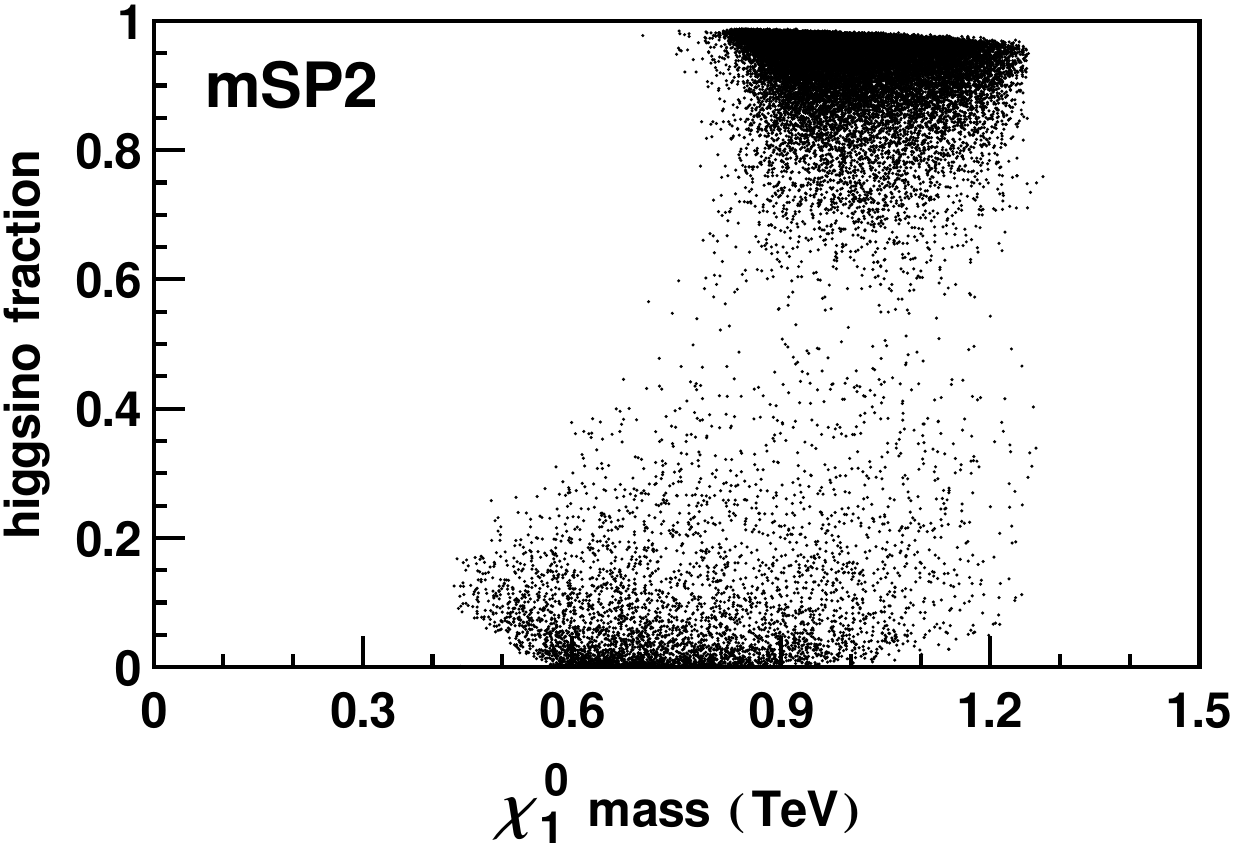} & \igraph{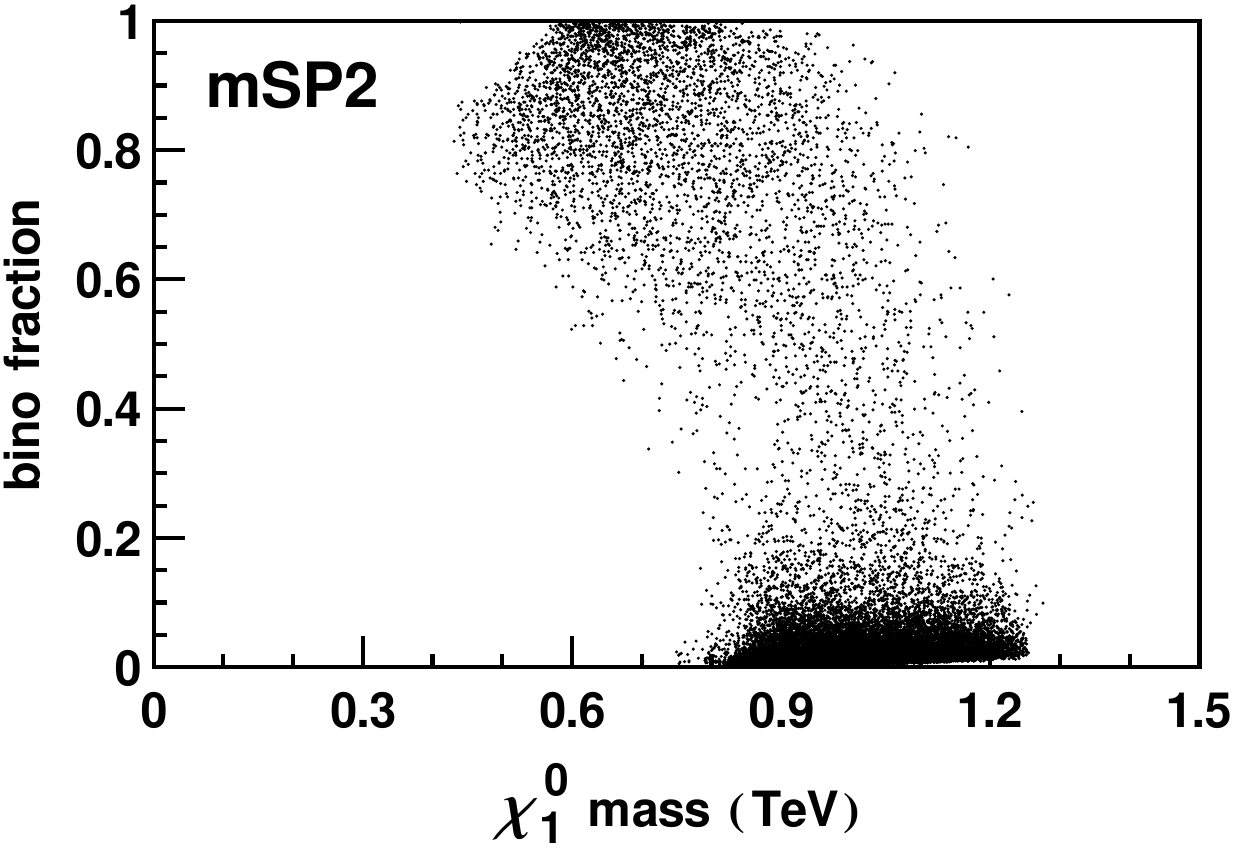}
\end{tabular}
\caption{\textbf{mSP2 pattern: higgsino and bino fraction}.
Two-dimensional histograms of the higgsino (left panel) and bino  (right panel) fraction of  $\chi_1^0$ as a function of the $\chi_1^0$ mass for the mSP2 pattern, with the direct and indirect limits of Tables 1-2 and the WMAP Preferred dark matter bound.}
\end{center}
\label{fig:mSP2higgsinobino}
\end{figure}

\subsubsection{DMM  patterns: mSP2 and mSP3$'$}
We now turn to the DMM mSP2 and mSP3$'$ patterns. The mSP2 pattern has low peak values of the mirage unification scale, and hence it shares many common features with the Higgs LSP pattern.  In Fig.~8, we show the masses  of the neutralino LSP and the heavy Higgs boson $H$ for the mSP2 pattern.  We see that the LSP is again of order the TeV scale, and that the heavy Higgses tend to be very close in mass to $\chi_1^0$.  As shown in Fig.~9, the LSP is most typically higgsino-dominated, with masses peaked at the TeV scale, though in this pattern there is a subset of models with a bino-dominated LSP that is correspondingly lighter.   The spectrum is relatively heavy but compressed, similar to the Higgs LSP pattern. The lightest chargino and second-lightest neutralino are
both highly degenerate with $\chi_1^0$, with $(m_{\chi_1^\pm}-m_{\chi_1^0})/m_{\chi_1^0}\sim
0.02$ and  $(m_{\chi_2^0}-m_{\chi_1^0})/m_{\chi_1^0}\sim 0.03$.
 Fig.~10 (which should be compared to Fig.~5) shows that the gluino mass distribution tends to peak at higher values but has a sharp cutoff around 2.5 TeV.   As seen in Fig.~11, the ratio
of the gluino mass to the LSP mass and the ratio of the first generation squark masses to the gluino mass are similar to that of the Higgs LSP pattern.  Indeed, the similarities beg the question of whether a separate grouping of the Higgs LSP and mSP2 patterns is in fact warranted.  Of course, this categorization in terms of the lightest four states is in many ways an artificial one to begin with.  The question of the distinguishability of benchmark models for the two cases is nevertheless an interesting one that we defer for future study.

%
\begin{figure}[]
\begin{center}
\begin{tabular}{cc}
\igraph{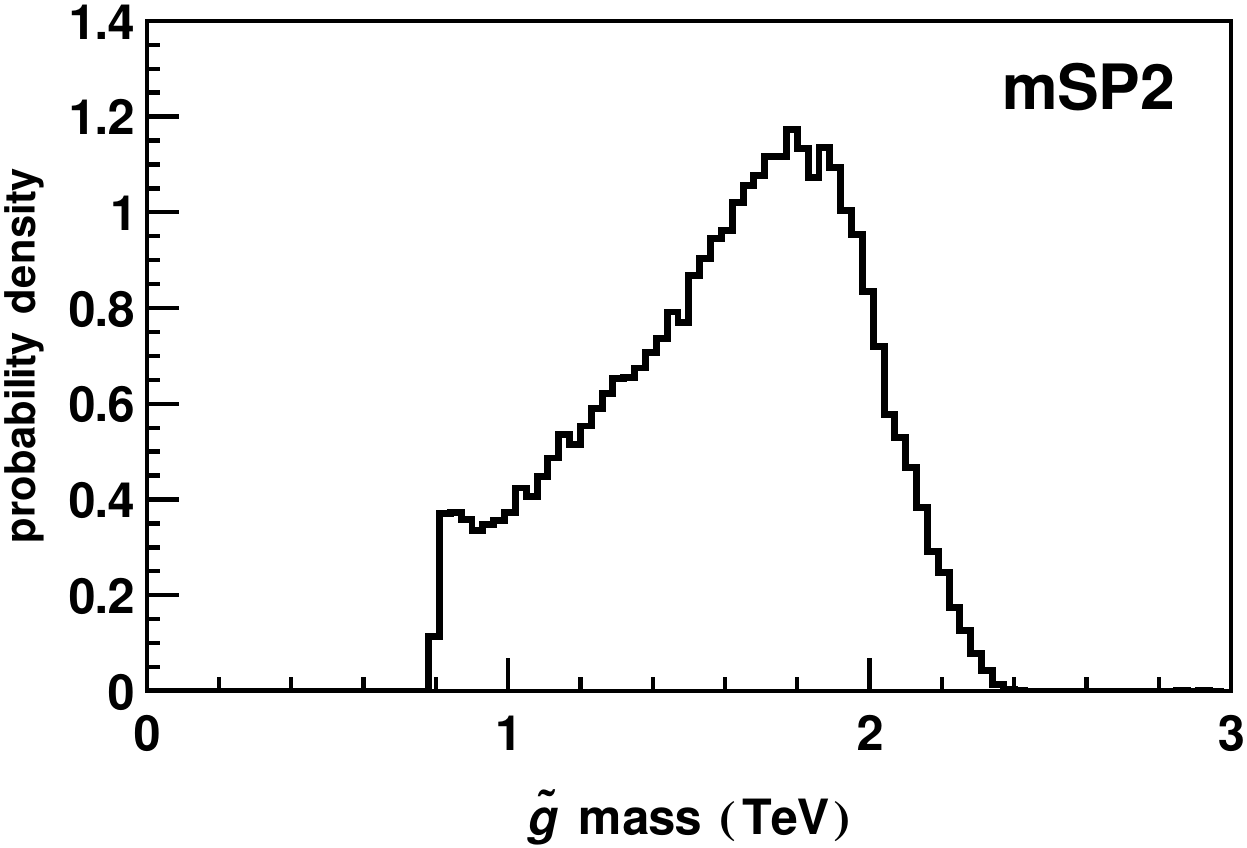} & \igraph{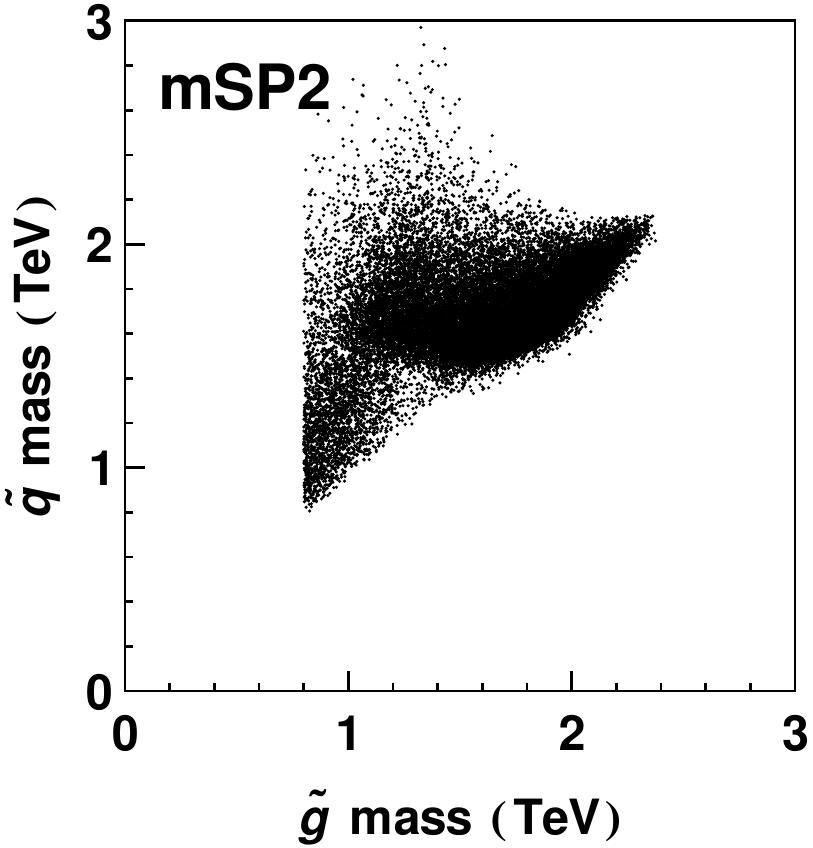}
\end{tabular}
\caption{\textbf{mSP2 pattern: gluino and squark masses}.
Histograms of the mass of the  gluino (left panel) and the squark versus  gluino masses (right panel) for the mSP2 pattern, 
with the direct and indirect limits of Tables 1-2 and the WMAP Preferred dark matter constraints. }
\end{center}
\label{fig:mSP2stuff2}
\end{figure}
\begin{figure}[]
\begin{center}
\begin{tabular}{cc}
\igraph{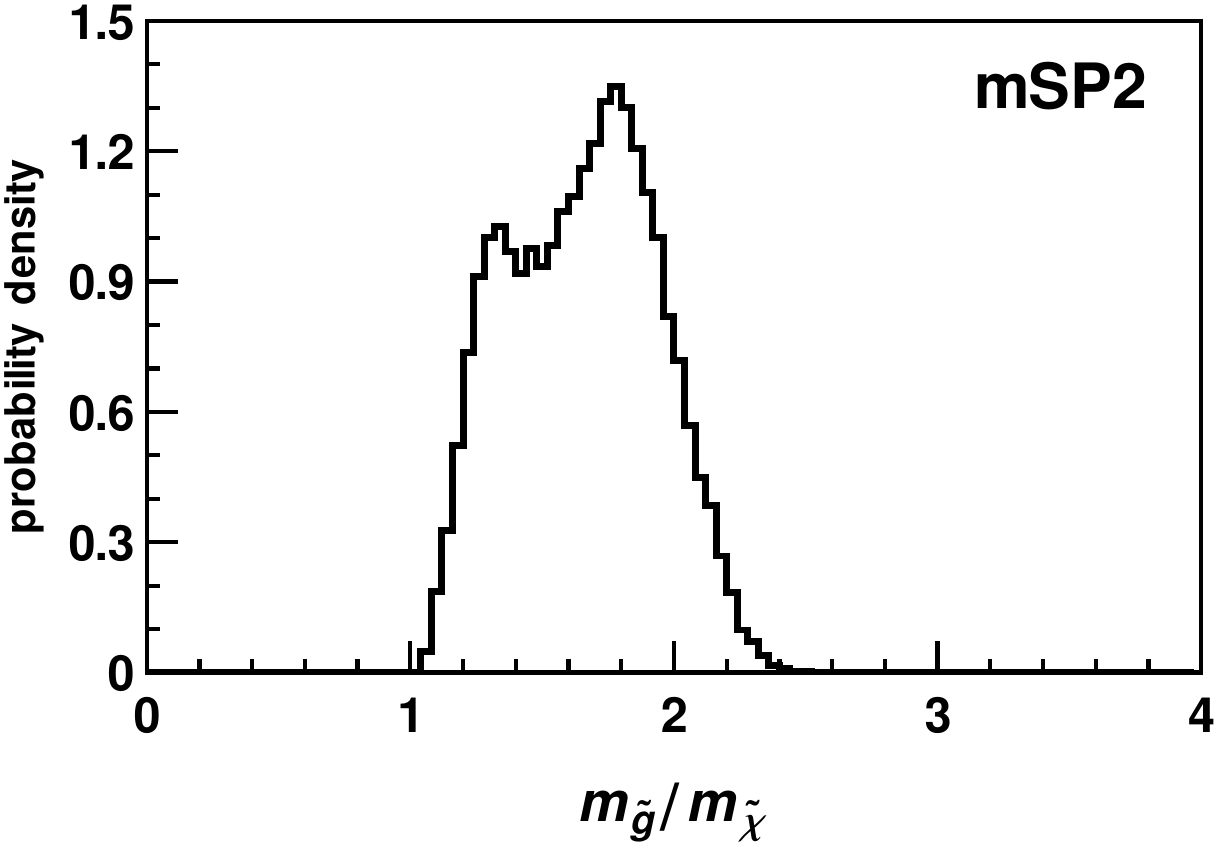} & \igraph{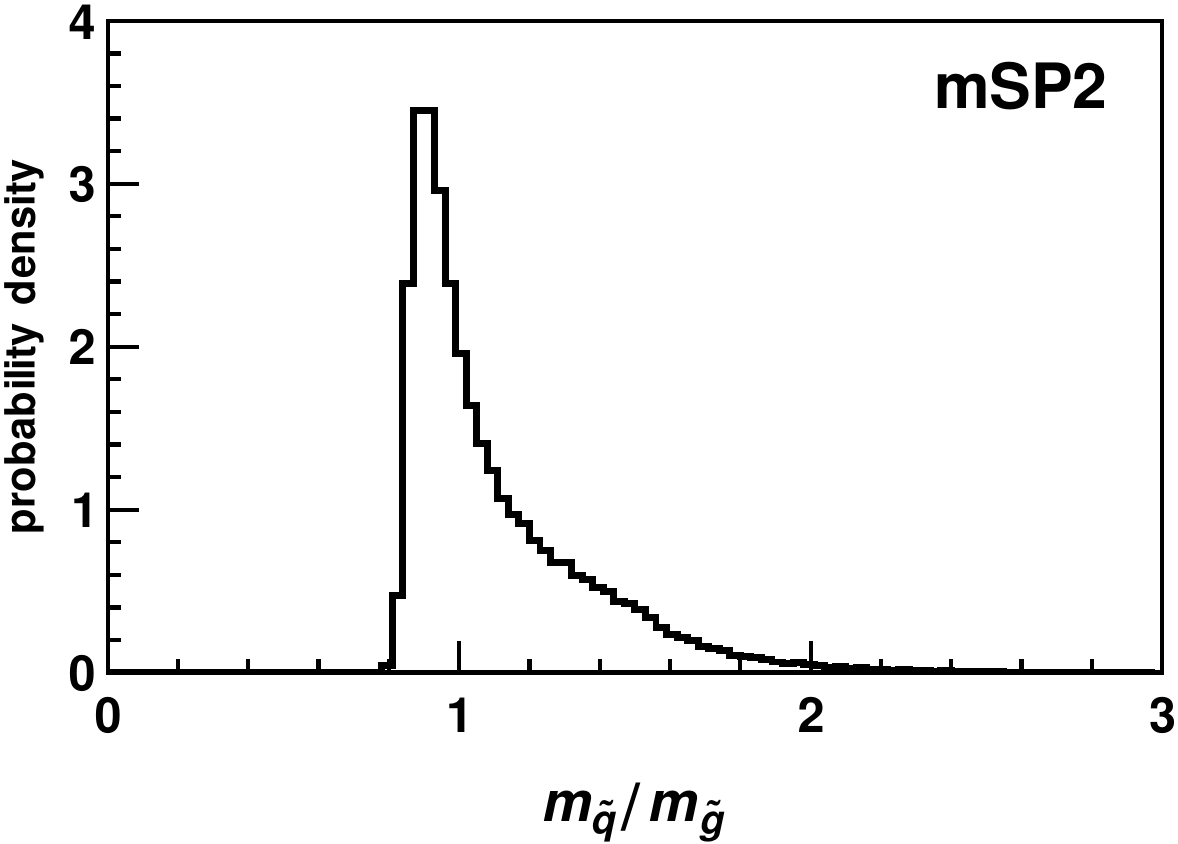}
\end{tabular}
\caption{\textbf{mSP2 pattern: gluino/neutralino mass ratio and squark/gluino mass ratio}.
Histograms of the ratio of the gluino mass to the mass of the lightest neutralino  (left panel) and the ratio of the first generation up-type squark mass to the gluino mass (right panel) for the mSP2 pattern, 
with the direct and indirect limits of Tables 1-2 and the WMAP Preferred dark matter constraints. }
\end{center}
\label{fig:mSP2stuff3}
\end{figure}
\begin{figure}[]
\begin{center}
\begin{tabular}{cc}
\igraph{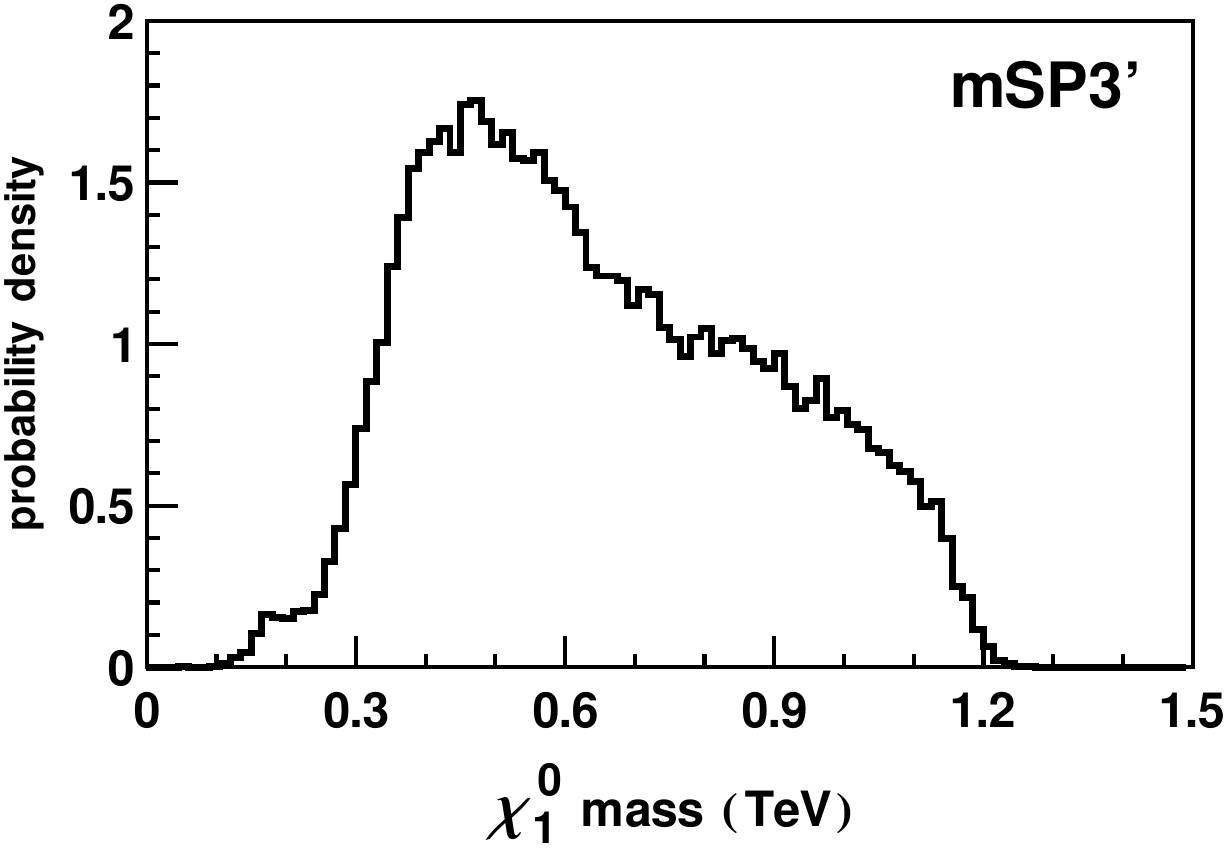} & \igraph{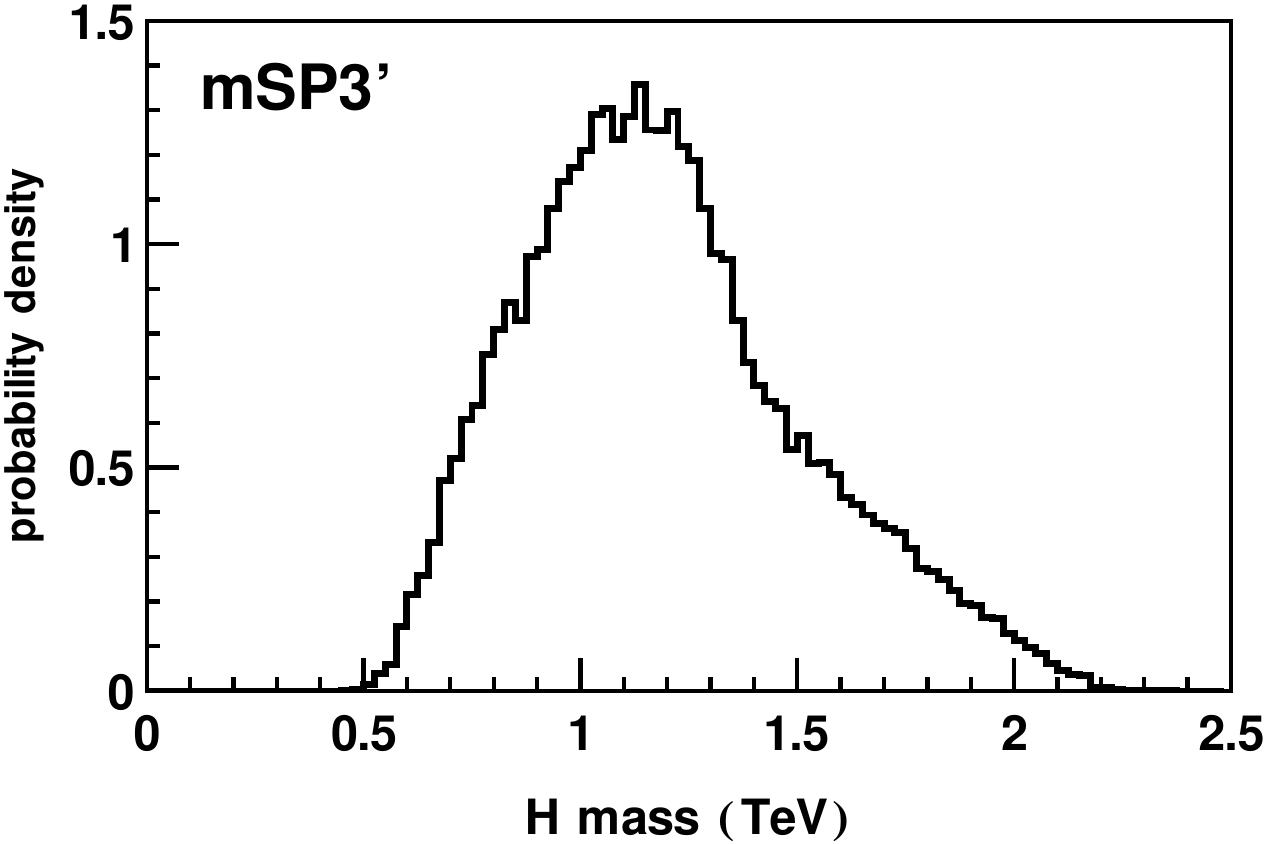}
\end{tabular}
\caption{\textbf{mSP3$'$ pattern: $\chi_1^0$ mass and heavy Higgs mass}.
 Histograms of the mass of the neutralino LSP $\chi_1^0$  (left panel) and the  heavy Higgs boson mass (right panel)
 for the mSP3$'$ pattern, with the direct and indirect limits of Tables 1-2 and the WMAP Preferred dark matter constraints.}
\end{center}
\label{fig:msp3primestuff1}
\end{figure}

For the mSP3$'$ pattern ($\chi_1^0<\chi_2^0<\chi_1^\pm<\tilde{\tau}$), we see significant differences from the Higgs LSP and mSP2 patterns, which can be traced back to the peak values of the mirage scale as shown in Fig.~1.  Recall that this pattern was dominated by higher values of the mirage unification scale, indicating that typical models in this category would more closely resemble mSUGRA/CMSSM models. We see from Fig.~12 that the LSP tends to be lighter than in the Higgs LSP and mSP2 patterns, while the peak value of the heavy Higgs mass has shifted to higher values.
\begin{figure}[]
\begin{center}
\begin{tabular}{cc}
\igraph{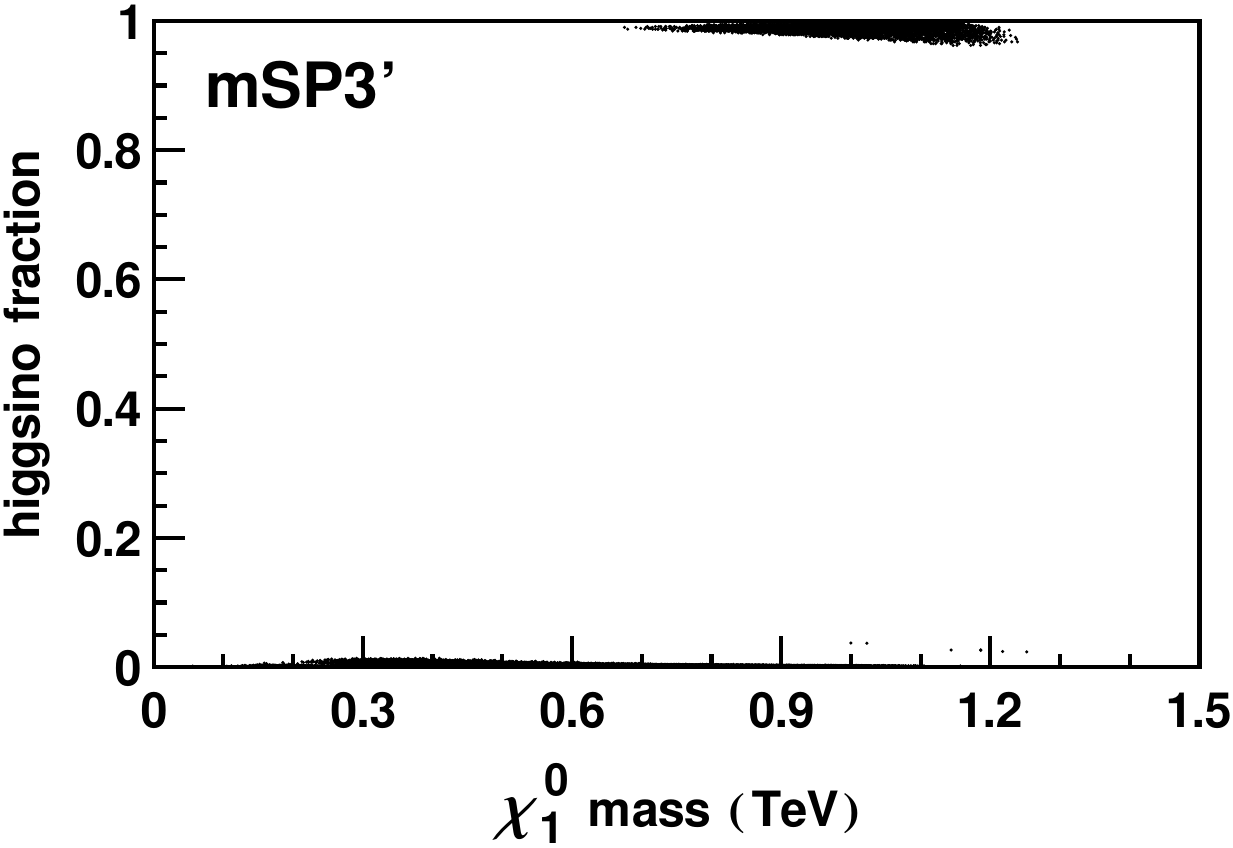} & \igraph{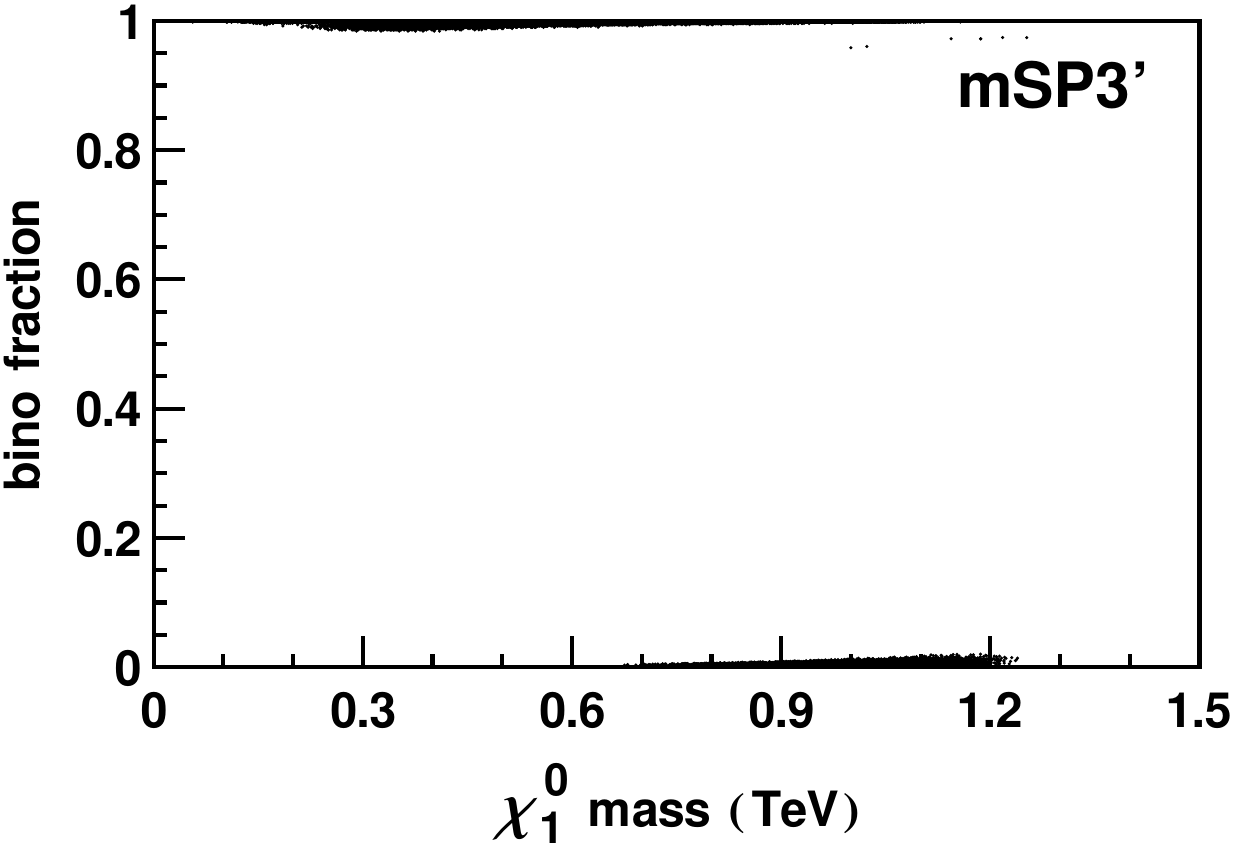}
\end{tabular}
\caption{\textbf{mSP3$'$ pattern: higgsino and bino fraction}.
Histograms of the higgsino (left panel) and bino  (right panel) fractions of $\chi_1^0$ as a function of the $\chi_1^0$ mass for the mSP3$'$ pattern, with the direct and indirect limits of Tables 1-2 and the WMAP Preferred dark matter constraints.}
\end{center}
\label{fig:mSP3phiggsinobino}
\end{figure}
Fig.~13 demonstrates that the LSP is most often bino-dominated, with a subset of models with heavier LSP's that are higgsino dominated.   For this pattern, $(m_{\chi_1^\pm}-m_{\chi_1^0})/m_{\chi_1^0}$ and  $(m_{\chi_2^0}-m_{\chi_1^0})/m_{\chi_1^0}\sim 0.26$ on average.  The lightest chargino and the second-lightest neutralino are typically quite degenerate, since for the majority of models there is a bino-like $\chi_1^0$ and a nearly degenerate  wino-like pair $\chi_1^\pm$, $\chi_2^0$, as is often the case in mSUGRA models.

\begin{figure}[]
\begin{center}
\begin{tabular}{cc}
\igraph{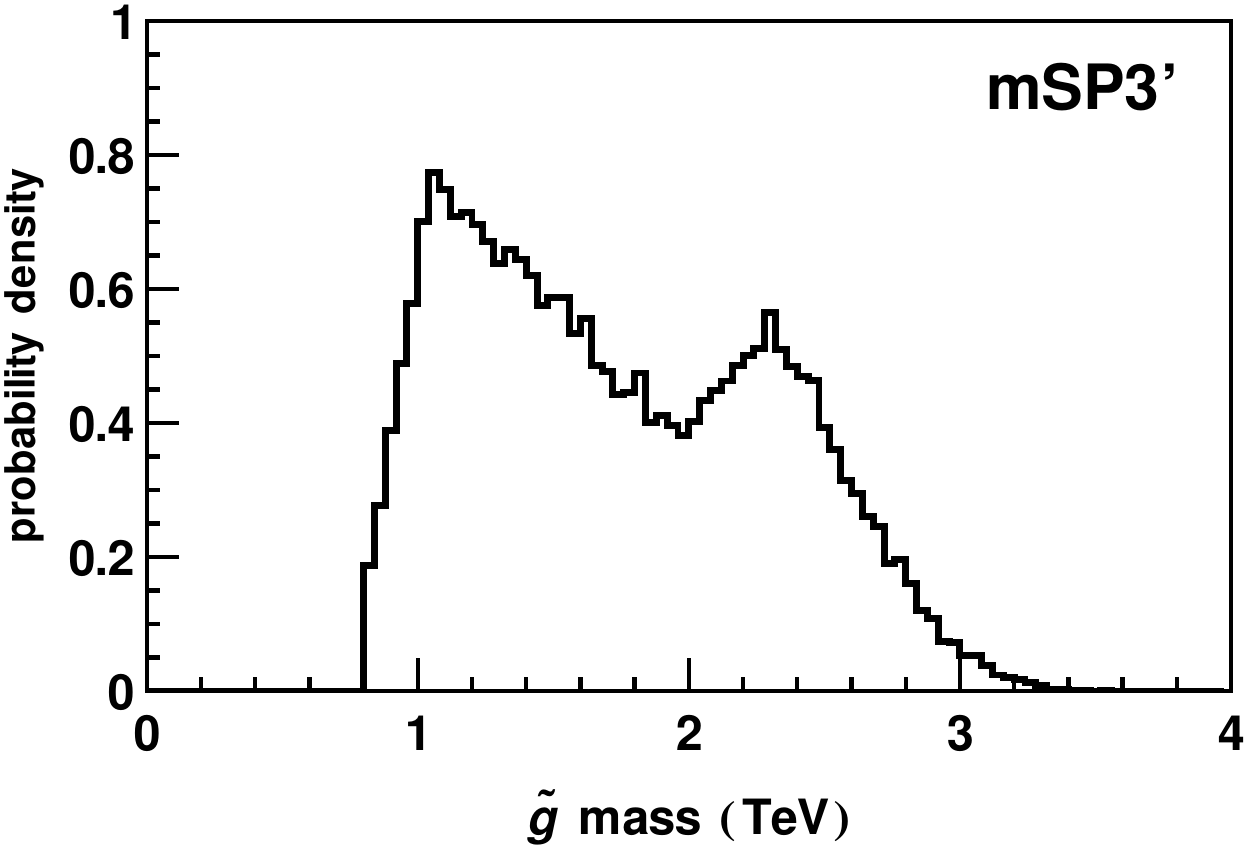} & \igraph{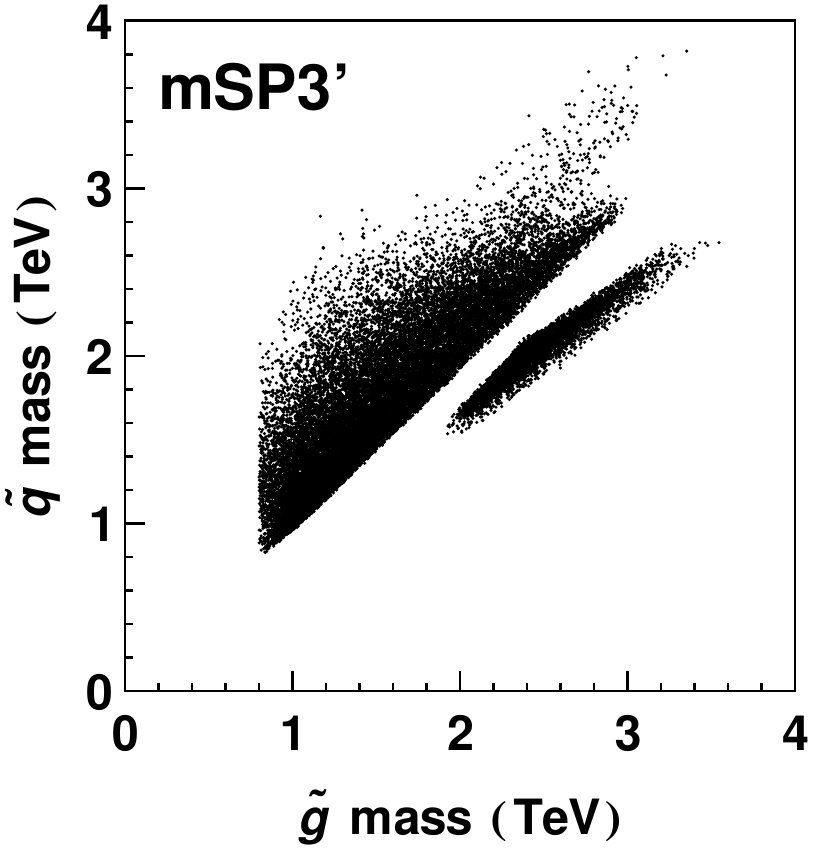}
\end{tabular}
\caption{\textbf{mSP3$'$ pattern: gluino and squark mass distributions}.
Histograms of the gluino mass (left panel) and the squark versus gluino (right panel) for the mSP3$'$ pattern, 
with the direct and indirect limits of Tables 1-2 and the WMAP Preferred dark matter constraints.}
\end{center}
\label{fig:msp3primestuff2}
\end{figure}
\begin{figure}[!h]
\begin{center}
\begin{tabular}{cc}
\igraph{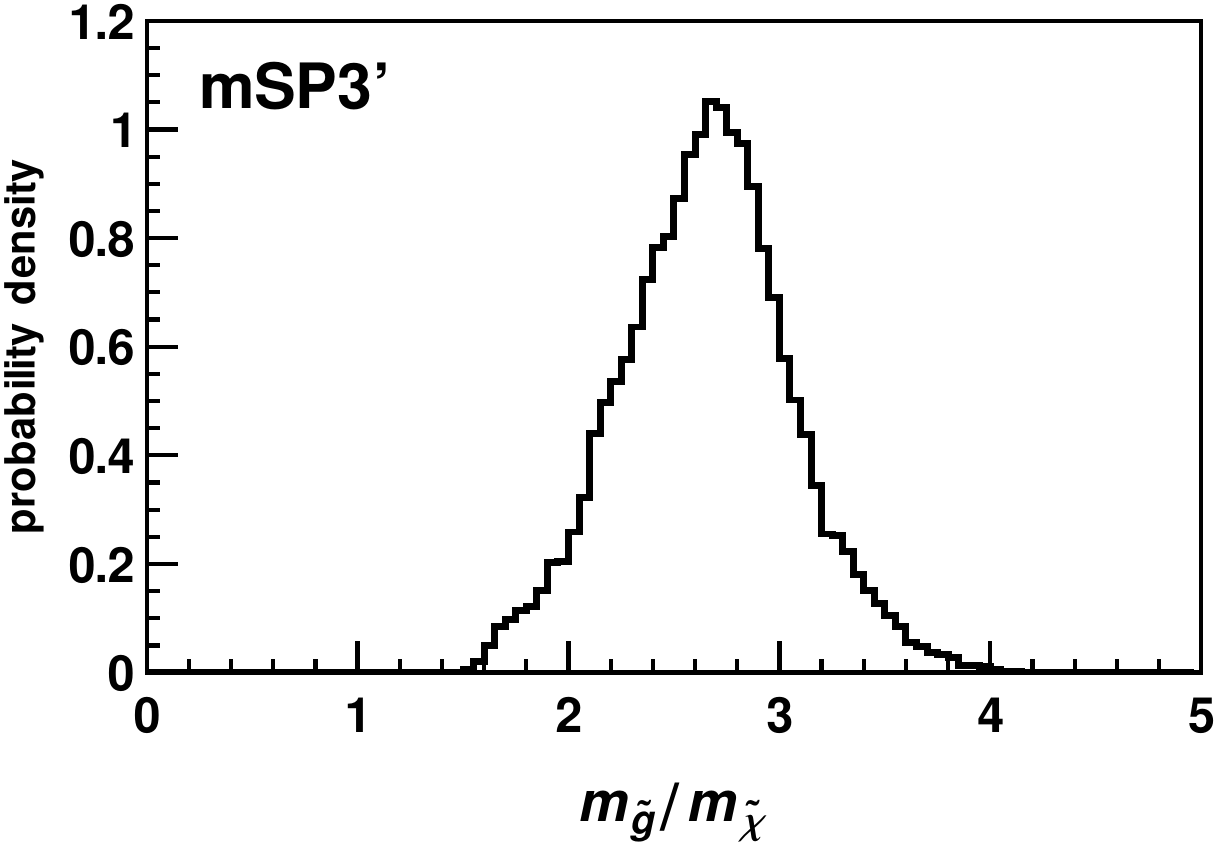} & \igraph{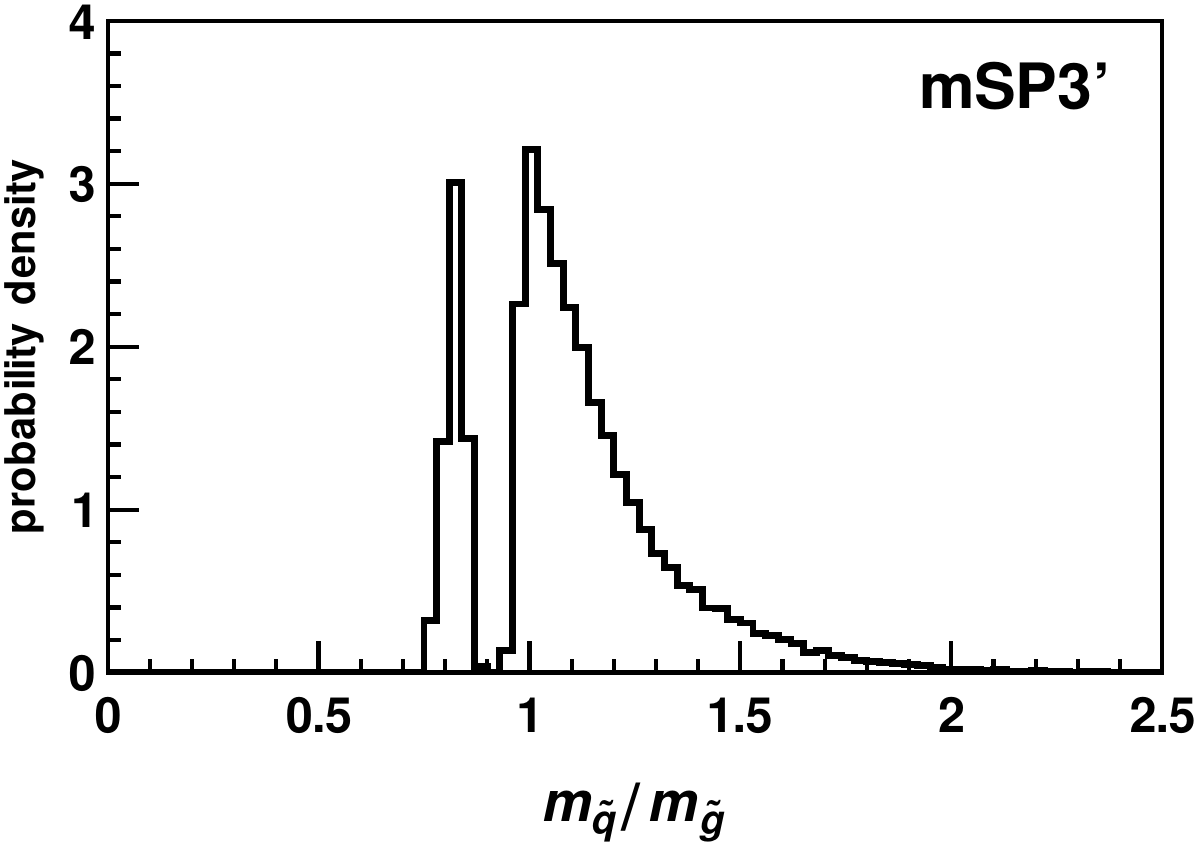}
\end{tabular}
\caption{\textbf{mSP3$'$ pattern: gluino/LSP mass ratio and squark/gluino mass ratio}.
 Histograms of the ratio of the mass of the gluino to the LSP mass  (left panel) and the ratio of the first generation up-type squark mass to the gluino mass (right panel) for the mSP3$'$ pattern, with the direct and indirect limits of Tables 1-2 and the WMAP Preferred dark matter constraints.}
\end{center}
\label{fig:msp3primestuff3}
\end{figure}

The distributions of the gluino and squark masses are shown in Fig.~14, and the $m_{\tilde{g}}/m_{\chi_1^0}$, and $m_{\tilde{q}}/m_{\tilde{g}}$ distributions are shown in Fig.~15.  Both distributions have bimodal features, with the gluino mass distribution that has the dominant peak at lower values than that of mSP2.  The squark versus gluino mass while the ratio of the gluino to LSP masses peaks at $\sim 3$, representing a shift toward mSUGRA/CMSSM-like features.  The one sharp peak at  $m_{\tilde{q}}/m_{\tilde{g}}<1$ corresponds to the low mirage scale behavior found also in the mSP2 case, but there is also a much larger peak at higher values that corresponds to models with a high mirage unification scale.

\subsubsection{mSUGRA-like DMM patterns:  mSP6$'$}
The final DMM hierarchy pattern we will consider is the mSP6$'$ pattern, in which the NLSP is the lightest stau.  As we will see, this pattern most strongly resembles typical hierarchy patterns found in mSUGRA/CMSSM models (significantly more so than the DMM mSP3$'$ pattern described in the previous subsection), though the spectrum will once again exhibit a (relatively) compressed feature that is characteristic of DMM models.

We begin by showing the distribution of the LSP mass and the bino fraction of the LSP as a function of its mass, which are given in Fig.~16.  This pattern is characterized by a bino-dominated LSP that is correspondingly much lighter than the higgsino-dominated neutralino LSP's found in the Higgs LSP pattern and the mSP2 pattern.   In fact, the mass pattern of the
charginos and neutralinos very much resembles that of mSUGRA/CMSSM models, with the lowest mass state given by bino-like
LSP and the next lightest states in the electroweak chargino/neutralino sector consisting of degenerate wino-like pair
consisting of $\chi_2^0$ and $\chi_1^\pm$.  The splitting of the $\chi_1^0$ mass compared to the $\chi_1^\pm$ and $\chi_2^0$ is more
substantial for this pattern,  with both $(m_{\chi_1^\pm}-m_{\chi_1^0})/m_{\chi_1^0}$ and
$(m_{\chi_2^0}-m_{\chi_1^0})/m_{\chi_1^0}\sim 0.6$ on average.

\begin{figure}[]
\begin{center}
\begin{tabular}{cc}
\igraph{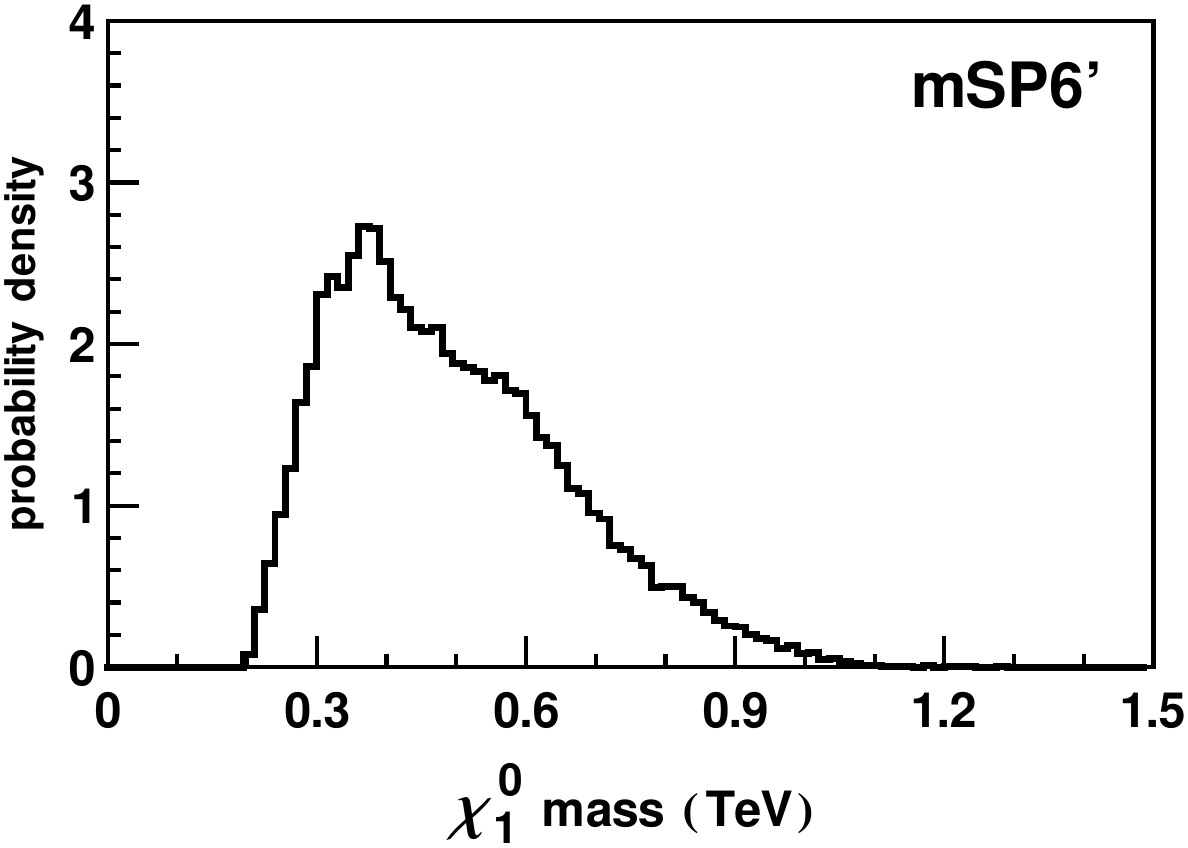} & \igraph{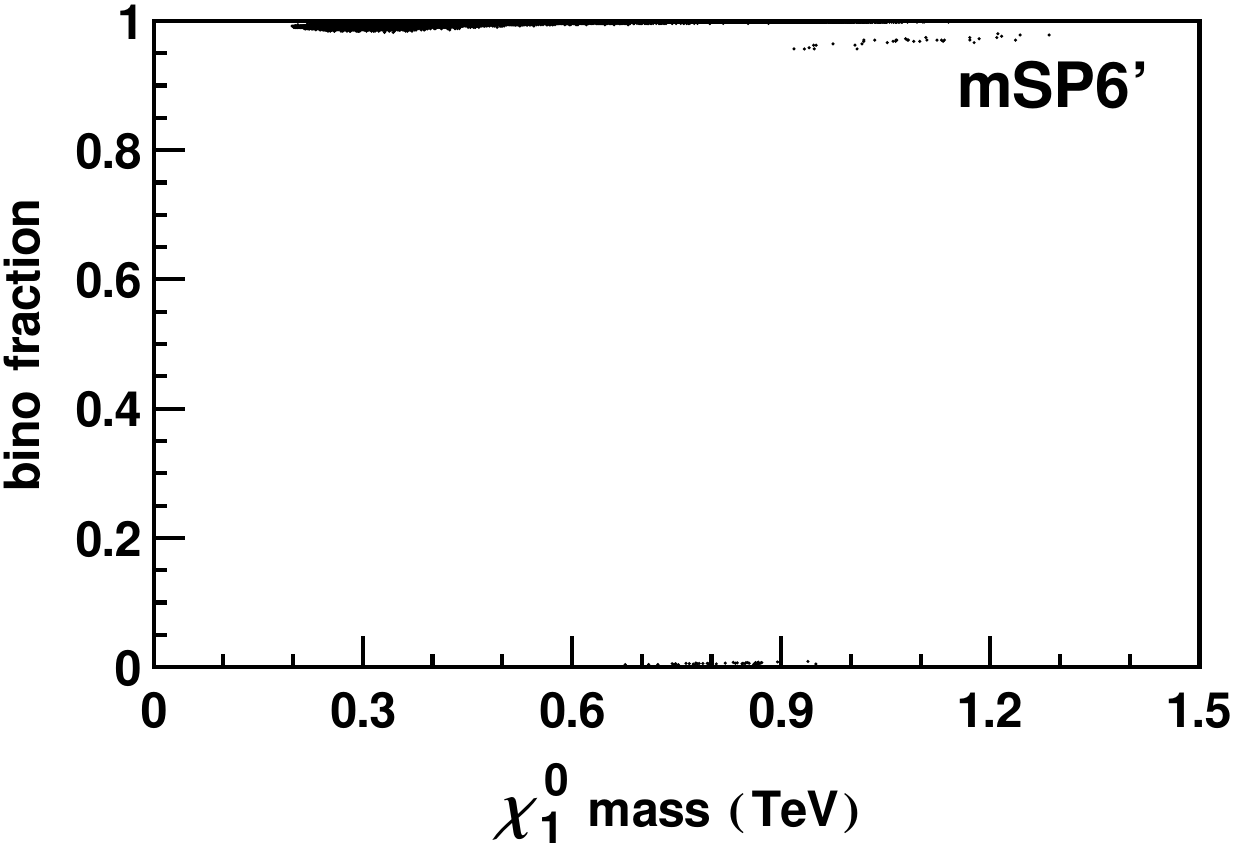}
\end{tabular}
\caption{\textbf{mSP6$'$ pattern: $\chi_1^0$ mass and bino fraction}.
Histograms of the mass of the neutralino LSP $\chi_1^0$  (left panel) and the bino fraction of $\chi_1^0$ as a function of the LSP mass (right panel) for the mSP6$'$ pattern, with the limits of Tables 1-2 and the WMAP Preferred dark matter constraints.}
\end{center}
\label{fig:mSP6neut1mass}
\end{figure}

\begin{figure}[!h]
\begin{center}
\begin{tabular}{cc}
\igraph{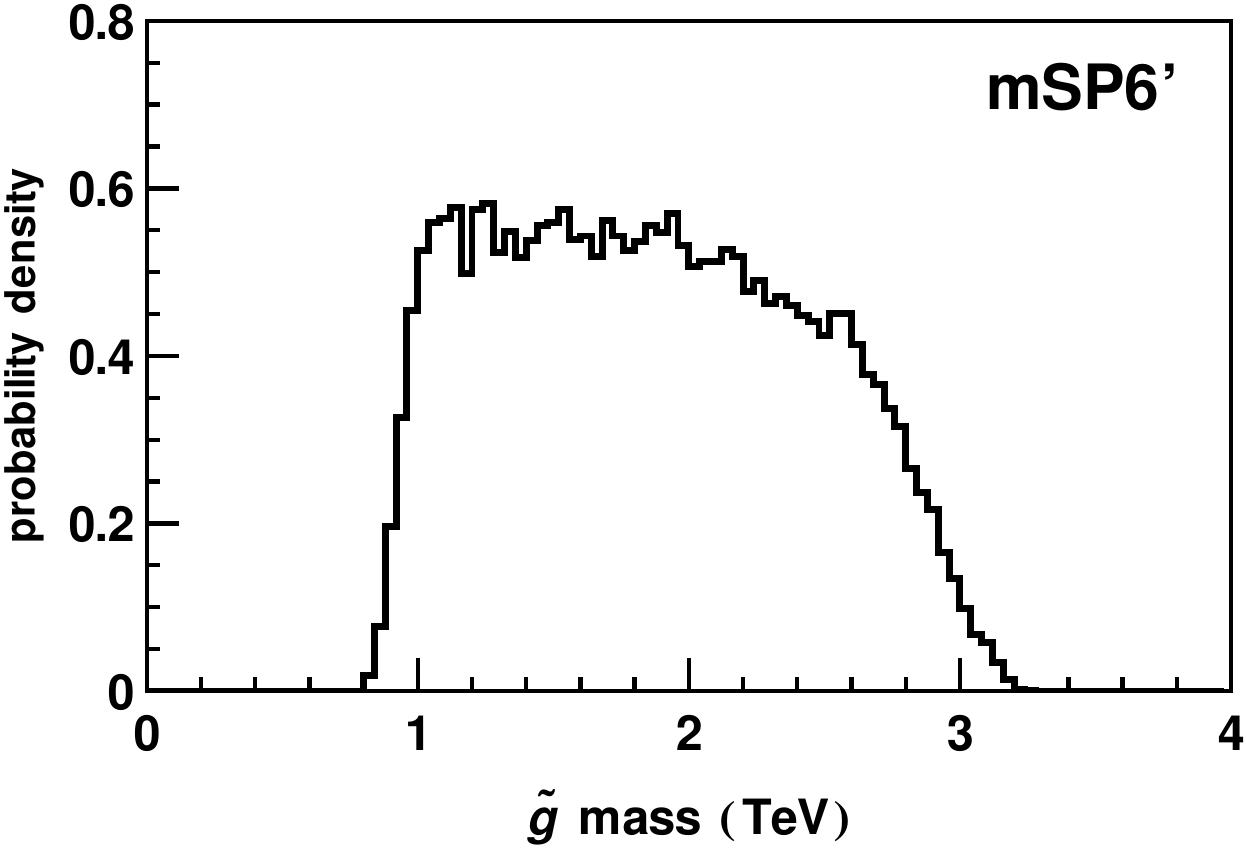}& \igraph{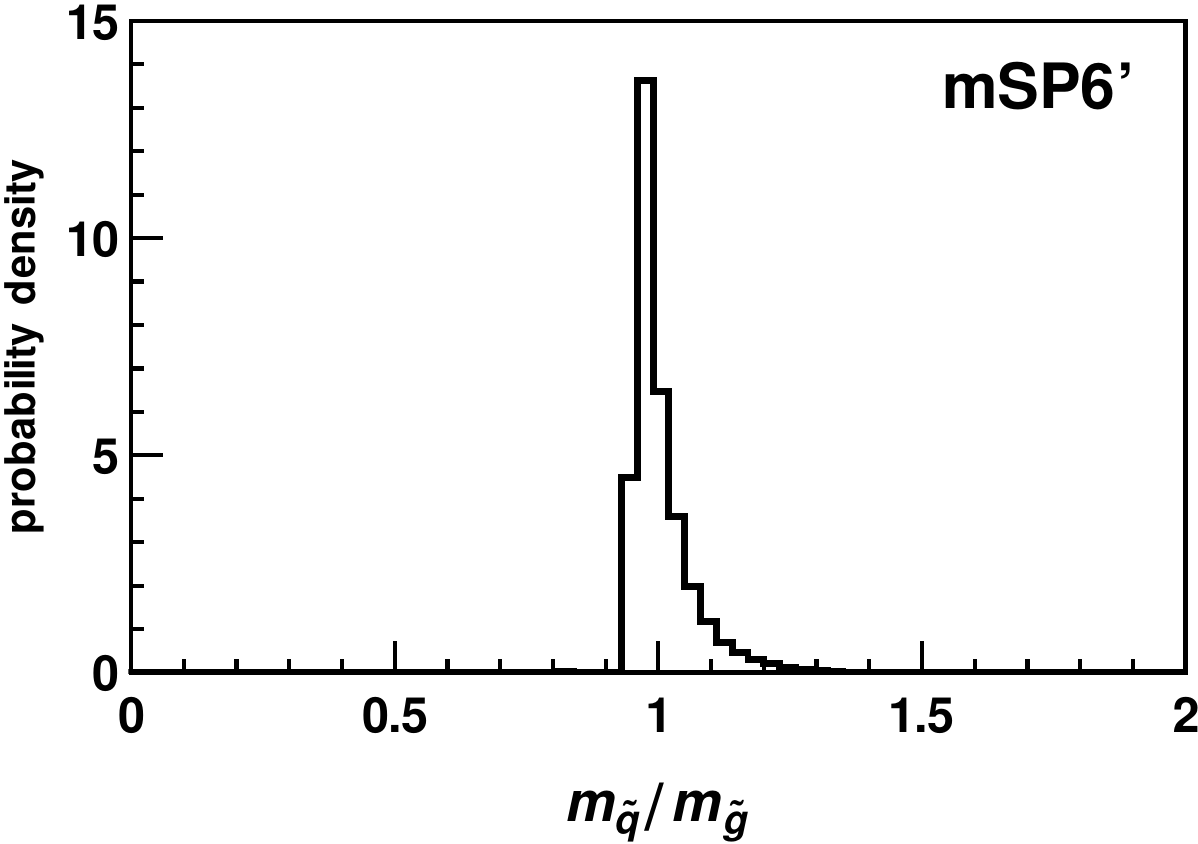}  
\end{tabular}
\caption{\textbf{mSP6$'$ pattern: gluino mass and
$m_{\tilde{q}}/m_{\tilde{g}}$ distributions}.  Histograms for the mSP6$'$ of the gluino mass (left panel) and the ratio
of the first generation squark mass to the gluino mass (right panel), 
with the bounds of Tables 1-2 and the WMAP Preferred dark matter constraints.}
\end{center}
\label{fig:msp6LSPstuff1}
\end{figure}

\begin{figure}[!h]
\begin{center}
\begin{tabular}{cc}
\igraph{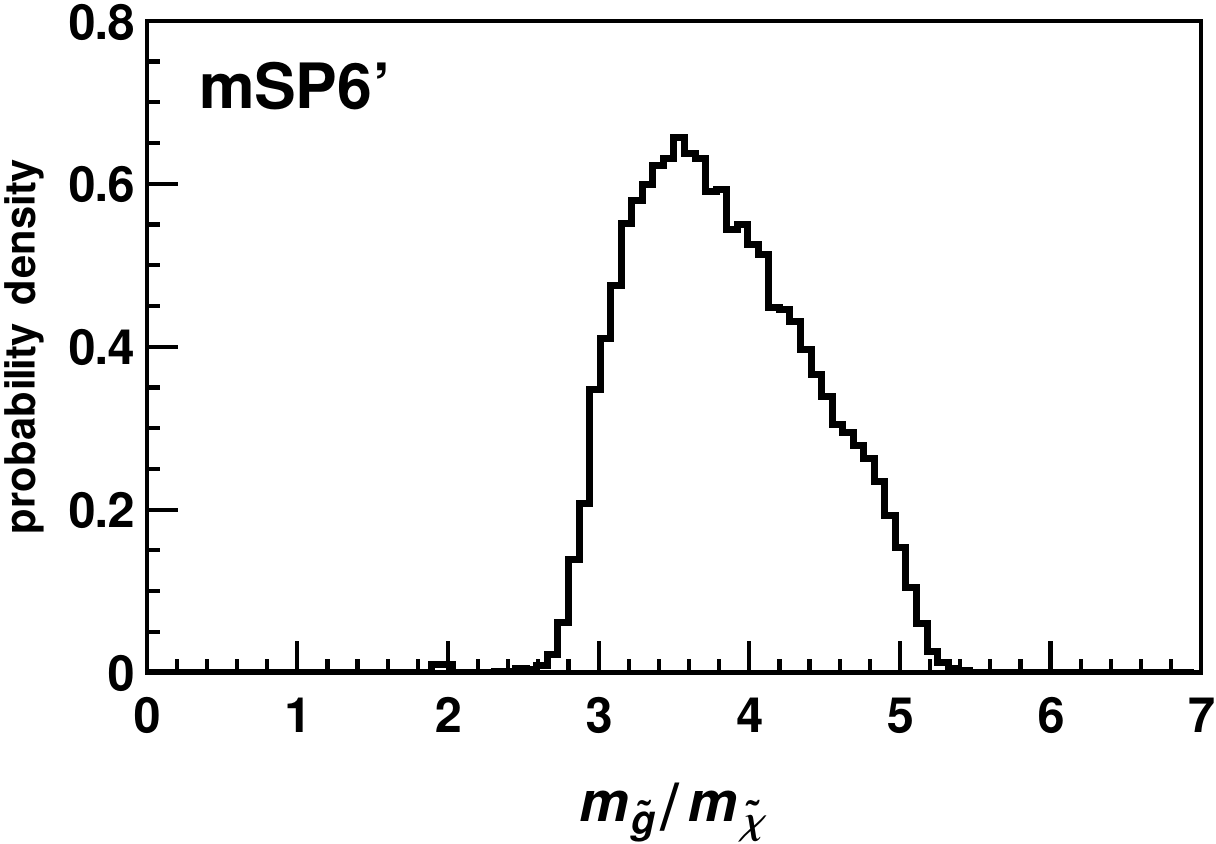} & \igraph{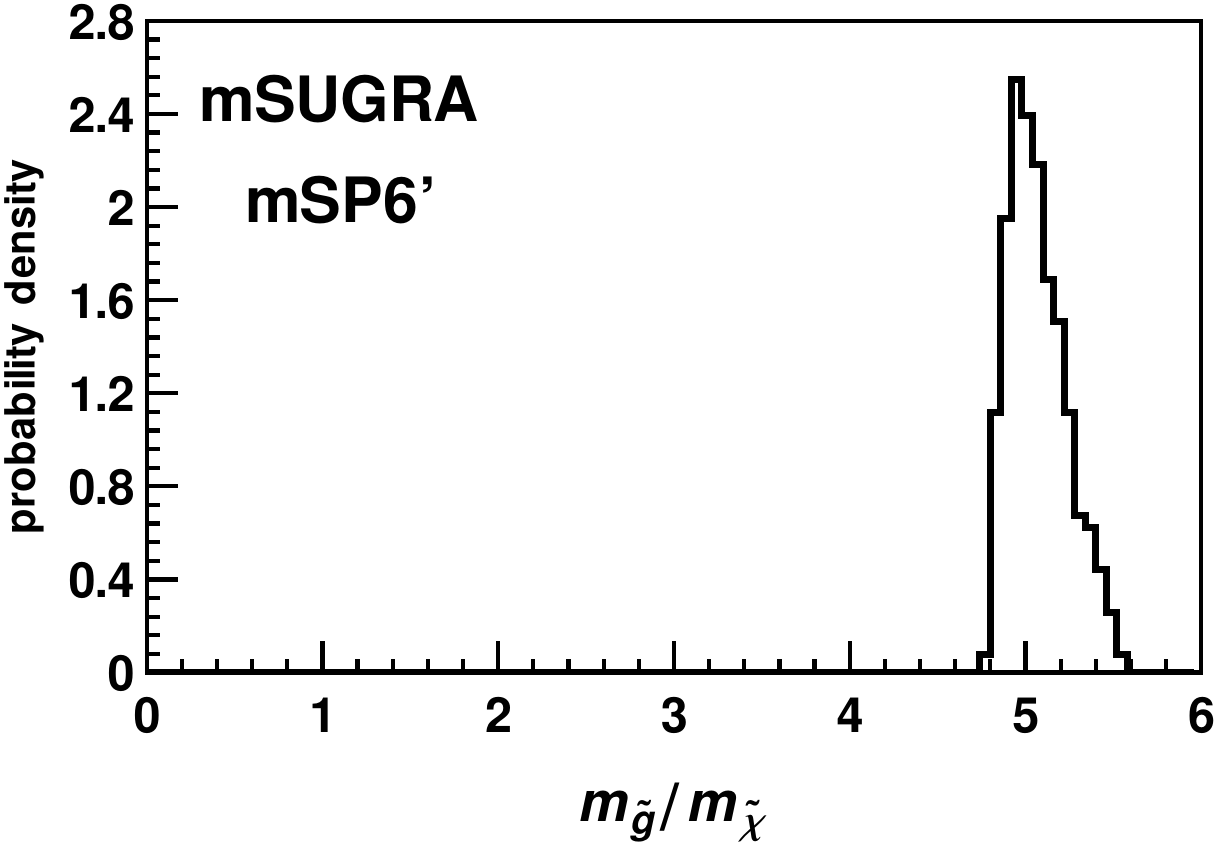}
\end{tabular}
\caption{\textbf{mSP6$'$ pattern: $m_{\tilde{g}}/m_{\chi_1^0}$ and comparisons with mSUGRA/CMSSM}.  Histograms for the mSP6$'$ of the
the ratio of the gluino mass to the LSP mass in DMM (left panel) and mSUGRA/CMSSM (right panel), 
with the direct and indirect limits of Tables 1-2 and the WMAP Preferred bound.}
\end{center}
\label{fig:msp6LSPstuff2}
\end{figure}


The spectrum of the mSP6$'$ pattern is less compressed on the average
than the other DMM pattern considered.  In Figs.~17 and 18, we plot histograms of the gluino mass (upper left panel) and the ratio of the
gluino mass to the LSP mass as well as the ratio of the
typical first generation squark mass to the gluino mass.  The $m_{\tilde{g}}/m_{\chi_1^0}$ distribution is peaked at
significantly larger values than the other DMM patterns, which is as
expected since the mSP6$'$ pattern favors high values of the mirage
unification scale.  In mSUGRA/CMSSM models,  the gluino to LSP mass ratio for the analogous mSP6$'$ pattern is peaked at still higher values, as shown in Fig.~18.  
The ratio of the squark to gluino masses in the DMM mSP6$'$ pattern, in contrast, is very sharply peaked at  $m_{\tilde{q}}/m_{\tilde{g}}
\sim 1$, indicating that on average the first generation squark and
gluino masses tend to be clustered and significantly heavier
than the LSP.

\begin{figure}[!h]
\begin{center}
\begin{tabular}{c}
\igraphnew{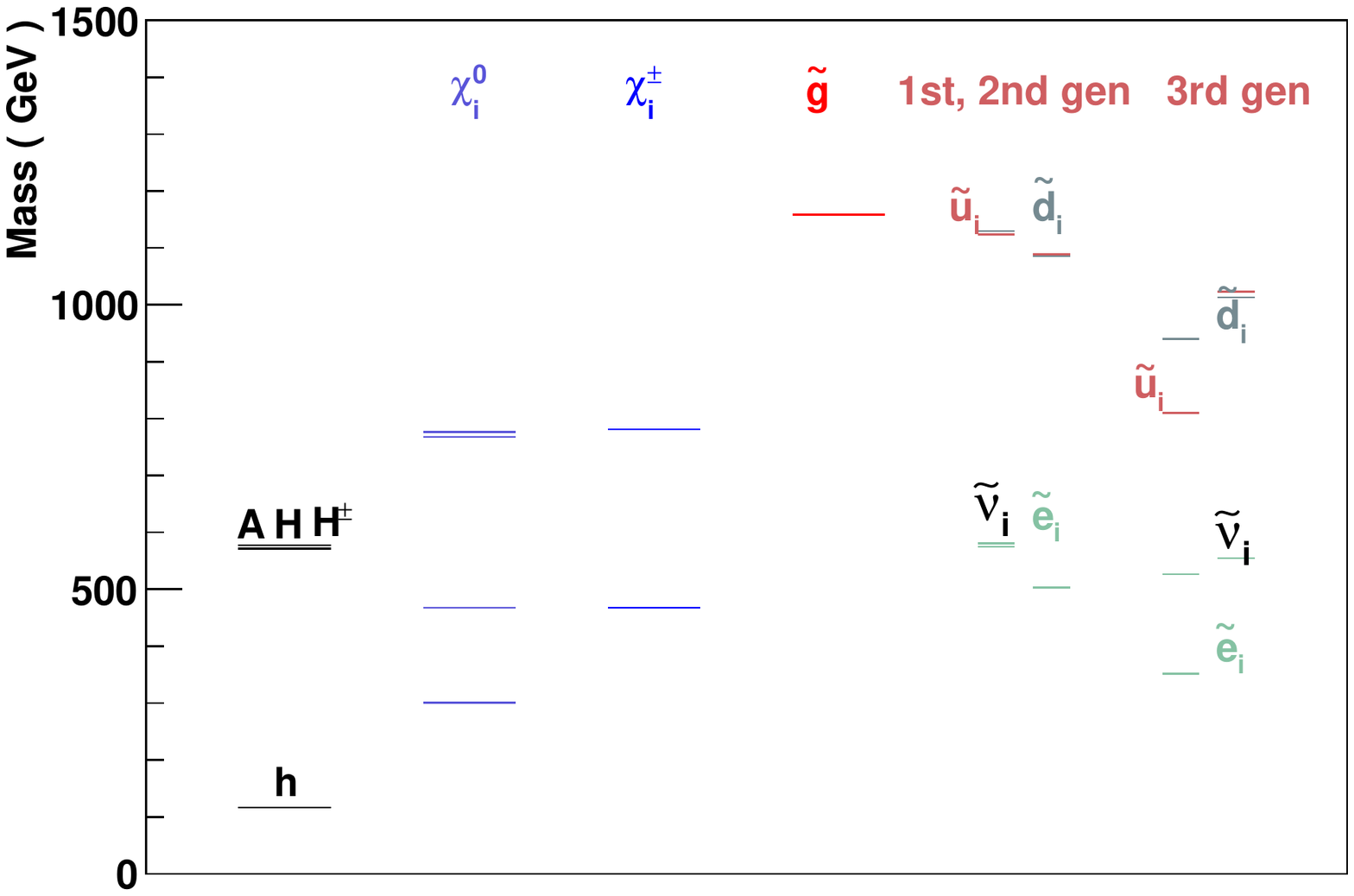} 
\end{tabular}
\caption{\textbf{DMM benchmark point \#2 (mSP6$'$ pattern): mass spectrum}.  The mass spectrum for DMM benchmark point \#2, as shown in Table~8.}
\end{center}
\label{fig:ms6pspectrum}
\end{figure}
\begin{table}
\begin{center}
\begin{tabular}{|cc|cc|cc|cc|}
\hline
\hline
Particle & Mass & Particle & Mass & Particle & Mass & Particle & Mass \\
\hline
$h$ & 116 & $H$ &571 &$\tilde{u}_L$ & 1123 & $\tilde{u}_R$ & 1089 \\
$A$ & 571 & $H^\pm$ & 577& $\tilde{d}_L$ & 1130 & $\tilde{d}_R$ & 1085 \\
$\chi^0_1$ & 301 & $\chi^0_2$ & 468 & $\tilde{\ell}_L$ & 581 & $\tilde{\ell}_R$ & 503 \\
$\chi^0_3$ & 768 & $\chi^0_4$ & 776 &$\tilde{t}_1$ & 810 & $\tilde{t}_2$ & 1023 \\
$\chi^\pm_1$ & 468 & $\chi^\pm_2$ & 781 & $\tilde{b}_1$ & 940 & $\tilde{b}_2$ & 1013 \\
$\tilde{g}$ & 1158 & $\tilde{\tau}_1$ & 351 & $\tilde{\tau}_2$ & 554 & &   \\
 \hline 
 \hline 
\end{tabular}
\end{center}
\caption{The particle masses (in GeV) for the  DMM benchmark point \#2, as shown in Fig.~19.}
\end{table}
To illustrate these points, we have selected a DMM benchmark point for this pattern (Point \#2), for which the mass spectrum is shown in Fig.~19 and the particle masses are shown in Table~8.  The parameter values for this point are $N=2$, $M_{\rm mess}=1.6\times 10^{12}$ GeV, $M_0=649$, $\alpha_m=0.33$, $\alpha_g=-0.013$, and $\tan\beta=42.2$.   We see that this spectrum is quite different than that of the Higgs LSP benchmark point of Fig.~7.   The overall scale is much lighter,  as reflected by a value of $M_0$ that is less than one TeV; this is possible with the dark matter constraints because the LSP is bino-dominated.   The gluino remains the heaviest superpartner, and is similar in mass with the first and second generation squarks, while the third generation squarks are lighter.

\section{Summary and Conclusions}
\label{sec:conc}

In this paper, we have taken a bottom-up approach to understanding
the phenomenology of deflected mirage mediation, a string-motivated
scenario involving comparable contributions to supersymmetry
breaking from gravity, anomaly and gauge mediation. Using the hierarchy of mass eigenstates for the superpartners as
an organizing principle, we have investigated the landscape of these
hierarchies for a class of DMM models, and compared the results with a model set of mSUGRA/CMSSM models.  Beginning with a
substantial data set, we have
applied progressive cuts motivated by phenomenology: radiative
electroweak symmetry breaking, a neutral lightest superpartner that
can be a dark matter candidate, direct search limits on superpartner
masses and indirect limits from rare processes. Of crucial
importance was the imposition of thermal relic density constraints
for the LSP. Requiring the LSP to account for all
non-baryonic dark matter restricts the parameter set in severe yet
interesting ways, while a more relaxed constraint (such as only
using the WMAP measurement as an upper bound) allows for a greater
variety of model possibilities. 
We have used this analysis to determine benchmark points, as shown in Figs.~7 and 19, that can be used for detailed collider studies. 
 Our benchmark point \#1 is a Higgs LSP pattern, characterized by a heavy but compressed spectrum and a higgsino-dominated LSP, while our benchmark point \#2 is more reminiscent of mSUGRA/CMSSM spectra with a stau NLSP, but with a slightly less stretched spectrum.  These points are quite different from prior DMM benchmarks that were motivated within the top-down approach in previous phenomenological studies.  Our new benchmark points are not intended to be viewed as ``smoking gun" signatures of DMM models: while patterns such as the Higgs LSP pattern appear to be more difficult to obtain in mSUGRA/CMSSM, it is certainly possible that Higgs LSP patterns can emerge even in simple one-parameter extensions of mSUGRA/CMSSM models.  Rather, our long-term goal is simply to get a more complete handle on the DMM parameter space and its phenomenological implications.  Indeed, this work represents just the first step toward that goal since we consider here only a restricted set of DMM parameters, as opposed to the full multidimensional parameter space.

\begin{figure}[!h]
\begin{center}
\begin{tabular}{cc}
\igraph{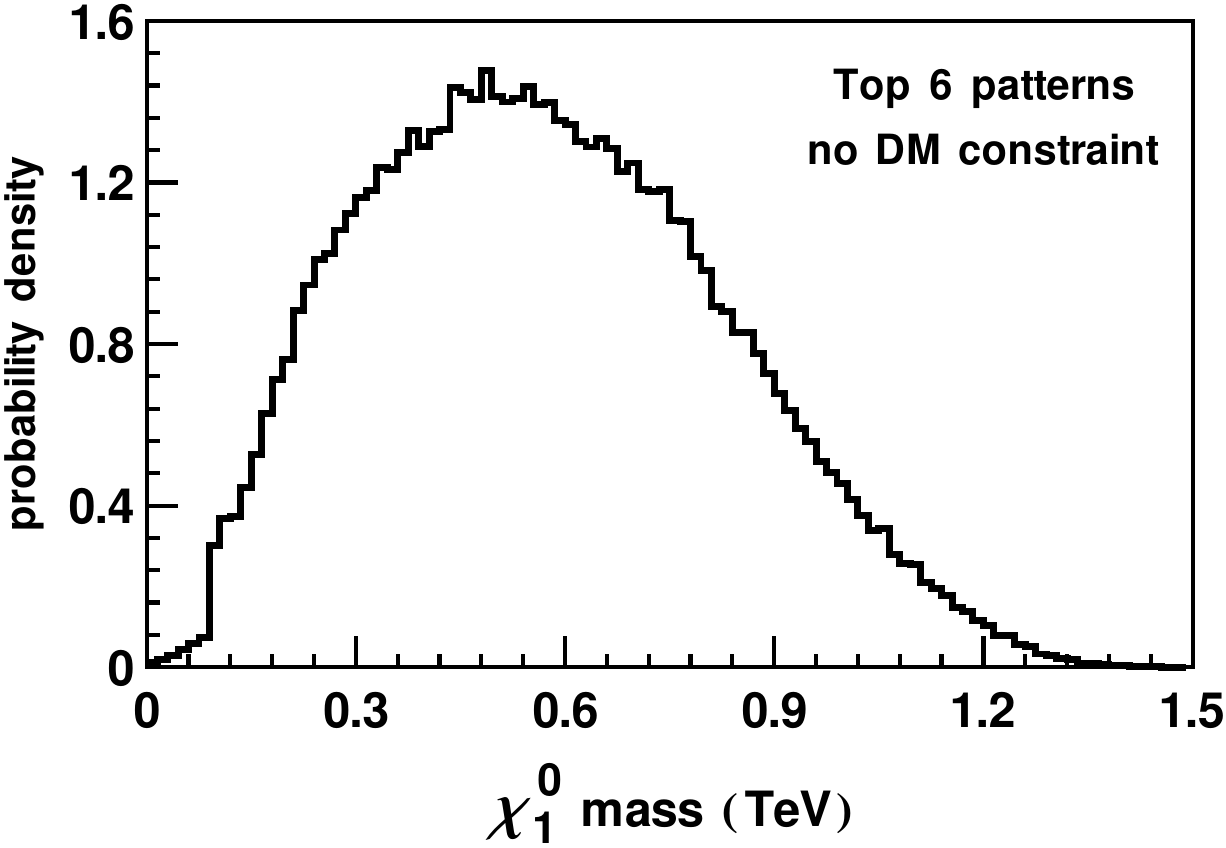} & \igraph{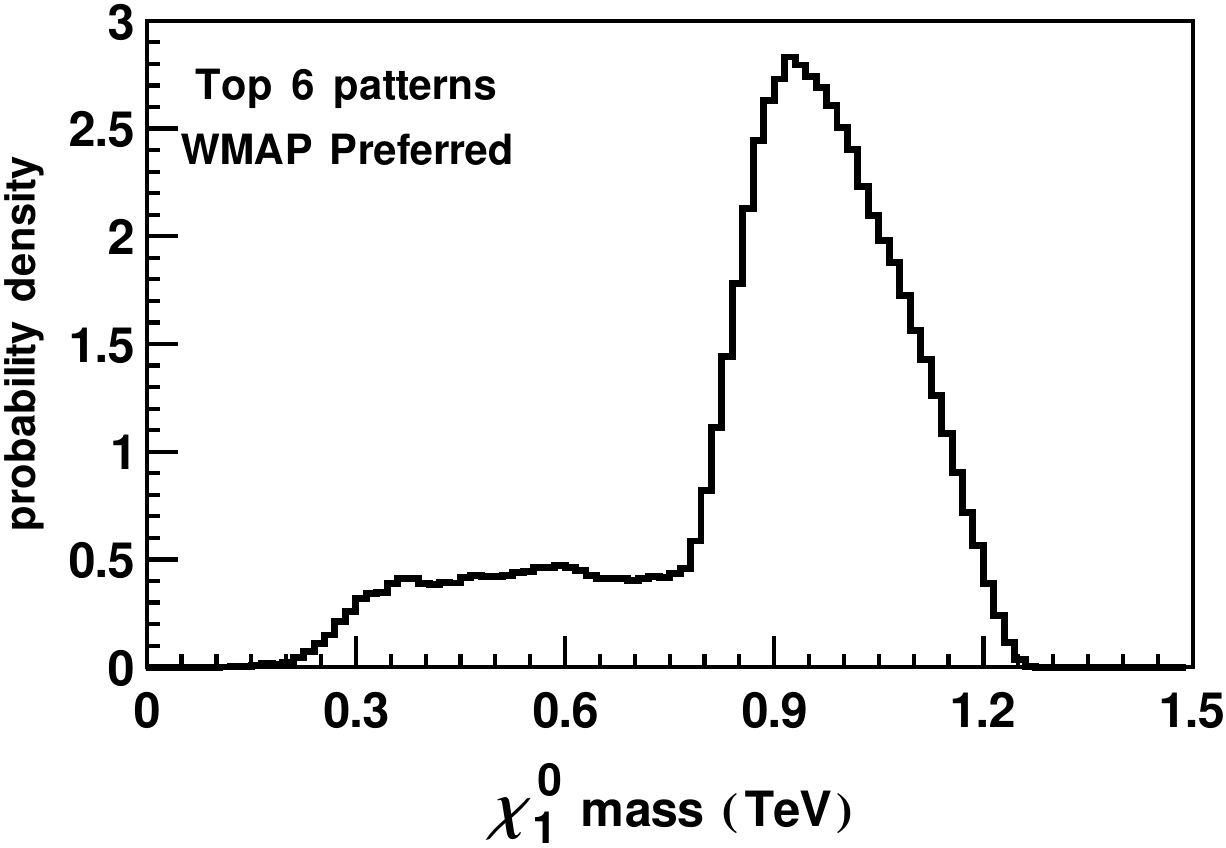}
\end{tabular}
\caption{\textbf{LSP mass distribution}.  We show the distribution of the neutralino LSP mass for the top six hierarchy patterns in DMM models, with no dark matter constraint (left panel) and the WMAP Preferred
constraint (right panel).}
\end{center}
\label{fig:summary_LSPmass}
\end{figure}

The significance of the dark matter requirement should not be
understated. Consider the case of Fig.~20, 
in which we plot the distribution in mass for the lightest
neutralino for the aggregate set of our top six hierarchy patterns
(Higgs LSP, mSP2, mSP3$'$, mSP2$'$, mSP6$'$, and mSP3). The left
panel shows the distribution with absolutely no dark matter
constraint imposed, revealing a general preference for masses of
order a few hundred~GeV with a wide distribution. In the right panel
we have required the strict condition that the thermal relic
abundance fall within the WMAP Preferred range. The probability
density immediately shifts to favor rather large masses for the LSP
on the order of 1~TeV. The bulk of the shape is accounted for by the
Higgs~LSP pattern which is the dominant pattern in DMM, representing  37\% of the total
models when this constraint is imposed. Patterns~mSP2 and~mSP2$'$
also have LSP masses that are heavily concentrated
near $m_{\chi^{0}_1} \simeq 1\TeV$. Most of the low-mass LSP models are those from mSP3, mSP3$'$ and~mSP6$'$ for which the
LSP is predominantly bino-like, while the high mass models will have
mixed neutralinos or predominantly higgsino LSPs. This has
large consequences for the direct and indirect detection of
neutralino dark matter~\cite{BirkedalHansen:2002am}.  As an example, according to this measure it is unlikely to have a wino-like
neutralino with a mass of 200~GeV or less which might explain the
excess in positrons observed by
PAMELA~\cite{Adriani:2008zr,Grajek:2008pg}.  Of course, one also must keep in mind that the strong dependence of our results on the dark matter constraints means that any modifications to the standard cosmological assumptions will drastically alter the outcome of this study (or any other analysis with dark matter constraints as a crucial ingredient).

\begin{figure}[!h]
\begin{center}
\begin{tabular}{cc}
\igraph{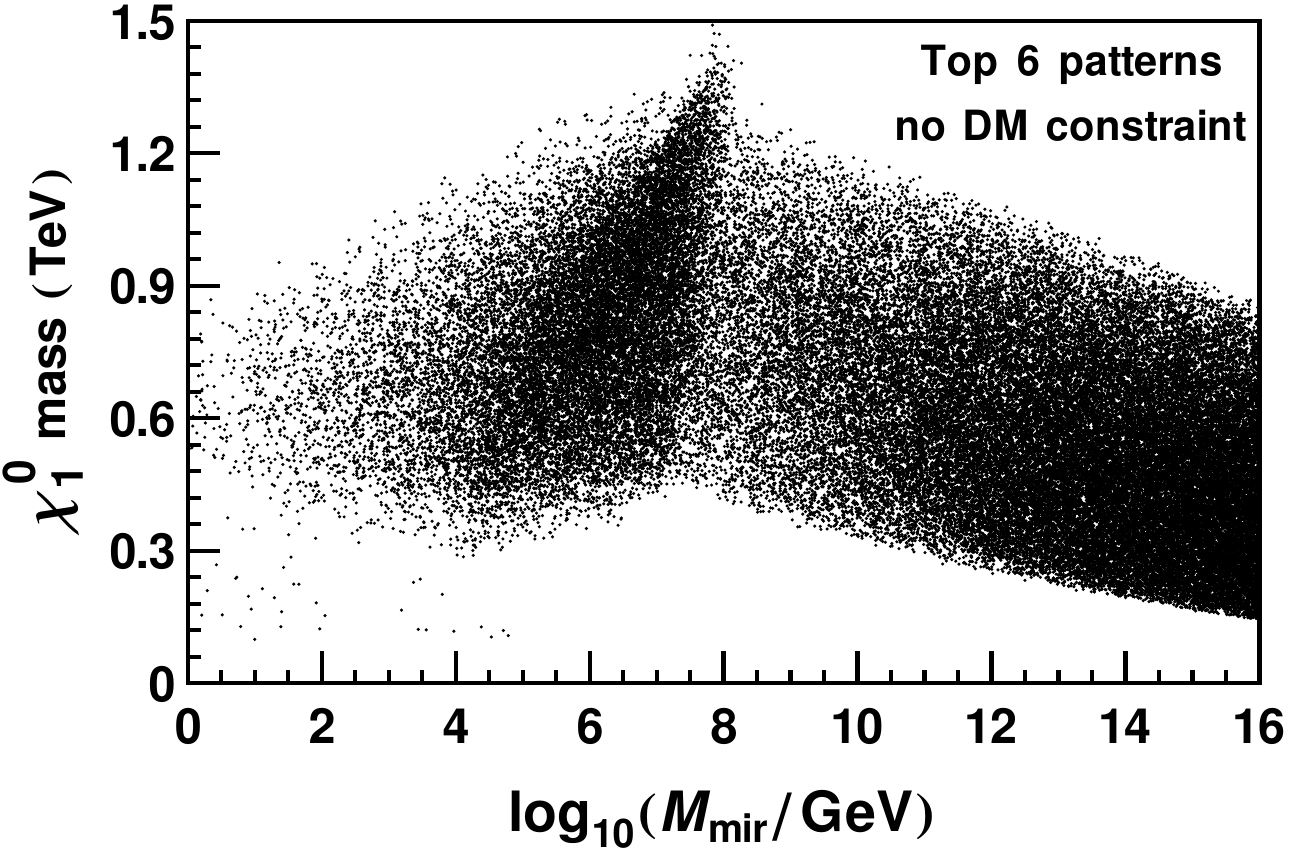} & \igraph{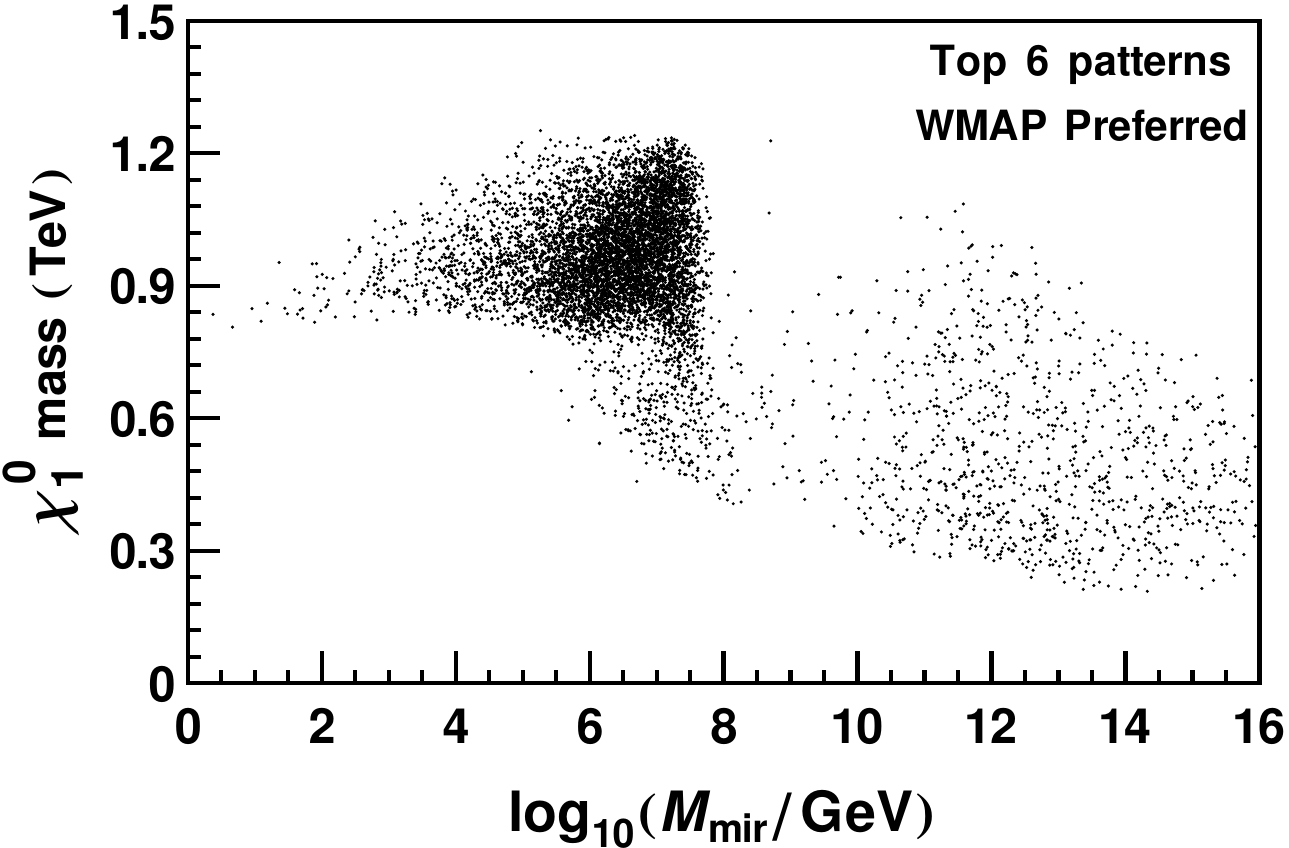}
\end{tabular}
\caption{\textbf{LSP mass versus mirage unification scale}. We show the distribution of the LSP mass as a function of the mirage unification scale for the top six hierarchy patterns in DMM models, with no dark matter constraint (left panel) and the WMAP Preferred
constraint (right panel).}
\end{center}
\label{fig:summary_mirscale}
\end{figure}

As shown in Section~\ref{sec:corr}, the LSP composition is strongly correlated with the mirage unification scale of Eq.~(\ref{miraged}). Hence, the imposition
of a tight dark matter density requirement will preferentially
select certain areas in the theoretical parameter space. This is
illustrated in Fig.~21 
for the combined
set of the top six  hierarchy patterns. Again, the left panel involves no
dark matter requirement at all, while the right panel imposes the
WMAP Preferred relic density constraint. While no strong
correlations are evident in the former case, the latter case shows
the clear transition at $M_{\rm mir} \simeq 10^8\,{\rm GeV}$ and the
absence of low-mass LSPs for lower mirage scales.
\begin{figure}[!h]
\begin{center}
\begin{tabular}{cc}
\igraph{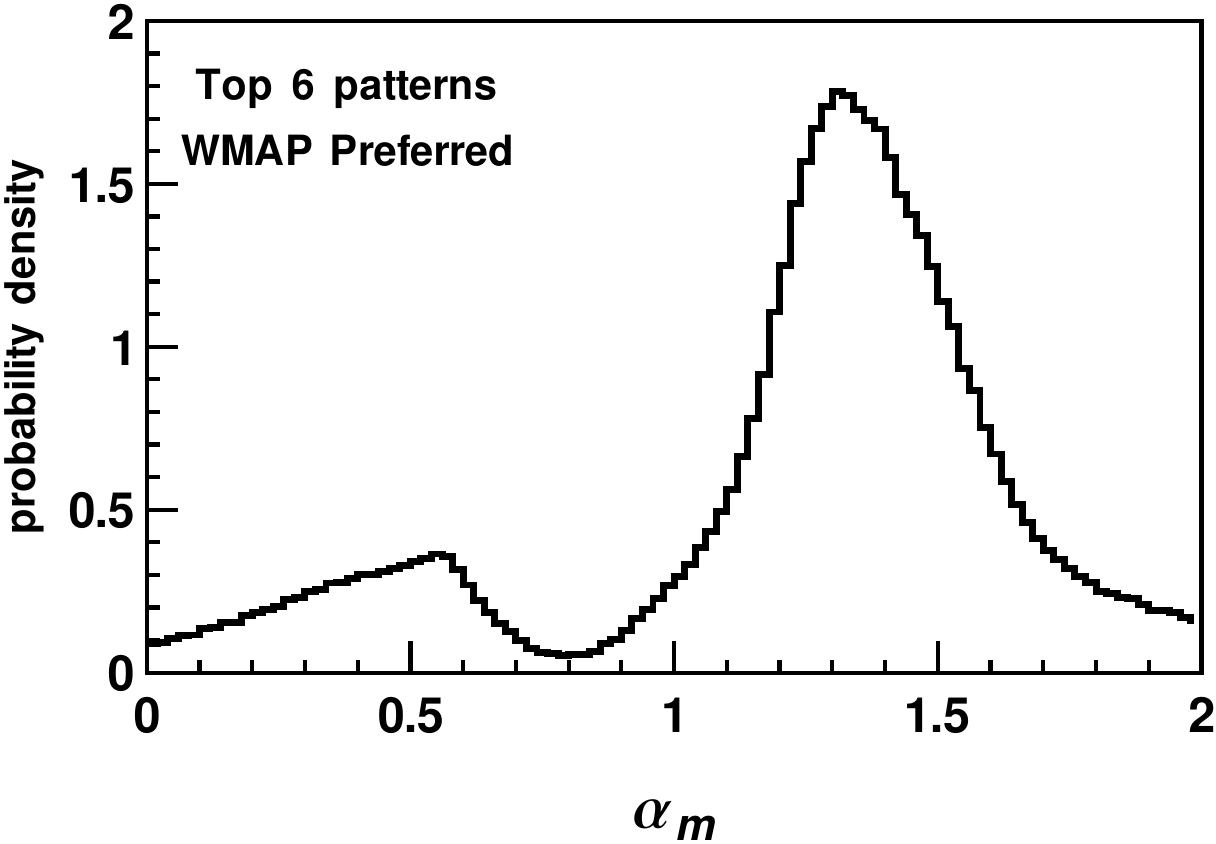} & \igraph{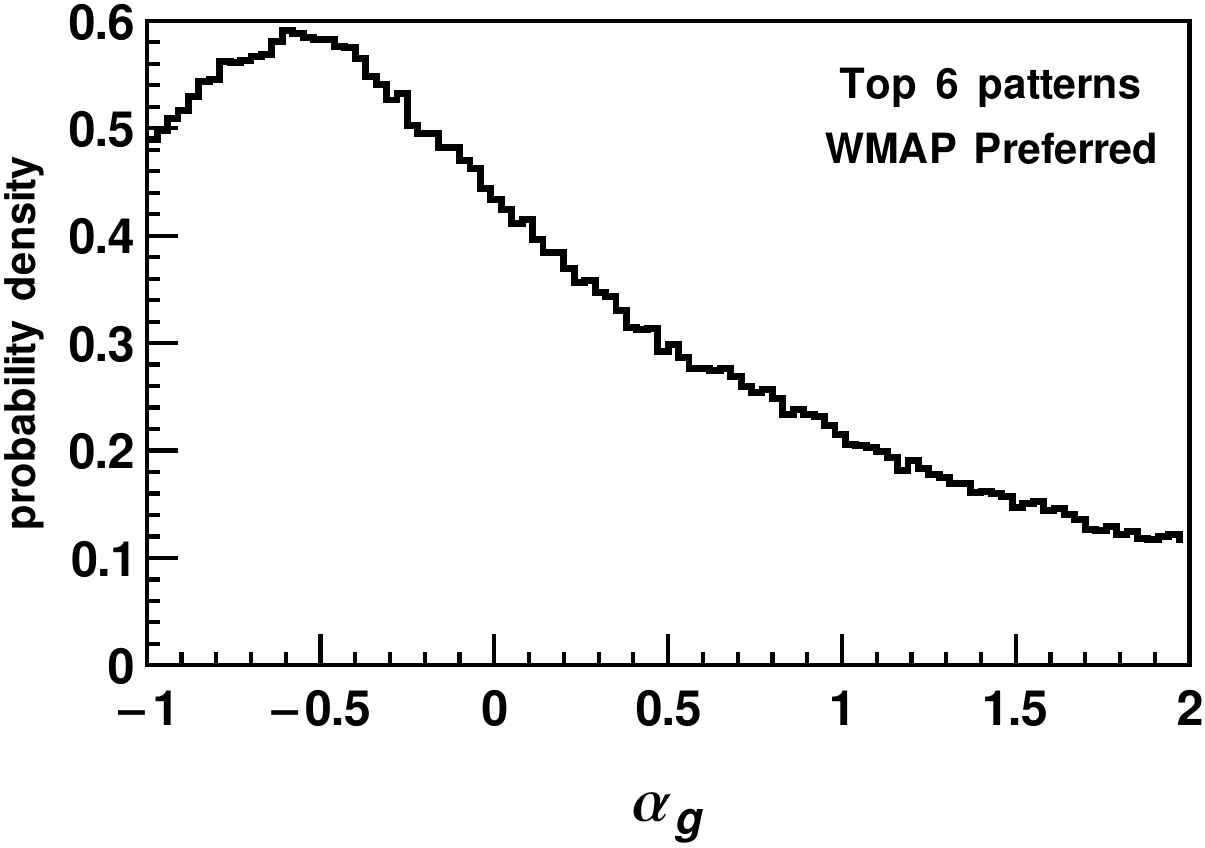}
\end{tabular}
\caption{\textbf{Distribution in input parameters $\alpha_m$ and
$\alpha_g$}. We show the distributions of $\alpha_m$ (left panel) and $\alpha_g$ (right panel) for the top six hierarchy patterns in DMM models with the WMAP Preferred
constraint.}
\end{center}
\label{fig:summary_alphas}
\end{figure}
The mirage scale itself is closely tied to the fundamental input
parameter $\alpha_m$ which measures the relative contributions of
anomaly and gravity mediation to the soft supersymmetry breaking.
This scale is much less sensitive to the value of $\alpha_g$
which compares the gauge mediation contribution to the anomaly mediation 
contribution. This is illustrated for our most prevalent hierarchy patterns
in Fig.~22 
The distribution in
$\alpha_m$ is the mirror image of the right panel in the previous
figure: higher mirage unification scales correspond to lower values
of $\alpha_m$, while the bulk of the models lie in the region
$1\lappeq \alpha_m \lappeq 2$ where the unification scale is below
$10^8$ GeV. It is interesting that one of the {\em least} favorable
regions for the model set is the KKLT-limit of $\alpha_m = 1$,
though lower values of $\alpha_m \sim 0.5$ which are typical of some
heterotic constructions~\cite{Gaillard:1999yb,Gaillard:2007jr}
continue to be viable. The overall distribution in the variable
$\alpha_g$ is rather broad, but clearly favors $\alpha_g <0$ once
the dark matter condition is imposed. This has model-building
implications: negative values  favor stabilization
mechanisms for the Standard Model singlet field $X$, which generates
masses for the gauge-charged messenger fields, which involve
higher-dimensional operators in the effective superpotential as
opposed to non-perturbative stabilization mechanisms.

The vast majority of viable DMM model points involve a mirage unification
scale below the (true) unification scale of $10^{16}\GeV$, which is where the gaugino masses unify in mSUGRA/CMSSM.
The lowering of the mirage unification scale from the true unification scale generally 
results in a more compressed spectrum in the gaugino sector which
distinguishes DMM models from mSUGRA/CMSSM models. The distinctive gaugino
mass ratios of mSUGRA/CMSSM will be perturbed and there arises a greater
possibility of mixed-composition lightest neutralino. Strategies for
measuring this deviation in the gaugino mass sector at the LHC have
been developed
elsewhere~\cite{Altunkaynak:2009tg,Altunkaynak:2010xe}. 
\begin{figure}[!h]
\begin{center}
\includegraphics[height=48mm]{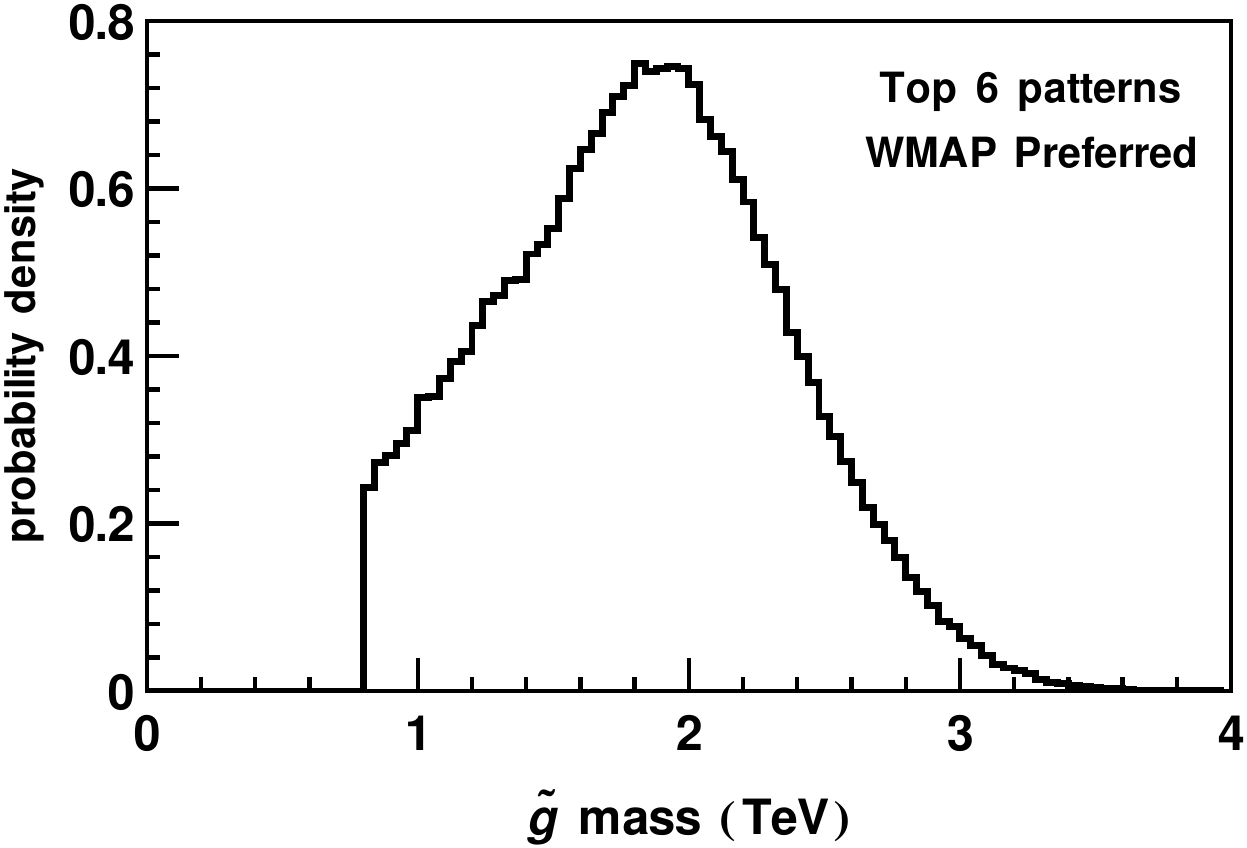}
\caption{\textbf{Distribution of gluino mass}. We show the distribution of the gluino mass for the top six hierarchy patterns in DMM models with the WMAP Preferred
constraint.}
\end{center}
\label{fig:gluino}
\end{figure}

A central question, however, is whether the relatively heavy yet compressed spectrum implied by
the large LSP mass as in our Higgs LSP benchmark point \#1 will be accessible at the LHC. In minimal
mSUGRA/CMSSM models one typically expects the physical gluino mass to be
5~to~6 times larger than the mass of the LSP. In the DMM framework,
this ratio can be significantly smaller, as seen in
Section~\ref{sec:corr}. The compression of the spectrum often
implies a gluino mass roughly a factor of two times larger than the
LSP mass, with gluinos and squarks comparable in mass. An LSP of mass on the order of
1~TeV in the mSUGRA/CMSSM limit would be a disaster for the LHC since even
1~ab$^{-1}$ of integrated luminosity at $\sqrt{s} = 14\TeV$ would
not be sufficient to discover a 5-6~TeV gluino~\cite{Baer:2009dn}.

Would such states in DMM be accessible at the LHC? The already impressive advances of the LHC in setting superpartner masses have not yet put a significant dent in the DMM parameter space considered here, as we have seen.  However, the typical gluino mass in DMM models is in the 1-2~TeV range, which is rapidly becoming more accessible.  For cases in which $m_{\tilde{q}} \simeq m_{\tilde{g}}$, as in the DMM paradigm, gluinos in the TeV range are detectable with 1-2~fb$^{-1}$ of integrated luminosity, even at $\sqrt{s} = 7\TeV$~\cite{Baer:2010tk,Altunkaynak:2010we}. Heavier gluinos in the range $2\TeV \lappeq m_{\tilde{g}} \lappeq 3\TeV$ are still
accessible at the LHC with 100~fb$^{-1}$ of data at $\sqrt{s} = 14\TeV$~\cite{Baer:2009dn}. As Fig.~23 
displays for our six most common hierarchy patterns, the gluino mass distribution peaks just below $m_{\tilde{g}} = 2\TeV$ and is well constrained to be below 3~TeV. This provides the exciting possibility that the rich structure of deflected mirage mediation could be probed at the LHC in the coming years.\\

The work of B.A. and B.D.N. is supported by the National Science Foundation Grant PHY-0653587.  L.E. and Y.R. are supported by the U.S. Department of Energy grant DE-FG-02-95ER40896.  The work of I.W.K. is supported by  the U.S. Department of Energy grants DE-FG-02-95ER40899 and  DE-FG-02-95ER40896.


\appendix
\section{DMM Parameters}
 The ${\theta_i}$ parameters, which appear in the mixed modulus-anomaly term in the soft scalar mass-squared parameters, are given by
\begin{eqnarray}
\theta_{Q,i} &=& \frac{16}{3} g_3^2 + 3 g_2^2 + \frac{1}{15} g_1^2
-2 ( y_t^2  + y_b^2) \delta_{i3},
\nonumber \\
\theta_{U,i} &=& \frac{16}{3} g_3^2 + \frac{16}{15} g_1^2
- 4 y_t^2 \delta_{i3},\;\; 
\theta_{D,i}=
\frac{16}{3} g_3^2 + \frac{4}{15} g_1^2
- 4 y_b^2  \delta_{i3}, \nonumber \\
\theta_{L,i} &=& 3 g_2^2 + \frac{3}{5} g_1^2
-2 y_\tau^2  \delta_{i3},\;\; 
\theta_{E,i} =
\frac{12}{5} g_1^2
- 4 y_\tau^2  \delta_{i3}, \nonumber \\
\theta_{H_u} &=& 3 g_2^2 + \frac{3}{5} g_1^2
- 6 y_t^2,\;\; 
\theta_{H_d} =
3 g_2^2 + \frac{3}{5} g_1^2
-6 y_b^2 -2 y_\tau^2.
\end{eqnarray}
The $\dot{\gamma}_i'$ parameters are:
\begin{eqnarray}
\dot\gamma_{Q,i}'
&=& \frac{8}{3} b_3^\prime g_3^4 + \frac{3}{2} b_2^\prime g_2^4 + \frac{1}{30} b_1^\prime g_1^4
- (y_t^2 b_t + y_b^2 b_b ) \delta_{i3} \nonumber \\
\dot\gamma_{U,i}'
&=& \frac{8}{3} b_3^\prime g_3^4 + \frac{8}{15} b_1^\prime g_1^4
- 2 y^2_t b_t \delta_{i3},\;\; 
\dot\gamma_{D,i}'=
\frac{8}{3} b_3^\prime g_3^4 + \frac{2}{15} b_1^\prime g_1^4
- 2 y^2_b b_b \delta_{i3} \nonumber \\
\dot\gamma_{L,i}'
&=& \frac{3}{2} b_2^\prime g_2^4 + \frac{3}{10} b_1^\prime g_1^4
 - y_\tau^2 b_\tau \delta_{i3},\;\; 
\dot\gamma_{E,i}' =
\frac{6}{5} b_1^\prime g_1^4
 - 2 y_\tau^2 b_\tau \delta_{i3} \nonumber \\
\dot\gamma_{H_u}'
&=& \frac{3}{2} b_2^\prime g_2^4 + \frac{3}{10} b_1^\prime g_1^4
 - 3 y^2_t b_t,\;\;
\dot\gamma_{H_d}' =
\frac{3}{2} b_2^\prime g_2^4 + \frac{3}{10} b_1^\prime g_1^4
 - 3 y_b^2 b_b - y^2_\tau b_\tau,  \label{dotgammaexp}
\end{eqnarray}
where $b_a^\prime=b_a+N$ ($a=1,2,3$), and $b_{3,2,1}=-3,1,33/5$ are the MSSM beta functions (our convention is that $b_a^\prime <0$ for asymptotically free gauge theories).

The threshold effects for the soft scalar mass-squared parameters (see Eq.~(\ref{msqrthresh})) are given by
\begin{eqnarray}
\Delta m^2_Q&=&M_0^2 \left (\frac{8}{3}g_3^4+\frac{3}{2}g_2^4+\frac{1}{30}g_1^4 \right) N \widetilde{\alpha}_m(1+\alpha_g)\nonumber \\
\Delta m^2_U&=&M_0^2 \left (\frac{8}{3}g_3^4+\frac{8}{15}g_1^4 \right ) N \widetilde{\alpha}_m(1+\alpha_g)\nonumber \\
\Delta m^2_D&=&M_0^2 \left (\frac{8}{3}g_3^4+\frac{2}{15}g_1^4 \right ) N \widetilde{\alpha}_m(1+\alpha_g)\nonumber \\
\Delta m^2_L&=&M_0^2 \left (\frac{3}{2}g_2^4+\frac{3}{10}g_1^4 \right ) N \widetilde{\alpha}_m(1+\alpha_g)=\Delta m^2_{H_u}=\Delta m^2_{H_d} \nonumber \\
\Delta m^2_E&=&M_0^2 \, \frac{6}{5}g_1^4  N \widetilde{\alpha}_m(1+\alpha_g),
\end{eqnarray}
where the gauge couplings are evaluated at the messenger scale, $M_{\rm mess}$.


\pagebreak

\end{document}